\documentclass{aa}  

\usepackage[utf8]{inputenc}
\usepackage{graphicx}
\usepackage{dcolumn}
\usepackage{lineno}
\usepackage{xtab,booktabs}
\usepackage{adjustbox}
\usepackage{todonotes}

\usepackage{xcolor}
\usepackage{bm}

\usepackage{color}
\usepackage{verbatim}
\usepackage{ulem}

\usepackage{amsmath}
\usepackage{hyperref}
\usepackage[hang]{footmisc}
\usepackage{subfig}

\def\nsurvs{13}
\def\nfilts{52}
\usepackage{txfonts}

\defcitealias{fragilistic}{Fragilistic}

\begin{document}

   \title{A Reassessment of the Pantheon+ and DES 5YR Calibration Uncertainties: Dovekie}
    \author{B.~Popovic\inst{\ref{ip2i}}\fnmsep\thanks{Corresponding author: \texttt{b.popovic@ip2i.in2p3.fr}}
    \and W.D. Kenworthy\inst{\ref{Sweden}}\fnmsep\thanks{Corresponding author: \texttt{darcy.kenworthy@fysik.su.se}}
    \and M. Ginolin\inst{\ref{ip2i}}
    \and A. Goobar\inst{\ref{Sweden}}
    \and P. Shah\inst{\ref{London}}
    \and B. M. Boyd\inst{\ref{Cambridge}}
    \and A. Do\inst{\ref{Cambridge}}
    \and D. Brout\inst{\ref{Boston}}
    \and D. Scolnic\inst{\ref{Duke}}
    \and M. Vincenzi\inst{\ref{Oxford}}
    \and S. Dhawan\inst{\ref{Birmingham}}
    \and D. O. Jones\inst{\ref{Hawaii}}
    \and M. Smith\inst{\ref{lancaster}}
    \and M. Rigault\inst{\ref{ip2i}}
    \and B. Racine\inst{\ref{Marseille}}
    \and E. E. Hayes\inst{\ref{Cambridge}}
    \and R. Chen\inst{\ref{Duke}}
    \and P. Wiseman\inst{\ref{Soton}}
    \and L. Galbany\inst{\ref{Spain}}
    \and M. Grayling\inst{\ref{Cambridge}}
    \and L. LaCroix\inst{\ref{ip2i}}
    \and C. Barjou-Delayre\inst{\ref{Clermont}}
    \and D. Kuhn\inst{\ref{Paris}}
    \and C. Lemon\inst{\ref{Sweden}}
    }
    \institute{
    Universite Claude Bernard Lyon 1, CNRS, IP2I Lyon / IN2P3, IMR 5822, F-69622 Villeurbanne, France
    \label{ip2i}
    \and
    The Oskar Klein Centre, Department of Physics, Stockholm University, SE - 106 91 Stockholm, Sweden \label{Sweden}
    \and
    Department of Physics \& Astronomy, University College London, Gower Street, London, WC1E 6BT, UK \label{London}
    \and
    Institute of Astronomy and Kavli Institute for Cosmology, Madingley Road, Cambridge CB3 0HA, UK \label{Cambridge}
    \and
    Departments of Astronomy and Physics, Boston University, Boston MA, 02215\label{Boston}  
    \and
    Department of Physics, Duke University Durham, NC 27708, USA \label{Duke}
    \and
    Department of Physics, University of Oxford, Denys Wilkinson Building, Keble Road, Oxford OX1 3RH, United Kingdom\label{Oxford}
    \and
    School of Physics and Astronomy \& Institute for Gravitational Wave Astronomy, University of Birmingham, Birmingham B15 2TT, UK\label{Birmingham}
    \and
    Institute for Astronomy, University of Hawai‘i, 640 N. Aohoku Pl., Hilo, HI 96720, USA \label{Hawaii}
    \and
    Department of Physics, Lancaster University, Lancs LA1 4YB, UK \label{lancaster}
    \and
    Aix Marseille Université, CNRS/IN2P3, CPPM, Marseille, France \label{Marseille}
    \and
    Institute of Space Sciences (ICE-CSIC), Campus UAB, Carrer de Can Magrans, s/n, E-08193 Barcelona, Spain.
    Institut d'Estudis Espacials de Catalunya (IEEC), 08860 Castelldefels (Barcelona), Spain \label{Spain}
    \and
    Université Clermont Auvergne, CNRS/IN2P3, LPCA, F-63000 Clermont-Ferrand, France \label{Clermont}
    \and
    School of Physics and Astronomy, University of Southampton, Southampton, SO17 1BJ, UK \label{Soton}
    \and
    Sorbonne Université, CNRS/IN2P3, LPNHE, F-75005, Paris, France \label{Paris}
    }
    \date{\today}

 
  \abstract
   {}
   {Type Ia Supernovae (SNe Ia) are crucial tools to measure the accelerating expansion of the universe, comprising thousands of SNe across multiple telescopes. Accurate measurements of cosmological parameters with SNe Ia require a robust understanding and cross-calibration of the telescopes and filters. A previous cross-calibration effort, `Fragilistic', provided 25 photometric systems, but offered no public code or ability to add new surveys. We provide an open-source cross-calibration solution with an eye to the future, available at https://github.com/bap37/Dovekie/ .}
   {Using the Pan-STARRs (PS1) and Gaia all-sky telescopes, we characterise the measured filters from 11 photometric systems, including CfA, PS1, Foundation, DES, CSP, SDSS, and SNLS, using published observations of field stars. For the first time, we derive uncertainties on effective filter transmissions and modify filters to  match the data. With the addition of direct observations of DA white dwarfs \citep{Boyd25}, we simultaneously cross-calibrate our zeropoints across photometric systems, and train a new SALT model to propagate to cosmology.}
   {With improved uncertainties from DA WDs, we find improvements to the calibration systematic uncertainty of $\times 1.5$ for the Pantheon+ \citep{Brout22} sample with a new systematic photometric uncertainty $\sigma_w(\rm{phot}) = 0.016$ for Flat$w$CDM, and modest improvements to that of the DES5YR analysis. We find good agreement with previous calibration, and show that even these small calibration changes can be amplified by up to a factor of $\times 6$ in the inferred SN\,Ia distances, driven by calibration sensitivity in the colour-luminosity relations and SALT training. Initial results indicate that these changes cause $d\mu/dz = 0.025$ across our redshift range, and change the recovered value of $\Omega_M$ given a LCDM parameterisation of $\sim0.01$. These may have a potentially larger impact in w0/wa space and inferences about evolving dark energy. We pursue this calculation in an ongoing full re-analysis of DES. }
   {}


    \titlerunning{Dovekie}
    \authorrunning{Popovic \& Kenworthy et al.}
   \maketitle
%

\section{Introduction}\label{sec:Introduction}

Type Ia supernovae (SNe\,Ia) are crucial in shedding light on the mysteries of the universe. Excellent measures of local and far distances, SNe Ia were the first method to discover the accelerating expansion of the universe \citep{Riess99, Perlmutter99}, and today are used for measuring the dark energy equation-of-state parameter $w$ \citep{Betoule14, Brout22, Amalgame, DES5YR, Rubin23}. Recent measurements by the DESI collaboration \citep{DESI_DR1}, in combination with these most recent analyses of supernova data, have shown discrepancy with the concordance cosmology model $\Lambda$CDM at significance varying from $2.8- 4.2 \sigma$. The calibration of the telescopes used to observe the SNe is quantified as one of the largest systematics in measuring cosmological parameters with SNe Ia. These calibration uncertainties depend on both the survey and the telescope filters, resulting in potentially redshift-dependent changes in the observed brightness of a SN Ia. Accordingly, developing our understanding of the calibration of SN Ia surveys is a critical step in both validating existing measurements and increasing precision for future measurements.

Competitive cosmology analyses with SNe\,Ia require  multiple surveys to increase both statistics and redshift range, combining multiple different photometric systems and potentially introducing additional calibration uncertainties. These unique telescopes, filters, and surveys must observe self-consistent brightnesses in order to measure cosmological parameters. Where SN measurements involve comparisons of photometry in similar or identical filters, calibration uncertainties tend to cancel; accordingly, measurements of $H_0$ have low sensitivity to calibration \citep{Dhawan2020,Brownsberger23Uncertainties}. However measurements of $w$ are \textit{a priori} particularly sensitive to calibration, as they involve comparing the brightness of $z\sim 1 $ SNe observed in the observer-frame blue to $z\sim 0.01$ SNe observed in the observer-frame red, requiring absolute colour calibration across thousands of angstroms. This sensitivity is amplified by a factor of $\sim 3$\footnote{Typical value for $\beta$, as used in the Tripp estimator}, as the colour of SNe is used to compensate for the effect of host-galaxy dust and/or intrinsic colour variation \citep{Tripp98}. 

The calibration systematic itself comes in two parts; the uncertainty in the absolute colour calibration of the \textit{HST} CALSPEC primary standards \citep{Bohlin14,Bohlin2020} used in all modern analyses, and the precision of the constraints that link each individual survey to CALSPEC. In this work, we are primarily focused on the latter process, tying together an ensemble of survey photometry to CALSPEC. In cosmology analyses, light curves from these surveys are used to first construct an empirical model that accounts for variation in SNe\,Ia, then fitted with that model to provide a reduced dataset for further analysis. Inconsistent calibration may lead to errors that accumulate in both stages, and consideration of the modelling stage is essential to capture the full sensitivity of measurements to calibration errors. Thanks to recent efforts by \cite{Taylor21,Kenworthy21}, NaCl, Osman, \textit{in prep.}, simultaneous calibration of the training and data sets is possible. Of the three most recent SN Ia analyses which measured $w$, the Fragilistic calibration (\citealp{fragilistic}; hereafter \citetalias{fragilistic}) was used in both the Pantheon+ \citep{Brout22} and DES5YR \citep{DES5YR} analyses. The Union3 analysis \citep{Rubin23} used their own calibration analysis with significant divergences in approach.

Improvements to measurements of $w$ in the future will require the use of new samples of thousands of supernovae from current and near-future surveys including the Zwicky Transient Facility \citep{Rigault24}, the Vera Rubin Telescope \citep{LSSTSRD}, and the \textit{Nancy Grace Roman Space Telescope} \citep{Hounsell2018}, which will need to be tied to existing samples or each other. To prepare for use of these surveys in cosmological analyses, and establish a consistent and open-source calibration solution, and validate existing measurements, we perform a new cross-calibration with 11 other surveys and the ability to add more in the future, building off of work from \citet{Scolnic15},\citet{Currie20}, and \citetalias{fragilistic}. Like the previous works, we use Pan-STARRS (PS1; \citealp{Chambers16}) photometry of tertiary stars to recalibrate photometric systems, alongside characterising telescope filter transformations of these photometric systems. Adding to previous works, we include observations from \textit{Gaia} \citep{GAIAONE, GAIATWO, GAIATHREE} and a novel model of DA white dwarfs \citep{narayan2019,axelrod2023,Boyd25}. Together with the PS1 photometry, Gaia spectroscopy, DA white dwarfs, and filter transformations, we simultaneously fit photometric magnitude zeropoints across all our telescopes and filters. The code, photometry, and filters when available, are all publicly released for future use at  \hyperlink{https://github.com/bap37/Dovekie}{https://github.com/bap37/Dovekie}. We then propagate the new recalibration through the light-curve model training, fitting, and cosmological inference.

Section \ref{sec:Data} provides an overview of the surveys and telescopes that are cross-calibrated in this work. Section \ref{sec:Method} presents the methodology we use to perform the cross-calibration. The calibration results themselves are presented in Section \ref{sec:CalResults}, which are then used to retrain a new suite of SALT surfaces and assess the impact on distances in Section \ref{sec:SALT}. We measure the estimated change to the inferred Pantheon+ cosmology in Section \ref{sec:PantheonPlus}, and the changes in photometric systematic uncertainties for the DES5YR analysis in Section \ref{sec:DES5YR}. Finally, Sections \ref{sec:Discussion} and \ref{sec:Conclusion} present the discussion and conclusions. We provide additional surveys, and an overview of the data therein, that are included in our software release and can be integrated into the Dovekie pipeline. 

\section{Data}\label{sec:Data}

\subsection{Calibration of Individual Surveys}\label{sec:Data:subsec:IndividualSurvey}

Modern analyses of cosmology with SNe\,Ia use the AB system \citep{Oke83, Fukugita96} for the absolute calibration of their photometric systems, moving away from the Vega flux standard \citep{Colina94} used in \cite{Riess98} and other early SN\,Ia analyses. The AB system defines a magnitude as 
\begin{equation}\label{eq:ABMag}
    m_{\rm AB} = -2.5 \times \log_{10}\frac{ \int (h\nu)^{-1}~p(\nu)~f(\nu)~d\nu }{ \int (h\nu)^{-1}~p(\nu)~f_0~d\nu },
\end{equation}
taking into account the energy transmission function of a given filter $p(\nu)$, the spectral flux density $f(\nu)$ of a given object in Jy, and the spectral flux density (constant in frequency) of a 0th magnitude celestial body $f_0 = 10^{-48.6/2.5} \textrm{ erg} \textrm{ s}^{-1} \textrm{ cm}^{-2} \textrm{ Hz}^{-1} \approx 3631 \textrm{ Jy}$. As these magnitudes are defined in energy units, while telescope observations are read out in analogue-to-digital counts from CCD's, the problem of calibration is to determine the response function between these quantities. The zero-point ${\rm ZP}^f_{\rm AB}$ of a photometric filter $f$ sets the fundamental scale, transforming between magnitudes and counts $N$ as
\begin{equation}
    m_{\rm AB}^f = -2.5 \log_{10}(N) + {\rm ZP}^f_{\rm AB}.
\end{equation}

For SN\,Ia cosmology, the absolute scale of calibration\footnote{e.g. A constant added to all zero-points of all filters under consideration} is not normally of interest as it is degenerate with the fiducial type Ia absolute magnitude $M_0$. However, the absolute colour calibration\footnote{e.g. Differences between zero-points as a function of wavelength} is a critical factor, as rest-frame comparison of low-$z$ and high-$z$ SNe requires comparing photometry in observer-frame bluer bands of high-$z$ objects against observer-frame redder bands measuring low-$z$ objects. The calibration path we take here is ultimately rooted in comparisons to the \textit{HST}/CALSPEC system \citep{Bohlin14,Bohlin21}, which determined the absolute colour calibration of \textit{HST} spectrophotometry by comparing observations to physically modelled white dwarf atmospheres, anchoring the absolute scale to Vega. 

As the CALSPEC primary standards are too bright to be observed by most survey telescopes under their normal observing strategies, these systems are typically anchored to secondary standards such as \cite{Landolt92} or \cite{Smith02} with catalogued values tied to CALSPEC. Observations of these secondaries are interleaved with observations of supernova fields; field stars are then used as tertiary standards to transfer calibration from the secondaries to each individual night of observations. Doing so typically involves comparisons of photometry measured using different filter bandpasses; such a comparison requires transformations from the to-be-calibrated natural system to that used in original measurements of the secondaries (e.g. $griz$ filters from PS1 to $UBVRI$ filters from \citet{Landolt92}). A linear transformation is typically given as
\begin{equation}\label{eq:Landolt}
    m_{{\textrm AB},f' } = m_{{\textrm{nat}},f } - C_f^{f'} \times c - \Delta_{\rm AB, f}^{f'}.
\end{equation}
The astronomical colour coefficient from filter $f$ to filter $f'$ is $C_f^{f'}$, which is multiplied by a colour $c$. This colour transformation accounts (at first order) for the differing shapes of the natural system photometric filters as compared to those originally used to measure the secondary standards. Zero-point offsets $\Delta_{\rm AB, f}^{f'}$ for each filter then translate from whichever calibration standard the natural system was set to the AB system. In the rest of this work, where this equation is used, the $f$ subscript will be dropped, as we simply use the filter in the PS1 system closest in wavelength to the calibrated filter $f'$.

Here we use comparisons of secondary standards from \citet{Boyd25} and tertiary standards from multiple surveys against the PanSTARRS DR2 and GaiaXP data releases with the goal of uniformly and simultaneously cross-calibrating all surveys to CALSPEC.

\subsection{Calibrator Stars}\label{sec:Data:subsec:Real}

Here we review the \nsurvs~different surveys and corresponding \nfilts~filters used in this calibration process. We provide an overview of the systems and filters for each survey in this subsection. The primary standard for all these surveys are the \textit{HST} CALSPEC standards from \citealp{Bohlin96,Bohlin21}. This spectral data was taken with NICMOS and STIS with an associated uncertainty of $\sim 1$ mmag/1000A from 3000A to 15000A \citep{Bohlin14}. Additional \textit{HST} spectrophotometry from \citet{Koleva12} is also used for filter characterization.

Following \cite{Currie20} and \citetalias{fragilistic}, the public PS1 photometric catalogue (DR2,  https://catalogs.mast.stsci.edu/panstarrs/) is used to cross-calibrate each survey. The PS1 photometric system is reported to be calibrated against CALSPEC with an accuracy of $\sim 5$mmag \citep{Schlafly16}. Additionally, we use GaiaXP \citep{GAIAONE,GAIATWO,GAIATHREE} and additional \textit{HST} observations of DA white dwarf stars \citep{narayan2019,axelrod2023} as alternative calibration paths.

A summary of the calibrator stars described in the following Sections is given in Tables \ref{tab:Photometries}, \ref{tab:Photometries2}, and \ref{tab:Photometries3}. 

\subsubsection{PS1 And Foundation}\label{sec:Data:subsec:Real:subsubsec:PS1}

We include stars from the PS1 SNe photometry and Foundation photometry data releases, and the previously mentioned public PS1 photometric catalogue, bringing the total number of photometric systems from the Pan-STARRS telescope to three. The zero-point of each system is treated independently, as the data reduction for each was conducted with divergent techniques.
 
\begin{itemize}
    \item PS1-Public: here we use the aperture magnitudes from the public PS1 DR2 as done in \cite{Currie20, fragilistic}. The aperture magnitudes are more robust to nonlinear effects than the PSF photometry used in SN\,Ia analyses \citep{Xiao23}; we make use of the aperture magnitudes to tie surveys together.
    \item PS1-SN: the original PS1 SN\,Ia images were calibrated with stars from \cite{Rest14, Scolnic14a}.
    \item PS1-Foundation: images for the Foundation SN Ia sample \citep{Foley18} were calibrated with a different set of stars than PS1-SN; we use these Foundation stars for Foundation.
\end{itemize}

\subsubsection{Center for Astrophysics (CfA3)}\label{sec:Data:subsec:Real:subsubsec:CfA3}

The CfA3 sample from \cite{Hicken09a} was taken across two telescopes at the F.L. Whipple Observatory using the 4Shooter and Keplercam cameras. We include observations with both telescopes. The Keplercam (CfA3K) $BVgri$ filters are published in \cite{Hicken09b}, and the 4Shooter (CfA3S) $BVRI$ filters are from \cite{Jha06}. \cite{Hicken12} states the CfA3K and CfA4p1 colour coefficients are reasonably consistent such that the unsubmitted C. Cramer et al. (2025, in preparation) CfA4p1 transmission functions can be used with the CfA3K natural light curves. As these transmission functions do not account for atmospheric transmission, we add the same MODTRAN atmospheric transmission \citep{StubbsTonry12} used in \citetalias{fragilistic}, with water vapour, aerosols, and an airmass of 1.2.

There are alternative transmission functions for 4Shooter and Keplercam that appear in the SNDATA\_ROOT environment \cite{SNDATA_ROOT} which appear to exactly match the $BVRI$ filters from \texttt{SNCosmo} \cite{SNCosmo}, but the latter attributes these functions to \cite{Betoule14} who neither provide copies of the filters for verification, nor describe the modifications made to the original functions.
For robust provenance, we choose to use the original functions, explicitly converting the CfA3S $BVRI$ filters from \cite{Jha06} to the photon-counting convention and applying a known atmospheric model. The original functions and our modifications are included in the Dovekie github repository.


\subsubsection{Carnegie Supernova Project (CSP)}\label{sec:Data:subsec:Real:subsubsec:CSP}

We make use of the CSP DR3 from \cite{Krisciunas17} with updates to the filter transmissions from \cite{Krisciunas20}. The DR3 includes tertiary stars used to calibrate the 134 SNe Ia in the CSP sample, with observations in the $ugriBV$ bands. We apply a calibration transformation to bring these magnitudes into the AB system. 

\subsubsection{SDSS, SNLS, DES}\label{sec:Data:subsec:Real:subsubsec:AB}

We use the updated CALSPEC offsets from \citetalias{fragilistic} as our starting point for the Sloan Digital Sky Survey (SDSS), Supernova Legacy Survey (SNLS), and Dark Energy Survey (DES). Following \cite{Betoule14}, we apply the same changes to the SDSS and SNLS offsets. We increase the sample size for the DES5YR photometry from \cite{Rykoff23}, and leave the DES3YR sample from \citetalias{fragilistic} untouched. 

\subsection{Synthetic Stars}\label{sec:Data:subsec:Synths}

Spectral measurements of flux standards are used in this analysis to characterise the filters of our photometric systems. Two libraries of synthetic spectra are used; the \textit{HST} CALSPEC library\footnote{https://archive.stsci.edu/hlsps/reference-atlases/cdbs/}, and the NGSL2 library\footnote{https://archive.stsci.edu/prepds/stisngsl/}. There is only one release of the NGSL2 library \citep{Koleva12}, but the 370 stars in the library are more extensive than the more-frequently updated CALSPEC library \citep{Bohlin21} at 64 stars. We use both libraries with equal weight in the fitting process.

We do not update the CALSPEC calibration for our surveys; there have been no major updates to CALSPEC standards since the release of \citetalias{fragilistic}. We therefore make use of the CALSPEC offsets from \citetalias{fragilistic}, accounting for internal changes to the CALSPEC standards \textit{before} cross calibration.

\subsection{White Dwarf Standards}
\label{sec:Data:subsec:WDs}
DA white dwarfs have pure-hydrogen atmospheres and can be modelled precisely as black bodies with distinct Balmer absorption lines in the optical wavelengths.
\cite{narayan2019} and \cite{axelrod2023} established an all-sky network of 32 dim DA white dwarfs (16.5 < $V$ < 19.5) as secondary standards that tied to the CALSPEC system \citep{bohlin2014a,bohlin2014b}.  As these standards are dimmer than the CALSPEC primaries, they can be directly observed by many wide field surveys in their survey configurations without saturating the detectors. Intrinsic and dust parameters were constrained for each white dwarf using six \textit{HST}/WFC3 photometric bands \citep{calamida2019} ranging from the UV to near-infrared wavelengths, as well as optical spectra from a variety of ground-based sources. Bayesian modelling by \cite{narayan2019} and \cite{axelrod2023} allowed synthetic SEDs to reproduce the whole network to sub-percent precision, with panchromatic photometric residuals consistent with 4.9 mmag RMS.

\citet{Boyd25} extended the DA white dwarf spectrophotometric analysis by jointly modeling the whole network simultaneously using hierarchical Bayesian modeling. This allowed the work to jointly infer changing zero-point systematics across \textit{HST} cycles, as well as the F160W count-rate nonlinearity. Further precision in this work came from using an updated intrinsic template with improved UV modeling and coverage with \textit{HST}/STIS \citep{Bohlin25}. The resulting SEDs yielded a photometric precision with residuals of 3.9 mmag RMS across the six \textit{HST} bands and 3.3 mmag RMS in the optical alone.

We use the \citet{Boyd25} white dwarf model SEDs in our work. The model infers posteriors over several parameters per white dwarf including a distance modulus $\mu$, effective temperature $T_{\text{eff}}$, natural logarithm of surface gravity $\log g$, V band extinction $A_\text{V}$ and dust law $R_\text{V}$. 100 posterior samples over each of these parameters were used to forward model 100 synthetic SEDs for each object. 

We query our catalogs of stellar calibrators from each survey to find photometric observations of these standards. We find matching photometry in these catalogs from PS1, DES, and SDSS, and accordingly include them in our data. 

\subsection{Gaia Spectrophotometry}\label{sec:Data:subsec:Gaia}

To facilitate a cross-calibration method independent of PS1, we make use of Gaia spectra \citep{GAIAONE, GAIATWO, GAIATHREE} where available. We use the \href{https://gaia-dpci.github.io/GaiaXPy-website/}{GaiaXPy} software to download spectra of stars that are available; these spectra are then integrated through the filters to provide synthetic photometry. We go into more detail in Section \ref{sec:Method}.

\subsection{Data Preparation}\label{sec:Data:subsec:Prep}

The PS1 aperture magnitudes (\texttt{MeanApMag}) for tertiary standards taken from each survey are queried via the PS1 DR2. We match on RA and DEC to $<1$ arcsecond, and choose only isolated stars; those with no other stars within a 15 arcsecond radius. Each star is corrected for any potential Milky Way extinction via IRSA\footnote{https://irsa.ipac.caltech.edu/} using data from \cite{Schlafly11}, and the corrections are applied at the mean wavelength of each filter. As in Fragilistic, we ensure linear responses within the PS1 system, by making the following magnitude cuts: $g>14.8, r>14.9, i>15.1, z>14.6$, and we also require that PS1 $g<19$ magnitudes to avoid Malmquist bias. Finally, we constrict our colour range to $(0.25 < g - i < 1.0)$.

\section{Calibration Method}\label{sec:Method}

The Dovekie codebase works in multiple stages. 
\begin{itemize}
    \item Use libraries of spectral standards,  integrating these spectra through the bandpasses of the desired photometric system (e.g, DES) and PS1 aperture photometry, to determine synthetic transformations of the PS1 system.
    \item Acquire PS1 aperture photometry of the tertiary standards in the desired photometric system, as well as Gaia spectroscopy, which is integrated through the desired photometric system bandpasses.
    \item Compare the colour-magnitude relationship $C^{b'}$ between data (tertiary standards) and spectral standards. Additionally, we simulate the data to determine any bias present and correct accordingly.  If necessary, bandpasses with $>3\sigma$ disagreement between spectral standards and data are modified to best match the data.
    \item The bandpass functions and colour slopes are fixed, and the zero-point offsets necessary to align the spectral standards and real data of all photometric systems are sampled over using an MCMC to derive an estimate of the mean offsets and associated covariance.
\end{itemize}

\subsection{Summary of Differences Between Dovekie and Fragilistic}

We list a brief, high-level overview of the changes between Dovekie and Fragilistic systems here.

\begin{itemize}
    \item \textbf{System Changes:} The original  \citetalias{fragilistic} code has been substantively rewritten, with improvements to robustness, speed, and data handling. A number of system-level changes were made to the software, including a new optimiser and updated error calculations.
    \item \textbf{Open Source:} The Dovekie code is publicly available, and structured to make the addition of new surveys simple. 
    \item \textbf{DA White Dwarfs:} \citetalias{fragilistic} used only PS1 aperture photometry for their cross-calibration. Here, we add DA white dwarfs from \cite{Boyd25}, opening an entirely new method of ZP calculation. 
    \item \textbf{Gaia:} We add Gaia spectroscopy for an additional constraint on filter characterisation.
    \item \textbf{Filter Uncertainty Calculations:} \citetalias{fragilistic} used quoted $1\sigma$ uncertainties for the filters in their analysis; here, we derive a new and self-consistent method of estimating filter uncertainties.
    \item \textbf{Novel Filter Changes:} We introduce a new method of weighting the transmission function of a filter by its effective wavelength as a potential change to the filter to ensure better agreement between synthetic standards and data. 
    \item \textbf{Simulation and Validation:} We simulate our data and apply our inference code to check the ability of our code to recover the ground truth.
\end{itemize}

\subsection{Filter characterisation}\label{sec:Method:subsec:Filters}

Due to the complexity of the simultaneous fit of the colour slope alongside the magnitude offsets, in the second stage of the Dovekie pipeline we ensure that our synthetic colour-magnitude slopes, $C^{b'}_S$, correspond to the slopes measured in the data. As in \citetalias{fragilistic}, we fix our colour slopes before fitting the zero-point offsets; however, we improve on this previous work by making use of two sources of data for filter characterisation, Gaia spectrophotometry and PS1 observations of tertiary standards.

\subsubsection{Filter constraints from PS1}

\begin{figure}
    \centering
    \subfloat[\centering Dovekie determines filter compliance with published field stars, filters, and multiple synthetic libraries. Here, we present a diagnostic plot for the DES $r$-band, using PS1 as our reference survey. The blue and gold points are the CALSPEC 23 and NGSLV2 libraries respectively, and the blue and gold lines are the associated slopes. Real data is shown in black. ]{{\includegraphics[width=7cm]{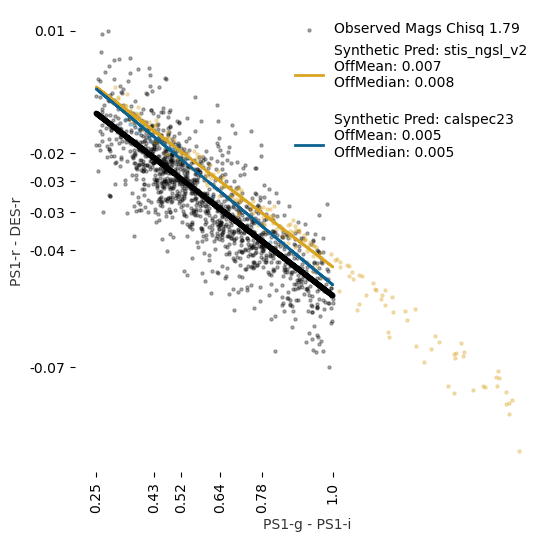}\label{fig:Dovekie-Process}}}
    \qquad
    \subfloat[\centering The filters are varied from $-50$\r{A} to $+50$\r{A} from their original central wavelength, using both PS1 and Gaia, and combined with a 50\r{A} prior to obtain the above posteriors.]{{\includegraphics[width=7cm]{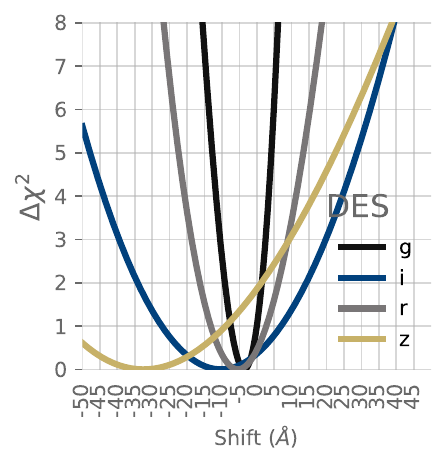} }}%
    \caption{A visual overview of the filter characterisation method.}
\end{figure}

For each filter $b'$ in a cross-calibrated surveys $S$, we choose the filter in PS1 public (e.g. the PS1 aperture magnitude) with closest central wavelength to serve as a reference $b1$. We also choose two filters in PS1 to serve as a colour index, $b2$ and $b3$. Generally, these colour indices were chosen as $\textrm{PS1-}g\ -\ \textrm{PS1-}i$, since this colour gave the a larger range of values than $g-r$ without using the noisier $z$ band values. However for the bluest filter of each survey we instead used $\textrm{PS1-}g\ -\ \textrm{PS1-}r$ to be more representative of the bluer end of the spectrum. We regress the residuals of the synthetic photometry between the  filter to be calibrated and the reference filter against the colour index, with intercept  ${ \Delta}_{\rm Int}^{b'}$ and slope $C^{b'}_S$,

\begin{align} \label{eq:MagDiff1}
    (\textrm{Synth}_{{\rm PS1}^{b1}} )   - (&\textrm{Synth}_{S_{b'}}  ) = \quad { \Delta}_{\rm Int}^{b'} +\\ &
    C^{b'}_S \times ((\textrm{Synth}_{{\rm PS1}^{b2}}-(\textrm{Synth}_{{\rm PS1}^{b3}} ))+ \epsilon \nonumber,
\end{align}
where $\epsilon$ is variation, assumed to be normal. The intercept  ${ \Delta}_{\rm Int}^{b'}$ represents the intrinsic differences between filters for a zero colour star, while the colour slope $C^{b'}_S$ accounts for the different transmission function shapes between them. As this is purely synthetic photometry, the $\epsilon$  term does not represent observational noise, but variation within the filter band-pass that doesn't extend across the spectral energy distribution. This linear regression is shown in Figure \ref{fig:Dovekie-Process} with DES$-r$ band as an example.

The observed data is somewhat more complicated to model in the presence of observational noise, cuts, outliers, and other sources of potential bias.  We use linear regression again, but iterate the fit up to three times, clipping $3\sigma$ outliers on each iteration\footnote{In practice, all fits on the data used here terminate after one reiteration.}. The model for linear regression then is 

\begin{align} \label{eq:obsmagdiff}
    (\textrm{Obs}_{{\rm PS1}^{b1}} )   - (&\textrm{Obs}_{S_{b'}}  ) = \quad { \Delta}_{\rm Int}^{b'} +\\ &
    \mathcal{C}^{b'}_O \times ((\textrm{Obs}_{{\rm PS1}^{b2}}-(\textrm{Obs}_{{\rm PS1}^{b3}} ))+ \epsilon \nonumber,
\end{align}
where $\epsilon$ includes spectral and observational variation. We can then compare the slopes derived from observed and synthetic data. In order to do so, we compare the difference between observed and synthetic slopes $\mathcal{C}^{b'}_S- {C}^{b'}_O $ to the mean and standard deviation of the same quantity as measured from the simulations discussed in Section \ref{sec:Method:sims}. From this, we derive a likelihood relative to the null hypothesis that the nominal bandpass correctly captures the spectral dependence of the observed magnitudes.

\subsubsection{Filter constraints from Gaia}
We compute synthetic photometry in the desired systems from Gaia spectrophotometry in each bandpass, denoting the integrated photometry as $\textrm{Obs}_{{\rm Gaia}^{b'}}$. We use a similar technique to that described above, regressing the Gaia synthetic photometry against observed photometry:

\begin{align} \label{eq:obsmagdiffgaia}
    (\textrm{Obs}_{{\rm Gaia}^{b'}} )   - (&\textrm{Obs}_{S_{b'}}  ) = \quad { \Delta}_{\rm Int}^{b'} +\\ &
    \mathcal{C}^{b'}_S \times ((\textrm{Obs}_{{\rm Gaia}^{\textrm{PS1-}g}}-(\textrm{Obs}_{{\rm Gaia}^{\textrm{PS1-}i}} ))+ \epsilon \nonumber.
\end{align}
However, in this case, as the null hypothesis is $ \{{ \Delta}_{\rm Int}^{b'}, \mathcal{C}^{b'}_S \}=0 $, and the dependent and independent variables regressed are uncorrelated, the simulation procedure described in Section \ref{sec:Method:sims} is unnecessary. We calculate the likelihood simply using the inferred slope and associated uncertainty of ordinary least-squares.

 \subsubsection{Joint constraints}

We modify our bandpasses in two ways. The first mirrors \citetalias{fragilistic}, shifting the wavelengths of the filter bandpass function by an amount $\lambda_{\rm Shift}$ such that
\begin{align}
    f_{\rm new}(\lambda) = f_{\rm old} (\lambda-\lambda_{\rm Shift}).
\end{align}
The second method of modifying the transmission function, absent from \citetalias{fragilistic}, is weighting the transmission by the effective wavelength of the filter as
\begin{equation}\label{eq:bandpassweighting}
    f_{\rm new}(\lambda) = f_{\rm old}(\lambda) * (\lambda_{\rm eff} / \lambda)^X,
\end{equation}
where $f$ is the transmission efficiency at a given wavelength $\lambda$, $\lambda_{\rm eff}$ is the effective wavelength of the filter, and $X$ is a factor of either $-1$ or 1, corresponding to a change between energy and photon counting units. We employ this methodology as an alternative method to the conventional filter shifts where the data lead us to believe a particular file's units may have been mislabelled. 

We use both Gaia and PS1 to measure the necessary $\lambda_{\rm Shift}$ values to align the colour-magnitude slopes of our filters with that of the data. For each filter, we compute slopes with shifts ranging from $-50$ to $50$\r{A}. The likelihoods of both reference systems for a given wavelength shift $\lambda_i$ are added together with a prior centred on zero with width $50$\r{A}
\begin{equation}
    \mathcal{L} = \sum^{{\rm Gaia, PS1}}_L \left( \frac{C^{b'}_O - C^{b'}_S(\lambda_i) - \Delta_C^{b'} }{\varsigma_{b'}} \right)^2 + \frac{\lambda_i^2}{(50\ {\AA})^2}.
\end{equation}
\noindent In the case of the PS1 constraint, the $\Delta_C^{b'}$ and $\varsigma_{b'}$ terms are the bias and scatter in the slope difference determined by running the analysis code on the simulated data, discussed in Section \ref{sec:Method:sims}. For Gaia, where the regression is simpler, the $\Delta_C^{b'}$ is 0 and $\varsigma_{b'}$ is taken as the slope uncertainty given by ordinary least-squares. In the case of $g-$bands across all of our surveys, we disregard the Gaia contribution, following advice from E. Rykoff (\textit{private communication}) who found discrepancies between Gaia spectra and PS1 and DES observations in the $g-$band. For all other bands, we weigh PS1 and Gaia equally, and shift bands until both the Gaia and PS1 determinations agree within 3 sigma. An example of this process is given in Figure \ref{fig:Dovekie-Process}. Where these quantities diverge by $3\sigma$ or more, we shift the transmission function of the filter in wavelength by the amount required to match the slope difference determined by simulations. We use the width of the posteriors to derive $1\sigma$ uncertainties on $\lambda_{\rm Shift}$ for each filter.

\subsection{Simulation of Stars for Bias Corrections} \label{sec:Method:sims}

The core of a cross-calibration analysis is the requirement to derive colour slopes between multiple filters of different wavelength responses. However, linear regression on data with noise in the independent variable, outliers, and/or cuts on the data is a biased estimator. We use simulated stars that mimic the distribution of observed data in order to estimate and correct the bias in the Dovekie pipeline. Each simulation consists of synthetically generated magnitudes of tertiary standard stars that mimic the distribution of PS1 and the desired system photometry. 

Each simulation is based on one of the two spectral libraries, NGSL2 or CALSPEC. All spectra from a given library with a PS1 colour $0<g-r<0.8$ are fitted to a dimensionality reduction model. We estimate the synthetic photometry in a given band from the $i$th stellar standard to be

\begin{align}
    {\rm Pred}^{b'}_{i}&= m_i + \alpha_{b'} \cdot c_i + \beta_{b'},
\end{align}
where $b'$ is an index which runs over both PS1 and the desired system filters. The reduced parameters $m_i, c_i$ are a magnitude and a colour parameter. $\alpha_{b'}$ are slope parameters that relate the filter responses in a given system to $c_i$. $\beta_{b'}$ is the mean stellar colour in each band. While similar in form to the colour transformations and zero-points of Equation \ref{eq:Landolt}, these terms model the stellar populations of a library, rather than characterizing a photometric system. To address degeneracies in this construction, the first and second $\alpha_{b'}$ values in each system are fixed to 0 and -1 respectively, and the first $\beta_{b'}$ is fixed to 0. We assume that the synthetic photometry are normally distributed about these predicted values with likelihood profile
\begin{align}
        \mathcal{L}_{\rm Synth} = \prod_{b'}^{\rm All\ filters} \prod_{i}^N  & (2\pi\sigma_{b'}^2)^{(-1/2)} \exp \left( \frac{-({\rm Synth}^{b'}_i-{\rm Pred}^{b'}_i)^2}{2 \sigma_{b'}^2} \right),
\end{align}
where the parameters $\sigma_{b'}$ represents the variation in standard spectra within each band. We maximise this likelihood to derive estimated colour responses and mean colours $\alpha_{b'}$ and $\beta_{b'}$, as well as nuisance parameters $\sigma_{b'}, m_i, c_i$. 

In order to construct simulated data, we then need to estimate the observational noise in each photometric quantity among the calibrator stars from each of our surveys. We use the same dimensionality reduction model on the observed data. However, while outliers are clipped in the analysis code (see Section \ref{sec:Method:subsec:Filters}), we are still required to generate outliers in the simulated data to mimic the actual data as accurately as possible for testing. We use a Gaussian mixture model to include these objects in the test set. This model has likelihood (omitting a constant factor of $2\pi$) defined 

\begin{align}
        \mathcal{L}_{\rm Obs} = \prod_{b'}^{\rm All\ filters} \prod_{i}^N  &(1-f_{\rm out})  \cdot  \sigma_{b'}^{-1} \exp \left( \frac{({\rm Obs}^{b'}_i-{\rm Pred}^{b'}_i)^2}{2 \sigma_{b'}^2} \right) \nonumber \\ &+ f_{\rm out}  \cdot   \Sigma^{-1}  \exp \left( \frac{({\rm Obs}^{b'}_i-{\rm Pred}^{b'}_i)^2}{2 \Sigma^2} \right),
\end{align}
where the parameters $\sigma_{b'}$ represent the variance in the population from within-filter spectral variance and photometric observation and the hyperparameter $f_{\rm out}$ represents the probability a star will be an outlier of the distribution. The width of the outlier distribution $\Sigma$ is filter independent, estimated manually for each survey to match the width of the observed outlier distribution. The mixture model likelihood is applied to the observational data and maximised to determine $\sigma_{b'}, f_{\rm out}$ for each survey. We found fitting for the width of the outlier distribution simultaneously with the other parameters lead to difficulties identifying the inlier and outlier distributions, so this parameter was fixed during this fit. Additionally, for each survey we fit an Ex-Gaussian distribution\footnote{A skewed distribution defined as the convolution of a Gaussian distribution with mean $\mu$ and dispersion $\sigma$ with a exponentially distributed variable with scale $\lambda$.} to the colour parameter distribution, labelling this distribution $\mathcal{C}_S$. The magnitude distribution is less significant, and we simply use a Gaussian kernel density estimator on the observed magnitudes to construct a magnitude distribution $\mathcal{M}$.

Using the slopes $\alpha_{b'}$ and $\beta_{b'}$ from the synthetic photometry and variance $\sigma_{b'}$ and outlier fraction $f_{\rm out}$ from the observed photometry, we can then generate simulated photometry in the PS1 and the desired photometric systems as

\begin{align}
    {\rm Simulated}_{b'}^{i}&= \alpha_{b'} \cdot c_i + \beta_{b'} + \Delta_b' + \epsilon \\
     c_i & \sim \mathcal{C} \nonumber\\
     m_i & \sim \mathcal{M} \nonumber\\
    \epsilon &\sim  \begin{cases} 
      N(0,\sigma_{b'})  &  U(0,1)>f_{\rm out} \\
      N(0,\Sigma)  &  U(0,1) \leq f_{\rm out}   \end{cases}\nonumber,
\end{align}
with the normal distribution $N(\mu,\sigma)$ and uniform distribution $U({\rm lower}, {\rm upper})$. $\Delta_b'$ are offsets randomly generated according to the priors we have on each system (see \ref{sec:Method:Prior}). 

We additionally generate simulated catalogue photometry for each white dwarf standard; in each simulation we select one of the sampled SEDs from \citet{Boyd25} at random, integrate them across the desired system filters, apply the generated offset, and generate magnitudes with noise given by the nominal uncertainties of the catalogue.

Generating data according  to the recipe above, we can then feed this data, with known input colour slopes and filter offsets, using the rest of the Dovekie pipeline to check for bias in the recovered parameters. We directly correct for slope biases found through this pipeline when considering if  wavelength shifts are necessary for each filter, but don't apply any derived zero-point offset biases. We generate 100 simulations for use in null-tests and bias correction.

\subsection{Offsets}\label{sec:Method:subsec:Actual}

With the colour slopes fixed from the previous step, we constrain the zero-point offsets required to match all photometric systems. We use the No U-Turn Sampler (NUTS, \citealp{NUTS}) to sample over the zero-point offset parameters for all surveys and bands simultaneously. The PS1-Public aperture magnitudes are used to tie together the zero-point offsets of our disparate surveys. We aim to sample over these magnitude offsets $\Delta_b$, of which there are \nfilts, including the 4 PS1-Public filters and associated offsets. This number is smaller than \citetalias{fragilistic}; we have not included miscellaneous low-redshift surveys such as SOUSA and LOSS in the fiducial Dovekie results\footnote{We do, however, include these surveys in the github, and a full re-calibration can be done with these samples included.}. The dependence on PS1 means that each system is correlated with PS1, and leads to covariance between the offsets. We define a three term posterior 

\begin{equation}\label{eq:chi2}
    \log p(\Vec{\Delta})= \log \mathcal{L}_{\rm WD}+ \log \mathcal{L}_{\rm Tertiary} + \log \mathcal{\pi}_{\rm Prior} .
\end{equation}
\subsubsection{Priors} \label{sec:Method:Prior}
For a given survey $S$, the prior on the system offsets are a zero-mean normal distribution with standard deviation $\sigma_{S}$. Thus, the prior PDF is 

\begin{equation}
    \log \mathcal{\pi}_{\rm Prior} = {\rm constant} - \sum_{S, b} \frac{\Delta_b^2}{2\sigma_{S}^2}.
\end{equation}
Table \ref{tab:priors} shows the summary of our calibration priors, which reflect our confidence on the original calibration of the system. All priors are centered at 0.

\subsubsection{White Dwarf Standards}

As discussed in Section \ref{sec:Data:subsec:WDs}, some of our surveys have observed faint white dwarf standards, allowing us to directly constrain the corresponding survey offsets. Measurements of the white dwarf standards are extracted from each of the catalogues of stellar calibrators (Section \ref{sec:Data:subsec:Real}) where available. Of the surveys cross-calibrated here, we have stellar photometry of the white dwarfs from PS1-Public, DES, and SDSS.

As we previously discussed, \citet{Boyd25} have produced multiple realisations of SEDs for each white dwarf modelled, that sample over uncertainty in astrophysical modelling and the Hubble instrumentation. We compute synthetic photometry using each sampled white dwarf SED for DES, PS1, and SDSS filters. Over all realisations of each SED, we take a mean of the synthetic magnitudes of the standard labelled $k$ in filter $b$ ${{\rm WD Synth}_{{\rm Surv}^b} }^k$. We compute the covariance over all filters for each object $\mathbf{\Sigma}_{b1,b2}^ {{\rm WD },k}$. To be conservative, we do not wish to assume independence of each white dwarf SED, considering that there may be systematics below the precision of \textit{HST} photometry. We make use of the mean of the covariances over all standards $\mathbf{\Sigma}_{b1,b2}^{\rm Synth }= \sum_k^{N_{\rm Standards}} \mathbf{\Sigma}_{b1,b2}^ {{\rm WD },k} / {N_{\rm Standards}} $ to represent the aggregated uncertainty in the modelling, effectively assuming a high degree of correlation between all standards. 

We define a vector of the WD standard residuals $\mathcal{R}^{{\rm WD}}_{i}$, where the index $i$ runs over all observations, from all standards, from all surveys. From the catalogued photometry of the standard $\textrm{Obs}_{{\rm S}^{b}_i}$ with uncertainty $\sigma_{{\rm Cat},i}$ values, we define the residuals, associated covariance, and likelihood to be

\begin{align}
    \Vec{\mathcal{R}}^{{\rm WD}}_{i}= & \textrm{Obs}_{{\rm S}^{b}_i} - {{\rm WD Synth}_{{\rm S}^{b}_i} } - \Delta_{S^{b_i}} \\
    \mathbf{\Sigma}^{\mathrm{WD}}_{ij}&= \delta_{ij} (\sigma^2_{{\rm Cat},i} \alpha_{{\rm S}^{b}_i}^2 + \sigma_{{\rm S}^{b}_i}^2) +  \mathbf{\Sigma}^{\rm Synth }_{b_i,b_j} \\
    -2 \log \mathcal{L}_{\rm WD} = & {\rm constant} -({\Vec{\mathcal{R}}^{{\rm WD}}_{i}}) ^T \left(\mathbf{\Sigma}^{\mathrm{WD}}_{ij}\right)^{-1} \Vec{\mathcal{R}}^{{\rm WD}}_{i}, 
\end{align}
with $\delta_{ij}$ the Kronecker delta,  ${\rm S}^{b}_i$ the label for survey and filter, and $\alpha_{{\rm S}^{b}_i}, \sigma_{{\rm S}^{b}_i}$ empirical parameters which inflate the uncertainties of catalogue photometry, which we find to be underestimated. These parameters are determined by likelihood maximisation and fixed during inference of the offsets.

\subsubsection{Tertiary Standards}

The second calibration path constraining zero-point offsets is through the tertiary standards observed both by PS1-Public and each individual survey. We compare the differences between the observed magnitudes to determine their consistency with predicted filter transformations.

\begin{figure}
    \centering
    \includegraphics[width=9cm]{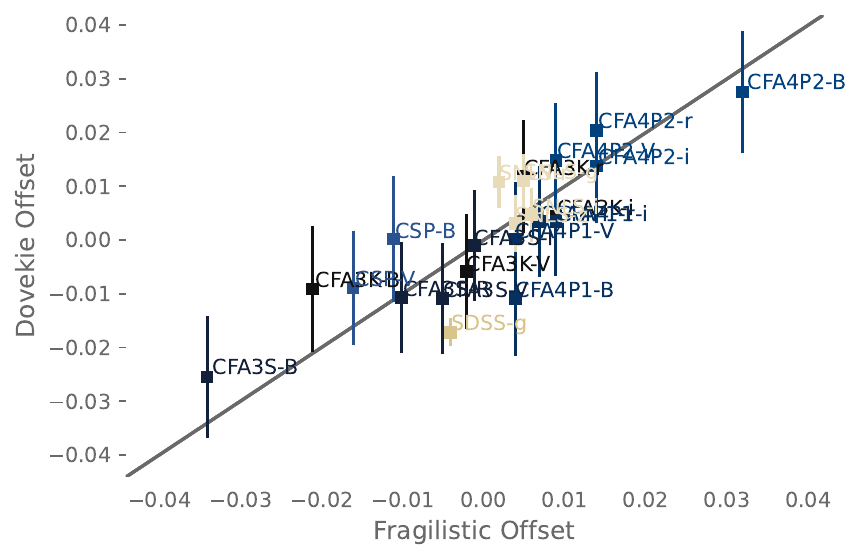}
    \caption{A comparison of the \citetalias{fragilistic}-derived offsets and the new Dovekie offsets for filters that were modified from their original published releases. For both Dovekie and \citetalias{fragilistic}, the mean offset of all surveys has been subtracted out, and the x-errors (similar in size to the y-errors) have been left out for visual clarity.}
    \label{fig:Frag-compare}
\end{figure}

For a given survey $S$ in filter $b$, we define the residual between filter $b$ and the PS1-Public filter nearest in wavelength ${b1}$ to be $\mathcal{R}_{{\rm Obs}^{b'}}=\textrm{Obs}_{{\rm PS1}^{b1}} - \textrm{Obs}_{S_{b'}}$. Using the parameters determined from one of the two spectral libraries of \textit{HST}, we predict that magnitude residual to be 

\begin{align}
    \mathcal{R}_{{\rm Pred}^{b'}}^{\rm Library}=  & \quad {{ \Delta}_{\rm Int}^{b'}}^{\rm Library}+ \Delta_{S^{b'}} - \Delta_{PS1^{b1}} +\\ & \mathcal{C}^{\rm Library}_{S^{b'}} \times ((\textrm{Obs}_{{\rm PS1}^{b2}} + \Delta_{{\rm PS1}^{b2}} -(\textrm{Obs}_{{\rm PS1}^{b3}} - \Delta_{{\rm PS1}^{b3}} )). \label{eq:MagDiff}
\end{align}
Within Equation \ref{eq:MagDiff}, we use filters $b2$ and $b3$ as a colour index to correct for the different colour sensitivities of the PS1 $b1$ and desired $b'$ band-passes. The colour-magnitude slope for filter $b$ in survey $S$ $ \mathcal{C}^{\rm Library}_{S^{b'}}$ and zero-point offset ${{ \Delta}_{\rm Int}^{b'}}^{\rm Library}$ have been discussed in Section \ref{sec:Method:subsec:Filters}. Finally, there are the magnitude offsets $\Delta_{S_{b'}}$ for a given filter in survey $S$ and $\Delta_{{\rm PS1}^{b1}},\Delta_{{\rm PS1}^{b2}},\Delta_{{\rm PS1}^{b3}}$ for the PS1-Public filters.

This predicted residual is dependent on the transformations derived from the spectroscopic libraries. 
We use a mixture model over the two spectral libraries with equal prior weights assigned to each library. The final likelihood is shown in Equation \ref{eq:tertiarylike}, with $\mathbf{I}$  the identity matrix, $\mathbf{1}$ the matrix with ones in each entry, and $N_{S^b}$ the number of standards observed in the given filter. The $\sigma^2_{S^{b}}$ term is the uncertainty, which we define as the standard deviation of the residuals, and the $f_{S}$ term is the error floor for the associated survey. Conservatively, we include a systematic uncertainty in the filter transformation for each filter, assigned by survey, labelled $f_{S}$.
\begin{figure*}
\begin{align}
   \mathcal{L}_{\rm Tertiary} =  \prod_{S^b}^{\rm All\ filters} \left[ \sum_{\rm Lib}^{\rm NGSL,\ Calspec}  \frac{\exp\left(-\frac{1}{2} \sum \left(\mathcal{R}_{{\rm Obs}^{b'}} - \mathcal{R}_{{\rm Pred}^{b'}}^{\rm Lib} \right)^T \left( \sigma_{S^{b}} ^2 \mathbf{I}  + f_{S}^2 \mathbf{1} \right)^{-1} \left(\mathcal{R}_{{\rm Obs}^{b'}} - \mathcal{R}_{{\rm Pred}^{b'}}^{\rm Lib} \right) \right)}{2\sqrt{(1 + N_{S^b}   f_{S}^2 / \sigma_{S^{b}} ^2) \cdot (2\pi \sigma_{S^{b}}^2)^{N_{S^b}}   }}  \right] \label{eq:tertiarylike}.
\end{align}
\end{figure*}

\section{Dovekie Calibration Results}\label{sec:CalResults}

Here we present the results of the cross-calibration procedure. The chains and corner plot are available in the Github link.

\begin{figure*}
    \centering
    \includegraphics[width=18cm]{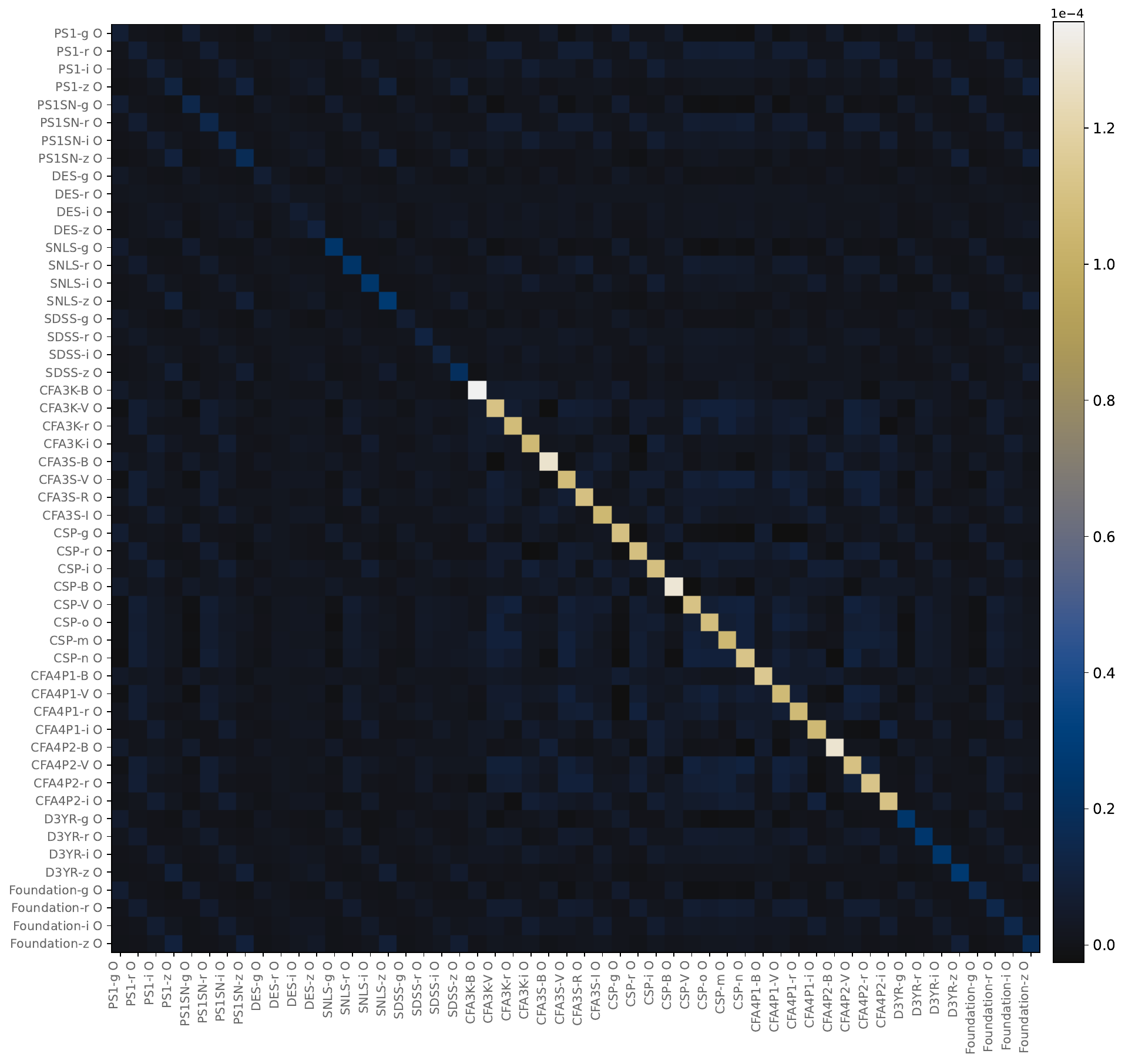}
    \caption{The posterior covariance matrix between best-fit Dovekie zero-point offsets for filters sampled by the MCMC.}
    \label{fig:COV}
\end{figure*}

\subsection{Updates to Bandpasses}\label{sec:CalResults:subsec:bandpasses}

We found it necessary to make changes to several published filters to ensure good agreement between the observed and synthetic colour-magnitude relationships as described in Section \ref{sec:Method:subsec:Filters}. These discrepant bands, and the changes applied, are summarised in Table \ref{tab:bandpasses}. 

\citetalias{fragilistic} performed a 30\r{A} shift to the PS1 $g-$band, which applied to Foundation, PS1SN, and PS1. We undo this shift; instead, we choose to modify, where appropriate, other bandpasses to ensure that our synthetic and observed colour-magnitude relationships are in agreement.

\begin{table}
    \centering
    \begin{tabular}{c|ccc}
         Filter & $\lambda_{\rm shift}$ & Weighting Factor & $\lambda_{\rm eff}$\\
         \hline 
         CfA4P1-B & $0$\r{A} & $X = -1$ & $4356$\r{A} \\
         CfA4P1-V & $0$\r{A} & $X = -1$ & $5410$\r{A}  \\
         CfA4P1-r & $0$\r{A} & $X = -1$ & $6242$\r{A}  \\
         CfA4P1-i & $0$\r{A} & $X = -1$ & $7674$\r{A}  \\
         \hline
         CfA4P2-B & $-20$\r{A} & $X = -1$ & $4356$\r{A} \\
         CfA4P2-V & $20$\r{A} & $X = -1$ & $5410$\r{A}  \\
         CfA4P2-r & $20$\r{A} & $X = -1$ & $6242$\r{A}  \\
         CfA4P2-i & $20$\r{A} & $X = -1$ & $7674$\r{A}  \\
         \hline
         CfA3S-B & $70$\r{A} & None & $4356$\r{A}\\ 
         CfA3K-B & $70$\r{A} & None & $5380$\r{A}\\ 
         CSP-B & $50$\r{A} & None & $4392$\r{A} \\ 
         CSP-V & $-50$\r{A} & None & $5358$\r{A} \\
         \hline
         SDSS-g & $+15$\r{A} & None & $4672$\r{A} \\
         SNLS-g & $+30$\r{A} & None & $4721$\r{A} \\
         SNLS-r & $+30$\r{A} & None & $6336$\r{A} \\
         SNLS-i & $+30$\r{A} & None & $7641$\r{A} \\
         SNLS-z & $+30$\r{A} & None & $8970$\r{A} \\
    \end{tabular}
    \caption{A summary of the bands modified in this paper. The $\lambda_{\rm eff}$ is provided where appropriate, but a summary of all available $\lambda_{\rm eff}$ can be found \href{http://svo2.cab.inta-csic.es/svo/theory/fps3/index.php?mode=voservice}{here}. Where "Weighting Factor" is marked with "None", we only apply the lambda shift as in \citetalias{fragilistic}.}
    \label{tab:bandpasses}
\end{table}

We find a number of changes to filters are required compared to \citetalias{fragilistic}, in spite of using the same data. These changes appear to be on account of improved errors and the addition of Gaia as an independent cross-check on our filter characterisation. We find a shift of $+30$\r{A} to SNLS is necessary across the board, along with a weighting of $X = -1$ applied to all CfA4P1 and CfA4P2. Beyond these survey-wide changes, we found changes to $B-$, $V-$, and $g-$ bands were the most frequent, changing CfA3K-$V$ ($-30$\r{A}), CfA3S-$B$ ($70$\r{A}) CSP-$B$ ($50$\r{A}), CSP-$V$ ($-50$\r{A}), SDSS-$g$ ($15$\r{A}); in addition to the re-weighting, we found shifts to CfA4P2-$B$ ($-20$\r{A}), CfA4P2-$V$ ($20$\r{A}), CfA4P2-$r$ ($20$\r{A}), and CfA4P2-$i$ ($20$\r{A}) were necessary.

\begin{figure*}
    \centering
    \includegraphics[width=18cm]{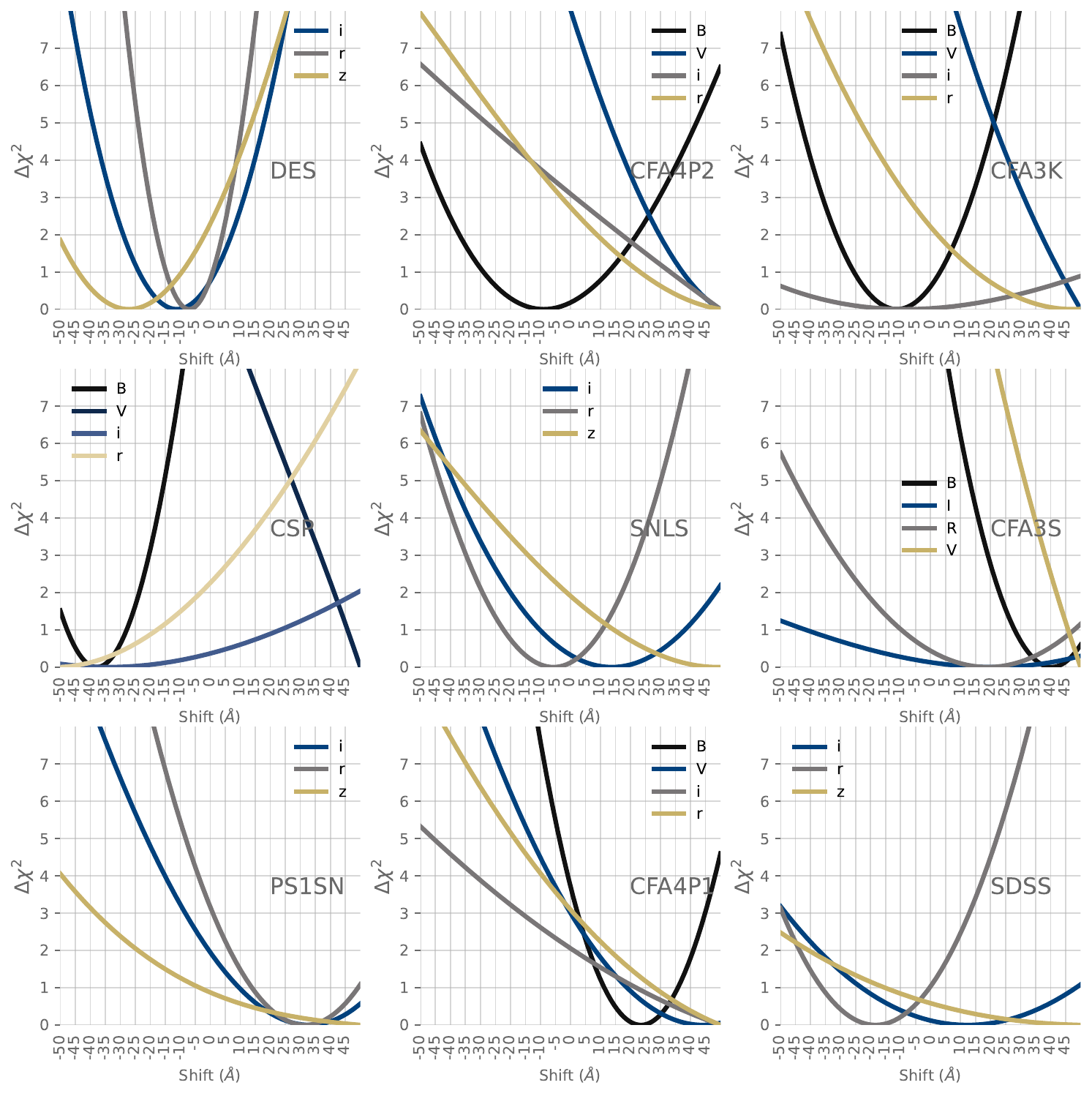}
    \caption{Posteriors for each filter in the surveys cross-calibrated in Dovekie, using Gaia as the reference survey.  }
    \label{fig:1sigmaGAIA}
\end{figure*}

Previous cross-calibration techniques have either used instrumental errors as the basis for their 1-sigma uncertainties \citep{fragilistic, Scolnic15} or used approximations that do not involve shifting the filters to determine systematic uncertainties \citep{Rubin23}. Figure \ref{fig:Filters1sigma} in the Appendix shows our joint constraints, and Figures \ref{fig:1sigmaPS1} and \ref{fig:1sigmaGAIA}  show the constraints from each of these reference surveys individually. A summary of these new uncertainties is given in Table \ref{tab:ShiftUncertainties}. We note that in many cases, these shifts are greater than the nominal uncertainty in previous works.

\begin{table}
    \centering
    \begin{tabular}{c|ccc}
        Survey & Filters & Uncertainty (\r{A}) & Fragilistic Vals. \\
        \hline
         CfA3K & $BVri$ & $15,15,20,20$ & $10,15,10,10$ \\
         CfA3S & $BVRI$ & $15,15,20,20$ & $10,10,10,10$\\
         CSP   & $BVgri$ & $8,10,10,15,20$ & $7,3,8,4,2$ \\
         SDSS  & $griz$ & $10,10,15,20$ & $8,6,6,6$ \\
         SNLS  & $griz$ & $5,20,15,20$ & $15,15,15,15$ \\
         PS1 (all)  & $griz$ & $5,10,15,20$ & $7,7,7,7$ \\
         DES  & $griz$ & $5,5,10,30$ & $6,6,6,6$\\
        CfA41 & $BVri$ & $20,15,10,15$ & $10,15,10,10$ \\
        CfA42 & $BVri$ & $20,15,15,15$ &  $10,15,10,10$
    \end{tabular}
    \caption{The output $1\sigma$ systematic uncertainties for each survey and filter in the calibration solution, for Dovekie and \citetalias{fragilistic}. These values are used during the SALT fitting process to assess the systematic uncertainty due to calibration. We update all values from \citetalias{fragilistic}.}
    \label{tab:ShiftUncertainties}
\end{table}

\subsection{Recovery of Simulated Offsets}\label{sec:CalResults:subsec:simreco}

We test the Dovekie pipeline for biases in fitted zero-points by generating 100 sets of simulated calibrator stars for each of our tested surveys. The biases on our recovered offsets are detailed in Table \ref{tab:simoffsetbias} in Appendix \ref{sec:App:subsec:simbiases}. In summary, we do not find biases with a significance of more than $1\sigma$ for any of our filters.

\subsection{Dovekie zeropoints}\label{sec:CalResults:subsec:data}

Table \ref{tab:model_params} presents our findings for the zero-point offsets from the Dovekie cross-calibration pipeline. We compare our results to \citetalias{fragilistic}, for changed bands, in Figure \ref{fig:Frag-compare}. We find our results to be individually consistent with \citetalias{fragilistic}, with only the DES-z filter exhibiting a ZP offset of more than $2\sigma$ ($2.1\sigma$), across all filters analysed in the Dovekie process.

\begin{center}
\begin{table}
    \centering
    \caption{Magnitude offsets (ZPs) from the Dovekie calibration solution, along with their errors.}
    \label{tab:model_params}
    \begin{tabular}{cc}
        \hline
PS1-g & $-0.0018 \pm 0.0027$ \\
PS1-r & $0.0062 \pm 0.0027$ \\
PS1-i & $0.0007 \pm 0.0027$ \\
PS1-z & $0.011 \pm 0.0034$ \\
PS1SN-g & $-0.0084 \pm 0.0038$ \\
PS1SN-r & $-0.0223 \pm 0.0038$ \\
PS1SN-i & $-0.0178 \pm 0.0038$ \\
PS1SN-z & $-0.013 \pm 0.0042$ \\
DES5YR-g & $0.0015 \pm 0.0027$ \\
DES5YR-r & $0.0014 \pm 0.0022$ \\
DES5YR-i & $0.004 \pm 0.0026$ \\
DES5YR-z & $-0.006 \pm 0.0032$ \\
SNLS-g & $0.0099 \pm 0.005$ \\
SNLS-r & $0.0097 \pm 0.0049$ \\
SNLS-i & $0.0036 \pm 0.005$ \\
SNLS-z & $0.002 \pm 0.0053$ \\
SDSS-g & $-0.0183 \pm 0.0026$ \\
SDSS-r & $0.0058 \pm 0.0034$ \\
SDSS-i & $0.001 \pm 0.0034$ \\
SDSS-z & $-0.0022 \pm 0.0045$ \\
CFA3K-B & $-0.0102 \pm 0.0117$ \\
CFA3K-V & $-0.007 \pm 0.0107$ \\
CFA3K-r & $0.0107 \pm 0.0105$ \\
CFA3K-i & $0.0033 \pm 0.0104$ \\
CFA3S-B & $-0.0266 \pm 0.0114$ \\
CFA3S-V & $-0.012 \pm 0.0104$ \\
CFA3S-R & $-0.0118 \pm 0.0104$ \\
CFA3S-I & $-0.0021 \pm 0.0103$ \\
CSP-g & $0.0009 \pm 0.0105$ \\
CSP-r & $-0.0049 \pm 0.0105$ \\
CSP-i & $-0.0206 \pm 0.0106$ \\
CSP-B & $-0.0009 \pm 0.0117$ \\
CSP-V & $-0.01 \pm 0.0106$ \\
CSP-o & $-0.0092 \pm 0.0105$ \\
CSP-m & $0.0016 \pm 0.0102$ \\
CSP-n & $0.0022 \pm 0.0106$ \\
CFA4P1-B & $-0.012 \pm 0.0107$ \\
CFA4P1-V & $-0.0009 \pm 0.0106$ \\
CFA4P1-r & $0.0023 \pm 0.0103$ \\
CFA4P1-i & $0.0023 \pm 0.0102$ \\
CFA4P2-B & $0.0264 \pm 0.0114$ \\
CFA4P2-V & $0.0137 \pm 0.0107$ \\
CFA4P2-r & $0.0193 \pm 0.0108$ \\
CFA4P2-i & $0.0127 \pm 0.0107$ \\
DES3YR-g & $0.0002 \pm 0.005$ \\
DES3YR-r & $-0.0082 \pm 0.005$ \\
DES3YR-i & $-0.001 \pm 0.005$ \\
DES3YR-z & $-0.0056 \pm 0.0052$ \\
Foundation-g & $0.0049 \pm 0.0038$ \\
Foundation-r & $0.0065 \pm 0.0038$ \\
Foundation-i & $0.0017 \pm 0.0038$ \\
Foundation-z & $0.0102 \pm 0.0042$ \\
\end{tabular}
\end{table}
\end{center}

Figure \ref{fig:COV} shows the associated covariance matrix from the Dovekie pipeline. We find that the historic low-redshift surveys CfA3, CfA4, and CSP contain the highest uncertainties of our surveys. To isolate the effects of our changes to the stellar catalogs and filters from the effects of our analysis changes, we run our code on the exact stellar catalogs with the original filters used in \citetalias{fragilistic}. 

\section{SALT Surface Retraining}\label{sec:SALT}

With the filter zero-points and transmission shifts from Dovekie, we train a suite of new SALT surfaces we label ``Dovekie'' to test the impact of calibration on cosmology. We use the \texttt{SALTShaker} code from \cite{Kenworthy21}, as updated in \citet{Kenworthy2025} to generate our fiducial SALT surface and 9 systematic surfaces representing the impact of calibration uncertainty. Due to the modifications we made to the training sample, we found that further upgrades to the code were necessary, in particular to the schema for regularization. These updates will be discussed in Kenworthy et al. in prep (Whimbrel). The configuration files used for this training, and the systematic surfaces, are available at \hyperlink{https://github.com/bap37/Whimbrel}{this link}. 

\begin{figure}
    \centering
    \includegraphics[width=10cm]{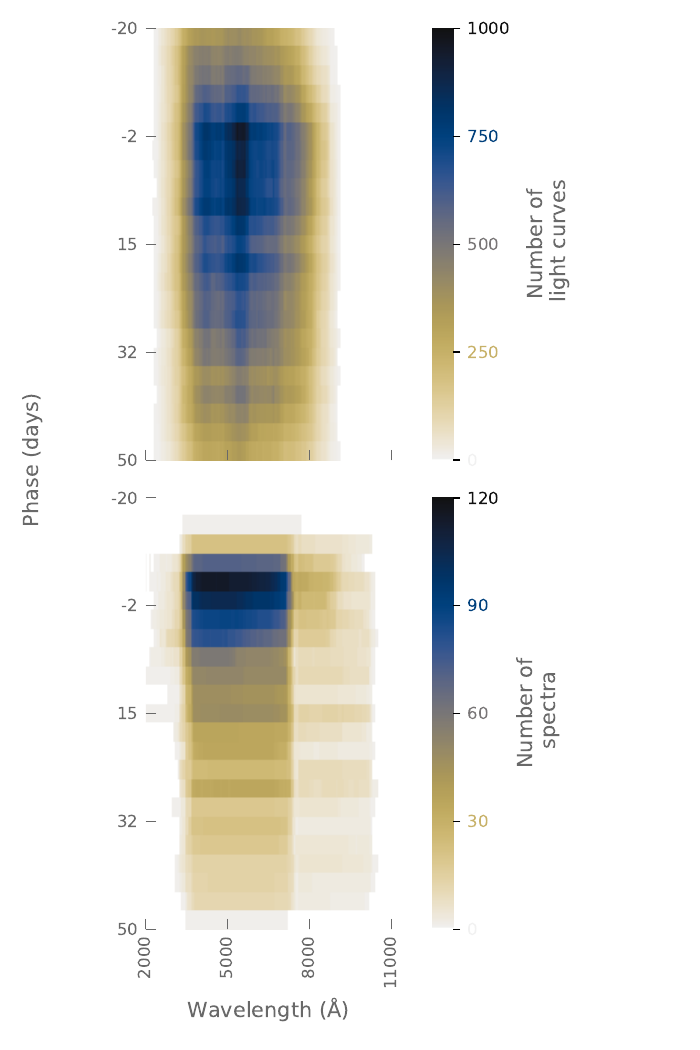}
    \caption{\textbf{Top:} Photometric coverage of the DOVEKIE training sample as a function of wavelength (\r{A}) and phase (MJD, relative to peak brightness MJD$=0$.) \textbf{Bottom:} Same as top, but for the spectral coverage of Dovekie. }
    \label{fig:DataDensity}
\end{figure}

SALT models the flux of a given SN Ia as:
\begin{align} 
\label{saltmodel}
\begin{split}
F(\rm{SN}, p, \lambda) = x_{0}^s &\times\left[M_{0}(p, \lambda)+x_{1}^s M_{1}(p, \lambda)+\ldots\right] \\
&\times \exp [c^s C L(\lambda)],
\end{split}
\end{align}
where the variables denoted with superscript $^s$ - $x_0^s, x_1^s, c^s$ - are fit for individual SNe Ia, and $M_{0}(p, \lambda),M_{1}(p, \lambda),$ and $C L(\lambda)$ comprise the `SALT surface' that is created during the training process. $M_{0}(p, \lambda)$, which is a function of the phase $p$ and wavelength $\lambda$, describes the average SED of the SNe Ia in the training sample, while $x_1^s$, $M_{1}(p, \lambda)$ encapsulates the phase-dependent variability of SN Ia SEDs. The SALT colour law, $CL(\lambda)$, is solely a function of wavelength.

Our training sample is taken from \cite{Kenworthy21}, with the exception of dropping the CfA1, CfA2, Calan-Tololo, and "Misc. Low-z" samples. This reduces the sample to 1009 SNe, with 755 spectra. We make the choice to remove these older surveys from cross-calibration and analysis to restrict the sample to better understood, more modern surveys whose systematics are more amenable to analysis, at a modest reduction to statistical power. The photometric and spectroscopic coverage of wavelength and phase is shown in Figure \ref{fig:DataDensity}. In short, we train on a select sample of SNe from SDSS, DES, SNLS, PS1, Foundation, CfA3, CfA4, and CSP. A more complete description of these surveys can be found in Section 4 of \cite{Kenworthy21}. s

\begin{figure}
    \centering
    \includegraphics[width=9cm]{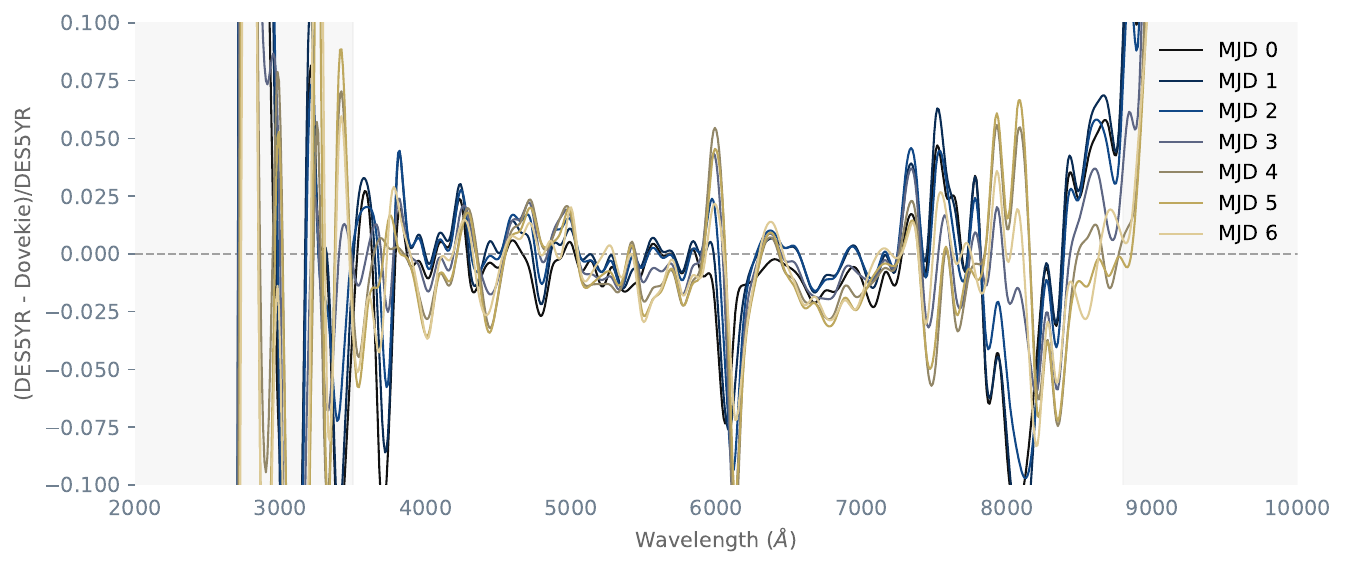}
    \caption{The M0 components for Dovekie and SALT3.DES5YR, with arbitrary offsets, relative to the peak brightness at MJD$=0$. A grey dashed line at 0 is presented for visual clarity. The wavelength ranges outside of the nominal Dovekie range ($3500-8800$\r{A}) are greyed out for reference.}
    \label{fig:DELTAM0}
\end{figure}

\begin{figure}
    \centering
    \includegraphics[width=8cm]{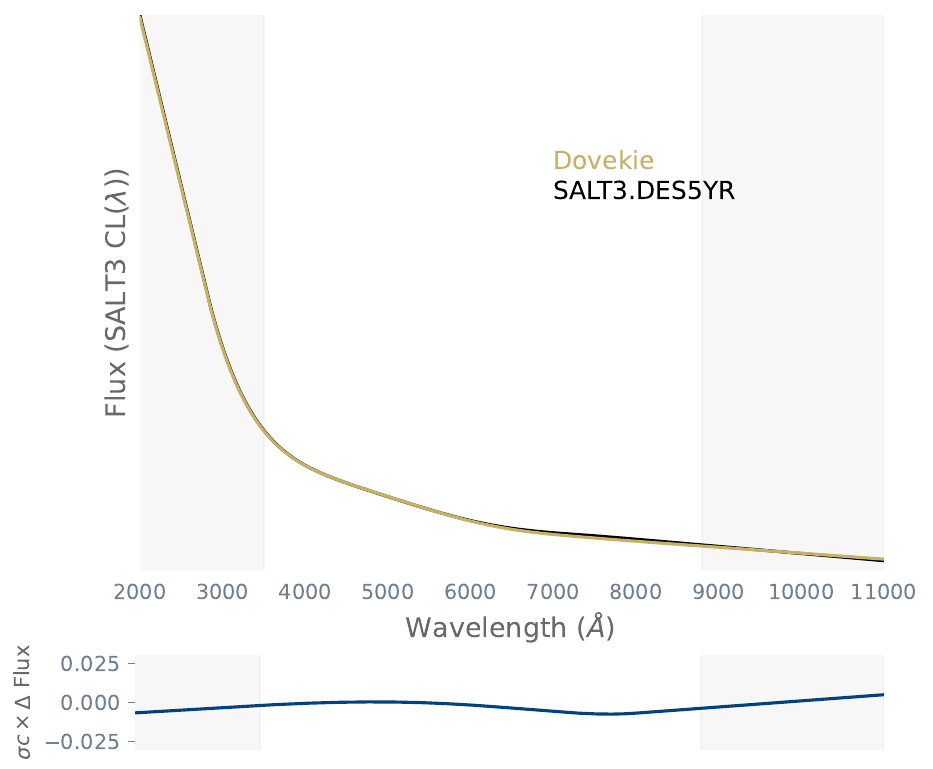}
    \caption{\textbf{Top:} The SALT3 colour law for Dovekie (gold) and SALT3.DES5YR (black). \textbf{Bottom:} The difference in SALT3 colour laws between Dovekie and SALT3.DES5YR, multiplied by the scatter in the colour distribution (0.1). The difference in this flux excess due to the colour law ranges from $0.0$ mags to $\sim 0.03$ mags. The wavelength ranges outside of the nominal range of central filter wavelengths ($3500-8800$\r{A}) are greyed out for reference.}
    \label{fig:CL}
\end{figure}

Figure \ref{fig:DELTAM0} shows the fractional difference ((SALT3.DES5YR - Dovekie)/Dovekie) between the M0 and M1 SALT components. There are no significant changes in the model between $4000-8000 $\r{A} in the M0 component; this covers most of the range of Dovekie ($3500-8800$\r{A}). The M1 component is noisier than the M0 component.

\begin{figure}
    \centering
    \includegraphics[width=8cm]{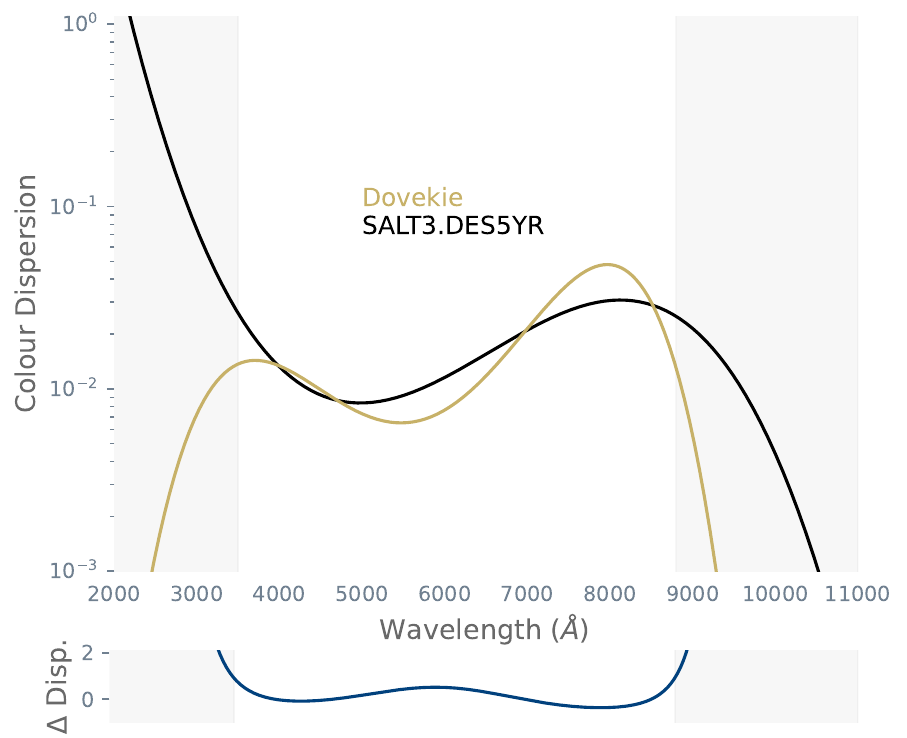}
    \caption{\textbf{Top:} The SALT3 colour dispersion for Dovekie (gold) and SALT3.DES5YR (black). \textbf{Bottom:} The fractional difference in SALT3 colour laws between Dovekie and SALT3.DES5YR. The difference in colour dispersion ranges from $0$  to $2$ mags. The wavelength ranges outside of the nominal Dovekie range ($3500-8800$\r{A}) are greyed out for reference.}
    \label{fig:DISP}
\end{figure}

Figure \ref{fig:CL} shows the colour law behaviour for Dovekie and SALT3.DES5YR as a function of wavelength and the difference between the two colour laws, multiplied by the standard deviation of the colour distribution $(\sigma c = 0.1)$. The two colour laws are similar, only significantly varying near the edges of the model boundary. 

Finally, we show the colour dispersion for Dovekie and SALT3.DES5YR in Figure \ref{fig:DISP}. The Dovekie colour dispersion is broadly similar to DES5YR, though performs slightly worse in the median-wavelength range ($4000-7000$\r{A}), and better above $7000$\r{A}. Below, we show the fractional difference between the two colour dispersions, which are peaked around $3000 $\r{A} (outside the model range) and $8000$\r{A}.

\subsection{Calculating Systematic Cosmological Uncertainty}\label{sec:SALT:subsec:Uncert}

Following \citetalias{fragilistic}, to quantify the effect of photometric calibration on distances derived with SNe Ia, we create 9 additional iterations of our Dovekie SALT surfaces in order to span the space systematic uncertainties. Each systematic surface is trained applying a set of $\Delta_b$ and $\lambda_{\rm Shift}$ values randomly sampled using the means and covariances from Section \ref{sec:CalResults} to the filter bandpasses. The light curves are fit with the surfaces and corresponding band-passes for each systematic iteration. The distance moduli from each iteration are then used to construct a covariance matrix to propagate the 9 Dovekie-derived Hubble Diagrams into a $N\times N$ covariance matrix as
\begin{equation}\label{eq:SysCosmo}
    C_{z_i, z_j} = \Sigma_k \frac{\partial \Delta \mu_{z_i} }{\partial k} \frac{\partial \Delta \mu_{z_j}}{\partial k} \sigma^2_k,
\end{equation}
where we sum over systematics $k$, $\Delta \mu_{z_i}$ is the distance residual for a given supernova between Dovekie surfaces, and $\sigma_k = 1/3$ is a weighting factor to ensure the quadrature sum of our systematic uncertainties is 1.

In Appendix \ref{sec:App:subsec:sysdiffs} we show the fractional differences between the fiducial Dovekie solution and the 9 systematic surfaces. Within the model range, the M0 variations are limited to within $0.025$ on either side, a slight improvement to the \citetalias{fragilistic} calibration solution. These variations, both in M0 and the colour law, are dictated by the spread in ZP offsets (Figure \ref{fig:COV}) and the $1\sigma$ uncertainties on the filters. 

\subsection{Impact on Distances}

To assess the impact of our systematic uncertainty arising from photometric calibration, we propagate our Dovekie SALT surface (and the systematic surfaces) to inferred distances. We fit a subset of the Pantheon+ sample: DES, SDSS, PS1, Foundation, CfA3, CfA4, SNLS, and CSP; a subsection of each was in our Dovekie training sample.

We estimate distances from the SALT parameters using the Tripp equation \citep{Tripp98}:
\begin{equation}\label{eq:Tripp}
    \mu = m_B - \beta c + \alpha x_1 - M_0,
\end{equation}
where $m_B=10.5-2.5\log_{10}(x_0)$ is the overall normalisation of the SN\,Ia, $c$ is the SALT colour parameter, and $x_1$ is the SALT stretch parameter, as discussed in Equation \ref{saltmodel}. $\alpha$ and $\beta$ are determined as global parameters in a SALT2mu \cite{Marriner11} fit\footnote{e.g. the `BBC' method without bias corrections.}, alongside the absolute magnitude of a fiducial $c=x_1=0$ SN\,Ia, $M_0$.

\begin{figure}
    \centering
    \includegraphics[width=9cm]{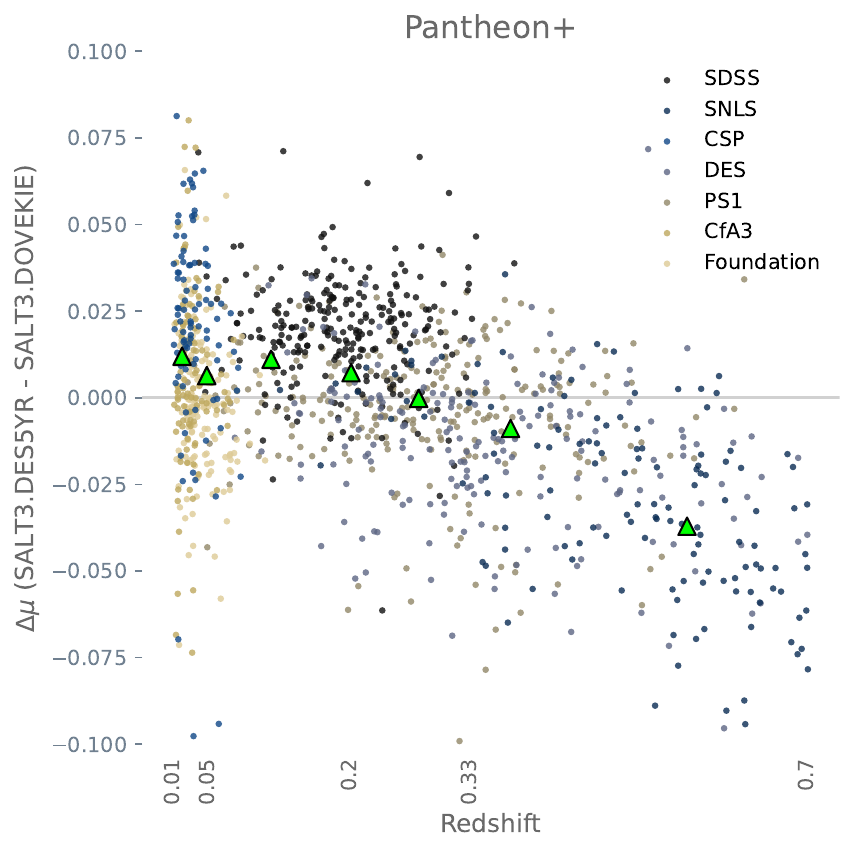}
    \caption{The difference in inferred $\mu$ values between SALT3.DES5YR (with Fragilistic ZP) and Dovekie (with Dovekie ZP), colour coded by survey. Green triangles denote the error-weighted average $\mu$ value.}
    \label{fig:DELTAMUDES}
\end{figure}

\begin{figure}
    \centering
    \includegraphics[width=9cm]{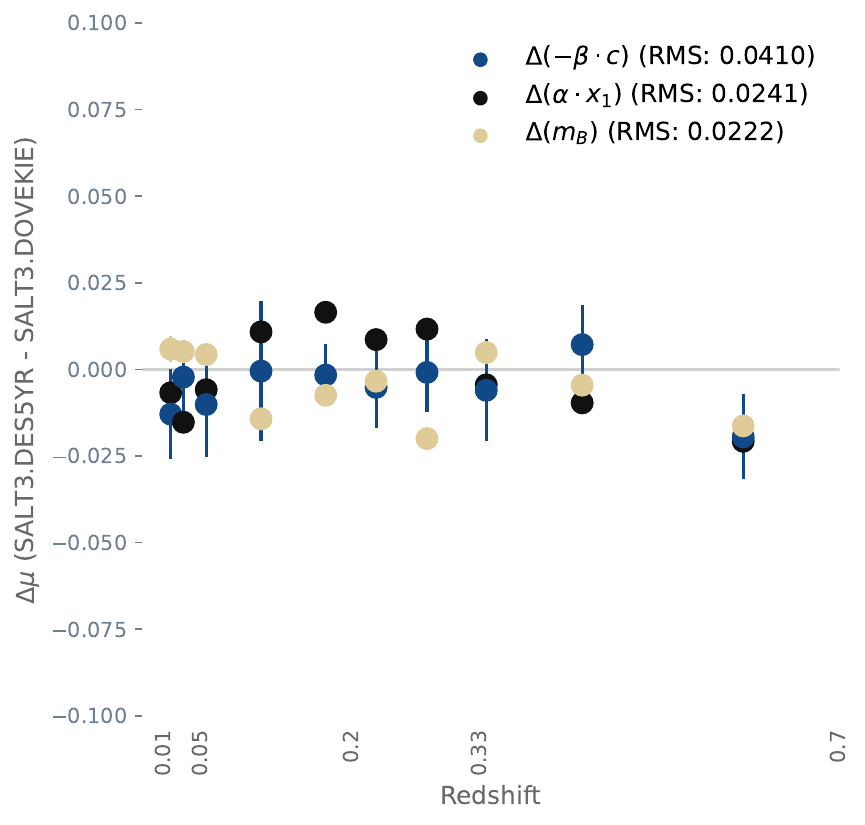}
    \caption{The difference in the individual Tripp components between SALT3.DES5YR and Dovekie, binned as a function of redshift. The $m_B$ component is shown in gold, the $\beta c$ in blue, and $\alpha x_1$ in black. The RMS of the differences over the entire sample is shown in the legend.}
    \label{fig:trippparamdifferences}
\end{figure}

\begin{figure}
    \centering
    \includegraphics[width=8cm]{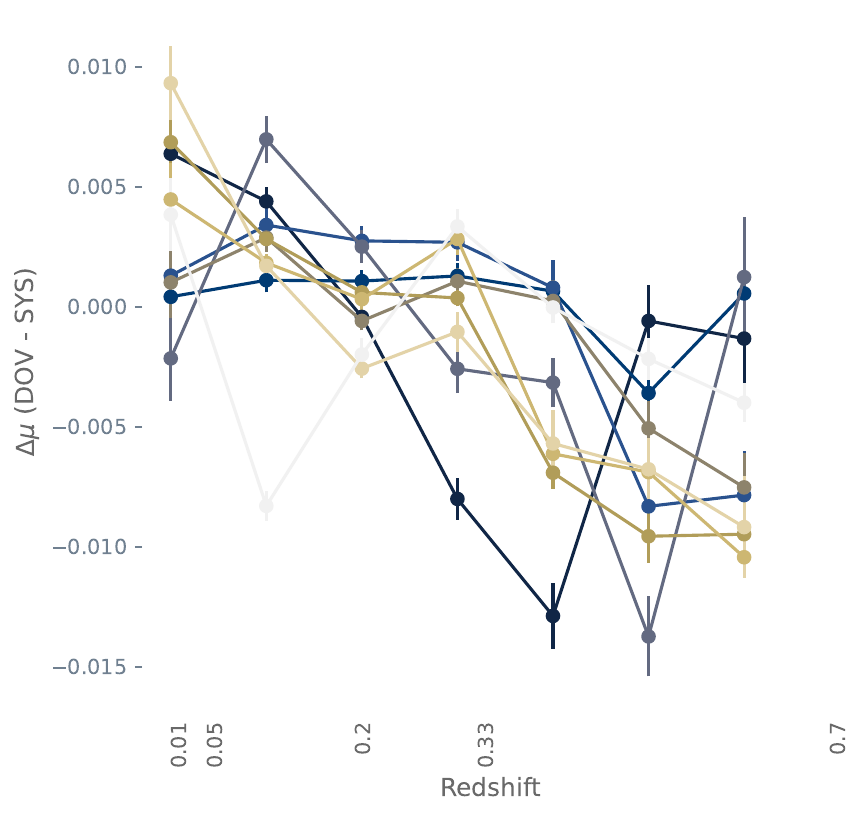}
    \caption{The binned, median differences between the Dovekie SALT surface and the 9 systematic uncertainties, including both changes to the SALT surface and ZP offsets. Each surface is colour coded as in previous figures; we find the increased filter uncertainties do not significantly impact the inferred distances.}
    \label{fig:DELTAMUSYS}
\end{figure}

In Figure \ref{fig:DELTAMUDES}, we compare the Dovekie distances to the DES5YR SALT distances. We find that the high-redshift distances are the most discrepant between Dovekie and SALT3.DES5YR. To understand how these differences are created, in Figure \ref{fig:trippparamdifferences}, we show how the distance changes are propagated through the three elements of the Tripp estimator. The RMS difference in distance modulus is most strongly affected by shifts to the $c$ parameter, while residuals binned as a function of $z$ show all three components contributing to the cosmological impact. As the final impact on cosmology will depend on how the calibration uncertainties affect distances, we show the difference in the binned distances produced using each systematic surface in Figure \ref{fig:DELTAMUSYS}.

To check the consistency of the Dovekie SALT surface, we compared the dispersion of $\mu$ values for each survey. Dovekie performed slightly better than SALT3.DES5YR for all surveys except SDSS, though these results are well within the error.


\subsection{Impact of New Filter Uncertainties on Cosmological Inference}

Here, we assess the change in cosmological systematic uncertainties from the use of these new filter uncertainties (as seen in Table \ref{tab:ShiftUncertainties}).

We estimate the systematic uncertainties as in Section \ref{sec:SALT:subsec:Uncert}. We generate 2 sets of 9 additional SALT surfaces; one with the \citetalias{fragilistic} $1\sigma$ uncertainties for our filters (Dovekie-Frag here), and one with the Dovekie $1\sigma$ uncertainties (Dovekie). 

For cosmology fitting to attain $w$ and $\Omega_M$, we use a simple and fast minimisation program in SNANA, \texttt{wfit}, that uses a Planck-like CMB prior based on the R-shift parameter, as shown in the $\sigma_R$-tuning discussion in Section 3 of \cite{Sanchez21}.

We provide the standardised SN\,Ia distances and an unbinned covariance matrix, following \citet{Binning}, that includes both the systematic and statistical uncertainties to \texttt{wfit}. 

We define the contribution of the systematic uncertainty by taking the full covariance and subtracting out the statistical-only uncertainty in quadrature:
\begin{equation}\label{eq:wsig}
    \sigma_w({\rm phot}) = \sqrt{\sigma_{\rm tot}^2 - \sigma_{\rm stat}^2}.
\end{equation}

We find that systematic uncertainty on $w$, $\sigma_w (\rm{phot})$, is lower for Dovekie-Frag ($\sigma_w(\rm phot) = 0.0117$) than for Dovekie: $\sigma_w(\rm phot) = 0.0155$, as summarised in Table \ref{tab:FragUncert}. 

\begin{table}
    \centering
    \begin{tabular}{c|c|c}
        Model & $\sigma_w (\rm{phot})$ & $\sigma_{\Omega_M}(\rm{phot})$\\
        \hline
        SALT3.DES5YR & 0.0230  & 0.007  \\
        Dovekie & 0.0155 & 0.004  \\
        Dovekie-Frag & 0.0117  & 0.003  \\
    \end{tabular}
    \caption{Summary of the systematic uncertainties for our P+ sample in $w$ and $\Omega_M$ when combined with a CMB-like prior.}
    \label{tab:FragUncert}
\end{table}

Given the increased filter uncertainties in the nominal Dovekie approach, this corresponding increase in systematic uncertainty is expected; however, both Dovekie models present a smaller systematic uncertainty than the SALT3.DES5YR model. We find no significant changes to the central $w$ or $\Omega_M$ value from changing these filter uncertainties. It is evident that both Dovekie and Dovekie-Frag present improvements on the systematic uncertainties from SALT3.DES5YR.


\section{Impact on Pantheon+ Cosmology}\label{sec:PantheonPlus}

Here we estimate the changes to the Pantheon+ cosmology results. We do not perform a re-analysis of the Pantheon+ analysis. Instead, we infer the distances from our SNe Ia using the SALT3.DES5YR and Dovekie SALT models, without bias corrections. The two sets of distances will have the same selection effects, and we expect the resulting changes to cosmology to be broadly comparable to a full reanalysis. We fit a cosmology to these samples and test the relative changes in our cosmological parameters and systematic uncertainties. This method of estimation is useful for qualitative comparisons, but as shown in \cite{Kessler16}, systematic uncertainties will change with greater distance from $\Lambda$CDM. A full reanalysis is left for {Popovic et al. 2025b}, which performs a re-analysis of the DES5YR cosmology with this updated calibration solution. 

Astute readers will note that the Pantheon+ sample was fit with SALT2.B21, rather than SALT3.DES5YR. We choose to refit the Pantheon+ data with SALT3.DES5YR for a more fair comparison of calibration changes, removing the effect of differing SALT platforms on our distance measurements. 

\begin{figure}
    \centering
    \includegraphics[width=8cm]{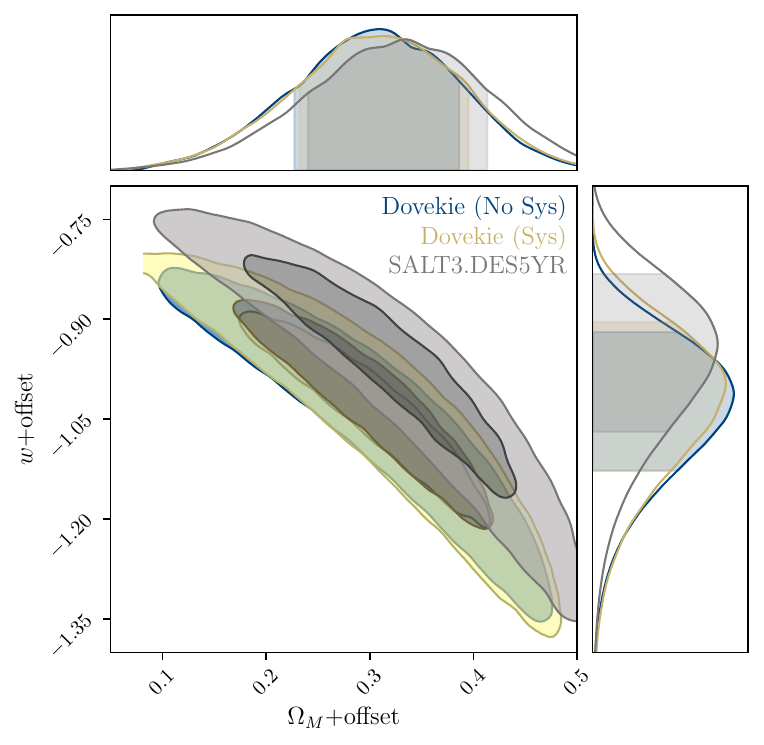}
    \caption{$w$/$\Omega_M$ contours from the Dovekie training sample. The nominal contour, with increased filter uncertainties, is shown in gold; statistical-only contour is shown in grey. For comparison, we show systematic uncertainties from the SALT3.DES5YR surface in blue. For visual clarity, we have added an offset to center the SALT3.DES5YR contour at the values found by Pantheon+; any shifts in Dovekie from this central value are reflective of real changes due to new calibration. The appearance of a green contour is from the strong overlap between the gold systematic and blue no-systematic contours. }
    \label{fig:DOVCONTOURS}
\end{figure}

Table \ref{tab:cosmoparamsPP} provides a layout of the changes in $w$ and $\Omega_M$ from Pantheon+ to Dovekie when considering SN+CMB. The contours in $w$ and $\Omega_M$ are shown in Figure \ref{fig:DOVCONTOURS}. We see the impact of Figure \ref{fig:DELTAMUSYS}; the systematics coherently shifting towards more negative magnitudes moves the systematic contour closer to the grey SALT3.DES5YR results. The change in systematic uncertainties arising from the new Dovekie calibration solution is given in Table \ref{tab:FragUncert}; in the rest of this Section we investigate the origin of these changes.

We also test time-varying dark energy with a present-day component $w_0$ and time-varying component $w_a$, according the parameterisation of \citet{Chevallier2001}
\begin{equation}
w(a) = w_0 + (1-a) w_a \;\; . 
\end{equation}
In addition to SN Ia data, we use DR1 results on baryon acoustic oscillations (BAO) from the Dark Energy Spectroscopic Instrument \citep[DESI]{DESI_DR1}, and the CMB-R parameter (a summary statistic which describes the positions of the peaks of the CMB and the small-scale damping, see \citealt{Bond1997}). We follow the procedure described in \cite{Camilleri24}, in which the BAO data is processed to describe distances as the relative angular size of the BAO compared to that on the surface of last scattering, and adopt the same prior for the CMB-R parameter.

This data combination is broadly consistent with the cosmological analysis of \citet{DESI_DR1}, but with somewhat weaker statistical power owing to the smaller numbers of SN Ia, the compression of the CMB likelihood, and the absence of a lensing likelihood. Nevertheless, it is sufficient for our purpose. Priors used are $\Omega_m \in (0.1,0.9)$, $w_0 \in (-10,5)$ and $w_a \in (-20,10)$. Chains are run using \texttt{Polychord} \citep{Handley15a, Handley15b}.

We find no notable changes to the $w_0$ and $w_a$ uncertainties, as shown in Table \ref{tab:w0waparamsPP}. The statistical uncertainties for $w_0$ and $w_a$ are still dominant. 

\begin{table}
    \centering
    \begin{tabular}{c|c}
         Parameter & P+ $-$ Dovekie Value \\
         \hline
         $\Delta w$ & 0.084 \\
         $\Delta \Omega_M$ & 0.028  \\
         $\Delta \Omega_M$ (Flat $\Lambda$CDM) &  0.009 \\
    
    \end{tabular}
    \caption{Estimated changes in cosmology parameters (SN+CMB) from Pantheon+ to Dovekie, including calibration systematics.}
    \label{tab:cosmoparamsPP}
\end{table}

\begin{table}
    \centering
    \begin{tabular}{c|c|c}
         Parameter & P+ Value & Dovekie Value \\
         \hline
         $\sigma_{w_0}$(stat+syst) & 0.059 & 0.058  \\
         $\sigma_{w_a}$(stat+syst) & 0.322 & 0.324 
    \end{tabular}
    \caption{Stat+Syst (photometric uncertainty) $w_0$ and $w_a$ uncertainties for SALT3.DES5YR and Dovekie when applied to the Pantheon+ like sample with BAO and CMB constraints.}
    \label{tab:w0waparamsPP}
\end{table}

\subsection{Impact of Changing SALT Model}

\begin{figure}
    \centering
    \includegraphics[width=9cm]{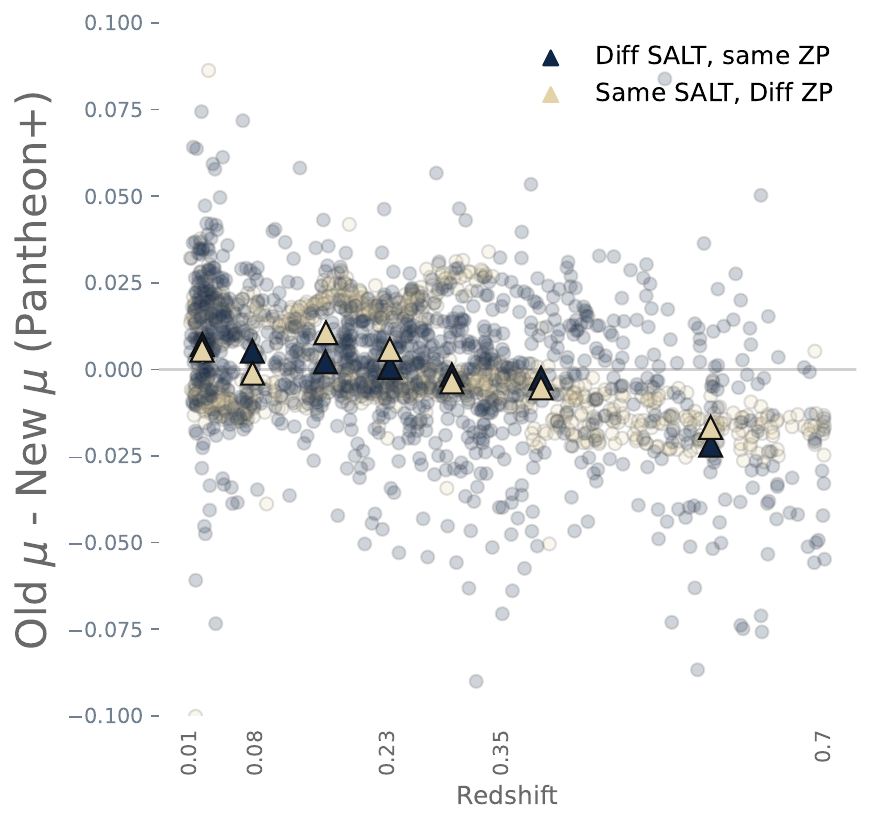}
    \caption{We show the impact of calibration in both model construction and light-curve fitting. Shown is the difference in inferred $\mu$ values from the Pantheon+-like dataset as a function of redshift; in gold, we apply Dovekie or \citetalias{fragilistic} zero-points during light-curve fitting with the SALT3.DES5YR model. In blue, we show the differences when fitting all light curves with the same zero-points but Dovekie/SALT3.DES5YR models}
    \label{fig:DOV-W-DES-ZP}
\end{figure}

Here, we keep the zero-points from \citetalias{fragilistic}, and only replace the SALT model when fitting the Pantheon+ Dovekie sample. Figure \ref{fig:DOV-W-DES-ZP} shows the difference in derived distances when using the same zero-points. The resulting $\Delta w = 0.037$ is roughly half of the total change in $w$ from the Dovekie analysis. 

Comparatively, we modify only the zero-points and keep the SALT surface as Dovekie; the difference in inferred distances is given in Figure \ref{fig:DOV-W-DES-ZP}.  The $\Delta w = 0.0429$ is complementary to the previously-found $0.037$, the two combined nearly entirely accounting for the observed shift of $\Delta w = 0.078$ between Pantheon+ and the full Dovekie calibration solution. Unsurprisingly, given the similarities between the SALT3.DES5YR and Dovekie SALT surfaces, the scatter in distances from varying zero-points is larger than that of varying SALT models. Of note in Figure \ref{fig:DOV-W-DES-ZP} is SDSS, with an average offset from the other surveys of $\sim 0.020$; these new offsets are from the improved DA WD constraints.

\section{Impact on DES5YR Cosmology}\label{sec:DES5YR}

For the DES5YR sample, we do not perform a full reanalysis, leaving this to {Popovic et al. 2025b, in prep}. Here, we give a short overview of the expected changes to the DES5YR photometric systematic uncertainty. We use the SuperNNova \citep{Moller19} from DES5YR \citep{DES5YR} to provide the SN\,Ia probability; however, we use the \cite{Hlozek12} core-collapse prior to model our non-Ia contamination. The Dovekie SALT model provides measurements with lower Hubble scatter than SALT3.DES5YR.

\begin{figure}
    \centering
    \includegraphics[width=9cm]{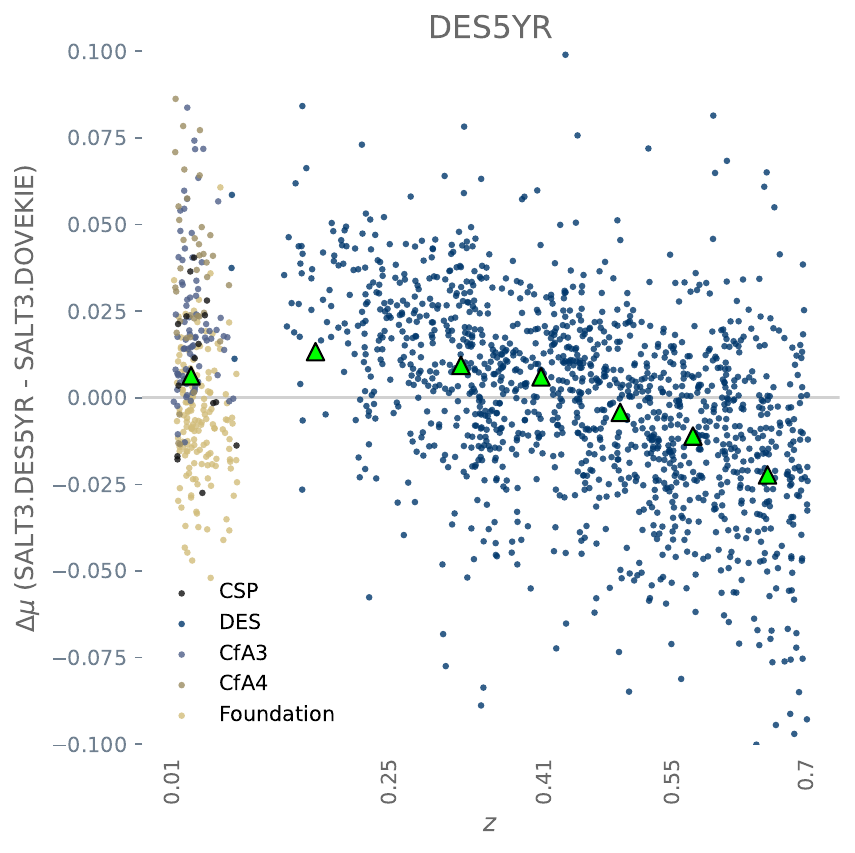}
    \caption{We apply the SALT3.DES5YR and Dovekie models to the DES5YR data set. Shown is the difference in inferred $\mu$ values as a function of redshift, colour-coded by the survey. We compare \textbf{different SALT surfaces and different ZP offsets.} }
\label{fig:DELTAMUDES5YR}
\end{figure} 

\begin{figure}
    \centering
    \includegraphics[width=9cm]{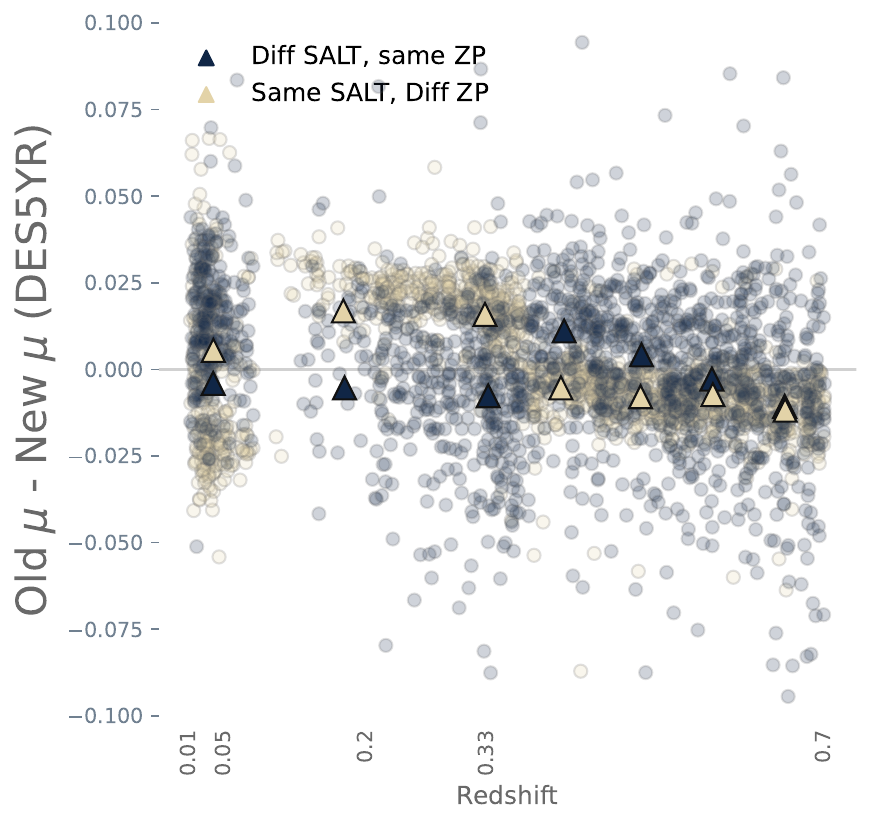}
    \caption{We show the impact of calibration in both model construction and light-curve fitting. Shown is the difference in inferred $\mu$ values from the DES5YR dataset as a function of redshift; in gold, we apply Dovekie or \citetalias{fragilistic} zero-points during light-curve fitting with the SALT3.DES5YR model. In blue, we show the differences when fitting all light curves with the same zero-points but Dovekie/SALT3.DES5YR models.}
    \label{fig:DELTAMUDES5YRZP}
\end{figure}

Figure \ref{fig:DELTAMUDES5YR} shows the difference in inferred $\mu$ values for the DES5YR data set. There is a definite slope in the DES survey as a function of redshift compared to the SALT3.DES5YR surface that will likely impact cosmological results. We also show how these differences in net distances compare to the effects of the calibration as applied to light-curve fits as compared to the new SALT surfaces in Figure \ref{fig:DELTAMUDES5YRZP}.

\cite{DES5YR} did not directly report a calibration systematic uncertainty when combined with a CMB prior; here we calculate $\sigma_w(\rm{phot}) = 0.023$ when using SALT3.DES5YR. In comparison, we find a new Dovekie systematic uncertainty of $\sigma_w(\rm{phot}) = 0.019$, a modest improvement to \citetalias{fragilistic}.

\begin{table}
    \centering
    \begin{tabular}{c|c}
        Survey & Average $\Delta \mu$  \\
        \hline
        DES (5YR) & $-0.008 \pm 0.001$ \\ 
        Foundation & $-0.015 \pm 0.001$ \\
        CSP & $-0.001 \pm 0.001$ \\
        CfA3 & $0.014 \pm 0.003$ \\
        CfA4 & $0.030 \pm 0.003$ \\
        \hline
        SDSS & $0.018 \pm 0.001$  \\
        SNLS & $ -0.035 \pm 0.002$ \\
        CSP &  $ 0.012 \pm 0.005$ \\
        DES (3YR) & $-0.032 \pm 0.001$ \\
        PS1 & $-0.012 \pm 0.001$ \\
        CfA3 &  $0.013 \pm 0.005$ \\
        Foundation &  $-0.004 \pm 0.002$ \\
    \end{tabular}
    \caption{The error-weighted average $\Delta \mu$ (SALT3.DES5YR - Dovekie) for the DES5YR and Pantheon+ data.}
    \label{tab:D5YROFFSETS}
\end{table}

Table \ref{tab:D5YROFFSETS} shows the error-weighted average distances difference between SALT3.DES5YR and Dovekie. Most changes are on the order of a milimag; however, the largest changes are for CfA4, SNLS, DES3YR, and CfA (Pantheon+). The CfA4 and SNLS changes are consistent with the new filters; however, the DES and CfA (Pantheon+) changes are due to strong outliers. In particular, the scatter in the CfA (Pantheon+) $\Delta \mu$ is 0.12, indicating the presence of outliers.

\section{Discussion}\label{sec:Discussion}

\subsection{Filter characterisation}

In order to minimise the number of filter changes necessary, \citetalias{fragilistic} change the PS1-g band by 30\r{A}, to bring, in particular, SNLS and SDSS in line. Philosophically, we take a different approach, choosing instead to preserve the band passes of the reference survey and modify other surveys as necessary to match.  This change in g-band definition between \citetalias{fragilistic} and Dovekie can explain most of the filter changes, with the exception of CfA4. We believe that the CfA4 filters were likely provided in energy-counting units rather than photon-counting units; the application of $X=-1$ in Equation \ref{eq:bandpassweighting} would convert this into the expected units. 

We find that a `re-calibration' of filter uncertainties is necessary; compared to the historical $1\sigma$ uncertainties used in JLA, DES, and Pantheon+, around 6\r{A} for modern AB surveys, we find noticeably larger values as we progress into redder wavelengths. Unlike previous works, which do not offer insight into the origins of these quoted uncertainties, we have presented a new method of estimating these uncertainties.

The increased uncertainties also point to the limits of the tertiary standard method employed here. The SALT training process requires spectral and phase coverage that can, today, only be provided by surveys such as CfA3 and SNLS, which have run their course. Because these surveys have concluded and cannot take any further measurements, any cross-calibration process to include them can only ever use the existing star observations and filter estimates that they have provided. 
Constraints on filter bandpasses depend strongly on both statistics and photometric repeatability. Even in the case of DES, with $\times4$ the number of stars used for filter characterisation as SDSS, SNLS, and PS1, we do not see significant improvements to the filter uncertainties. The precision of laboratory-based filter measurements do not appear to accurately predict the colour responses of the tertiary standards we measure here. For example, we shift the CSP-$B$ band by $70\AA$, nearly 10 times the nominal uncertainties from \cite{Krisciunas20} of $<8\AA$. We consider that such shifts are likely due to elements of the survey-mode telescope observing conditions not directly measured under laboratory conditions. Accordingly we believe that the shifts to bandpasses we apply are better understood without reference to laboratory filter measurements.

\subsection{Zero-Point Offsets}

Our zero-point offsets are individually consistent with \citetalias{fragilistic}. Of interest is the comparison of our quoted uncertainties on the ZP offsets. For low-redshift surveys, we have maintained the same error floor of 0.01 magnitudes; both \citetalias{fragilistic} and Dovekie are entirely error-floor dominated for low-z surveys. 

Comparatively, those surveys with DA WD observed by \cite{Boyd25} see improvements of $\times 2$ on the associated ZP offset uncertainty. These reduced uncertainties, from well-modelled and direct observations of novel calibrator stars, represent an exciting avenue for future calibration efforts in LSST and Roman, which should be able to gather sufficient phase and spectral information to eschew older surveys, and further winnow the systematic uncertainties associated with calibration. 

\subsection{Cosmology}

The new Dovekie SALT model matches or exceeds the previous SALT3.DES5YR model across the board when comparing Hubble Residual scatter. Due to the improved calibration constraints, the overall distance scatter across the nominal + 9 SALT surfaces is reduced compared to SALT3.DES5YR, even considering the increased filter uncertainties. 

These increased filter uncertainties from the Dovekie pipeline more accurately reflect the systematic uncertainty arising from the cross-calibration of telescopes; however, it does not present significant reductions to the systematic uncertainty. 

Both Dovekie approaches -- the increased filter uncertainties and the historical ones -- represent a significant improvement to the Pantheon+ systematic uncertainties, around a $\times 1.5$ reduction to $\sigma_w$(phot). This reduction comes from a decrease in the Hubble residual scatter, both within the nominal SALT surface, and across the nominal + 9 systematic SALT surfaces. 

The reductions in systematic uncertainty for $w_0$ and $w_a$ from Dovekie are not notable, indicating that our constraints on $w_0$ and $w_a$ are still strongly statistically-dominated. 

We have not performed a reanalysis of the DES5YR data set in this paper, nor any initial estimations of changes in cosmological parameters. Because of the recency of the DES5YR results, and the relative ease of the analysis chain only containing 3 distinct surveys, compared to the $\sim20$ of Pantheon+, we instead leave the reanalysis of DES5YR to {Popovic et al. 2025b, in prep}. However, we note that \cite{Tang25} shows that shifts in $\Omega_M$ can propagate to changes in $w_a$, potentially showing the impact of this new calibration. 

However, from estimates of the systematic photometric calibration uncertainty, we see only modest changes from \citetalias{fragilistic} ($\sigma_w(\rm{phot}) = 0.023$) to Dovekie ($\sigma_w(\rm{phot}) = 0.019$) in DES5YR.

\section{Conclusion}\label{sec:Conclusion}

We present Dovekie, a cross-calibration of 11 different photometric systems, with open-source code and the ability for users to add more surveys, some of which are included in the associated software release. Through the use of PS1 and Gaia photometry and spectroscopy, we provide new filter uncertainties and adjust published filters to best match the data.

Using novel DA white dwarf modeling from \cite{Boyd25}, we cross-calibrate our photometric systems and derive a covariance matrix between our zero-points. We find acceptable agreement between \citetalias{fragilistic} and Dovekie.

We retrain SALT with the new Dovekie calibration, and an additional 9 realisations to construct an estimate of the systematic calibration uncertainties. Fitting the Pantheon+ SN, we compare our distances to the previous state-of-the-art SALT model and propagate these new distances to cosmology.

We find significantly improved constraints on the systematic uncertainty associated with calibration: $\sigma_w(\rm{phot}) \approx 0.015$, an improvement of $\times1.5$ over Pantheon+. This increased precision is accompanied by a Flat $w$CDM $\Delta w = 0.0758$, explained by updated zero-points and the new SALT surface, which would result in a Pantheon+ $w = -1.06$ when combined with a CMB prior. 

Fitting the DES5YR sample, we find modest improvements to $\sigma_w(\rm{phot})$. We leave further analysis of the DES5YR sample to {Popovic et al 2025b}, where we redo the entire DES5YR analysis with our improved calibration.

\section{Acknowledgements}

This work has made use of data from the European Space Agency (ESA) mission Gaia (https://www.cosmos.esa.int/gaia), processed by the Gaia Data Processing and Analysis Consortium (DPAC, https://www.cosmos.esa.int/web/gaia/dpac/consortium). Funding for the DPAC has been provided by national institutions, in particular the institutions participating in the Gaia Multilateral Agreement.
This job has made use of the Python package GaiaXPy, developed and maintained by members of the Gaia Data Processing and Analysis Consortium (DPAC), and in particular, Coordination Unit 5 (CU5), and the Data Processing Centre located at the Institute of Astronomy, Cambridge, UK (DPCI). This project has received funding from the European Research Council (ERC) under the European Union’s Horizon 2020 research and innovation program (grant agreement n 759194 - USNAC). \cite{PIPPIN} was used, and is good. BP would like you acknowledge you, dear reader, and ongoing conservation efforts in Iceland \href{https://www.government.is/topics/environment-climate-and-nature-protection/national-parks-and-protected-areas/}{here}, and thank Tamara Davis for discussions over the paper.
This work has been supported by the research project grant “Understanding the Dynamic Universe” funded by the Knut and Alice Wallenberg Foundation under Dnr KAW 2018.0067 and the Swedish Research Council, project 2020-03444.
L.G. acknowledges financial support from AGAUR, CSIC, MCIN and AEI 
0.13039/501100011033 under projects PID2023-151307NB-I00, PIE 20215AT016, CEX2020-001058-M, ILINK23001, COOPB2304, and 2021-SGR-01270. BMB is supported by the Cambridge Centre for Doctoral Training in Data-Intensive Science funded by the UK Science and Technology Facilities Council (STFC). AD and MG are supported by the European Union’s Horizon 2020 research and innovation programme under ERC Grant Agreement No. 101002652 and Marie Skłodowska-Curie Grant Agreement No. 873089. EEH is supported
by a Gates Cambridge Scholarship (\#OPP1144).

\newpage
\
\newpage

\section{Appendix}

\subsection{Additional Diagnostics for SALT Surfaces}

\begin{figure}
    \centering
    \includegraphics[width=8cm]{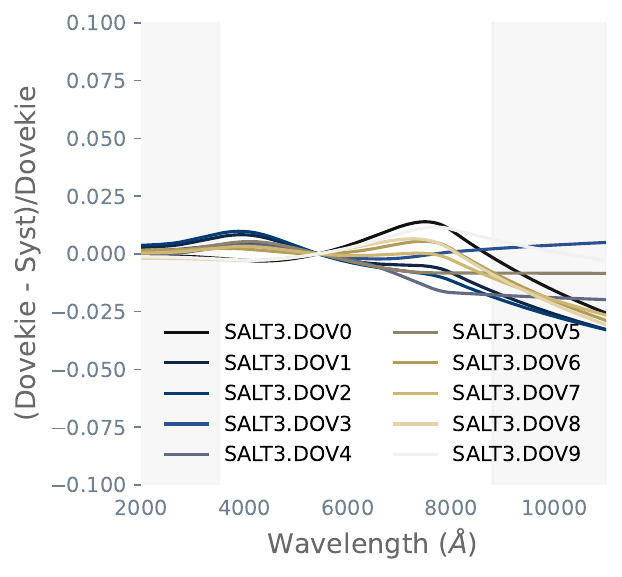}
    \caption{The fractional differences of the SALT3 colour law CL$(\lambda)$ of the systematic Dovekie surfaces (DOV0-9) as compared to the nominal Dovekie surface, as a function of wavelength.}
    \label{fig:SYSTCL}
\end{figure}

Here we present additional information on our systematic surfaces for Dovekie. Figure \ref{fig:SYSTCL} shows the fractional differences of the SALT3 colour law CL$(\lambda)$ compared to the nominal surface. The changes to the colour law are comparatively small, constrained to less than 0.025 mags within the valid range of the Dovekie model. 

\begin{figure*}
    \centering
    \includegraphics[width=18cm]{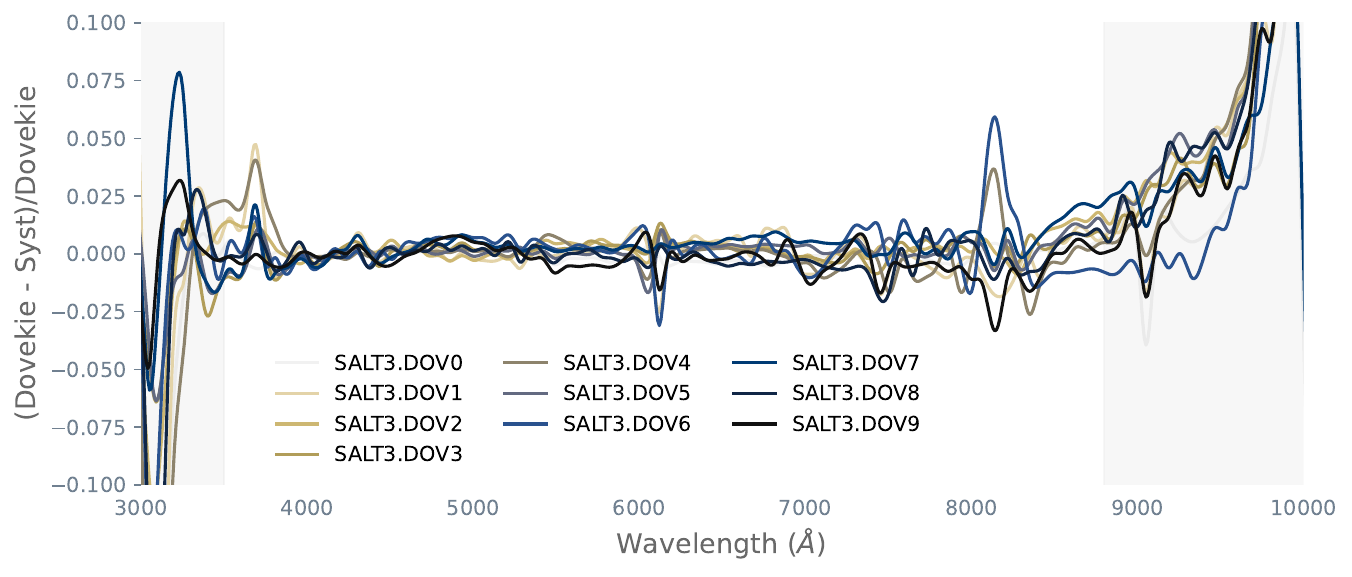}
    \caption{The fractional differences of the M0 component of the systematic Dovekie surfaces (DOV0-9) as compared to the nominal Dovekie surface, as a function of wavelength.}
    \label{fig:SYSTM0}
\end{figure*}

Similarly we present Figure \ref{fig:SYSTM0}, the fractional changes to the M0 values across the Dovekie surfaces. The M0 values are less well-constrained than the SALT colour law, exhibiting some larger changes where the photometric data no longer spans the entire wavelength range. 

\subsection{Changes to Bandpasses and Photometric Uncertainties}

\begin{table}
    \centering
    \begin{tabular}{c|c|c}
        Survey & Prior & Floor \\
         \hline
         PS1-Public & $\mathcal{N}(0, 0.01)$  & N/A  \\
         PS1-SN & $\mathcal{N}(0, 0.01)$ & 0.003 \\
         PS1-Foundation & $\mathcal{N}(0, 0.01)$ & 0.003 \\
         CfA3K & $\mathcal{N}(0, 0.1)$  & 0.01\\
         CfA3S & $\mathcal{N}(0, 0.1)$ & 0.01\\
         CfA4p1 & $\mathcal{N}(0, 0.1)$ & 0.01\\
         CfA4p2 & $\mathcal{N}(0, 0.1)$ & 0.01\\
         CSP & $\mathcal{N}(0, 0.1)$& 0.01 \\
         DES & $\mathcal{N}(0, 0.01)$ & 0.005\\
         SDSS & $\mathcal{N}(0, 0.01)$ & 0.005\\
         SNLS & $\mathcal{N}(0, 0.01)$& 0.005
    \end{tabular}
    \caption{Priors and error floors for each survey in Dovekie. Low-redshift surveys have a higher error floor.}
    \label{tab:priors}
\end{table}

\begin{figure*}
    \centering
    \includegraphics[width=18cm]{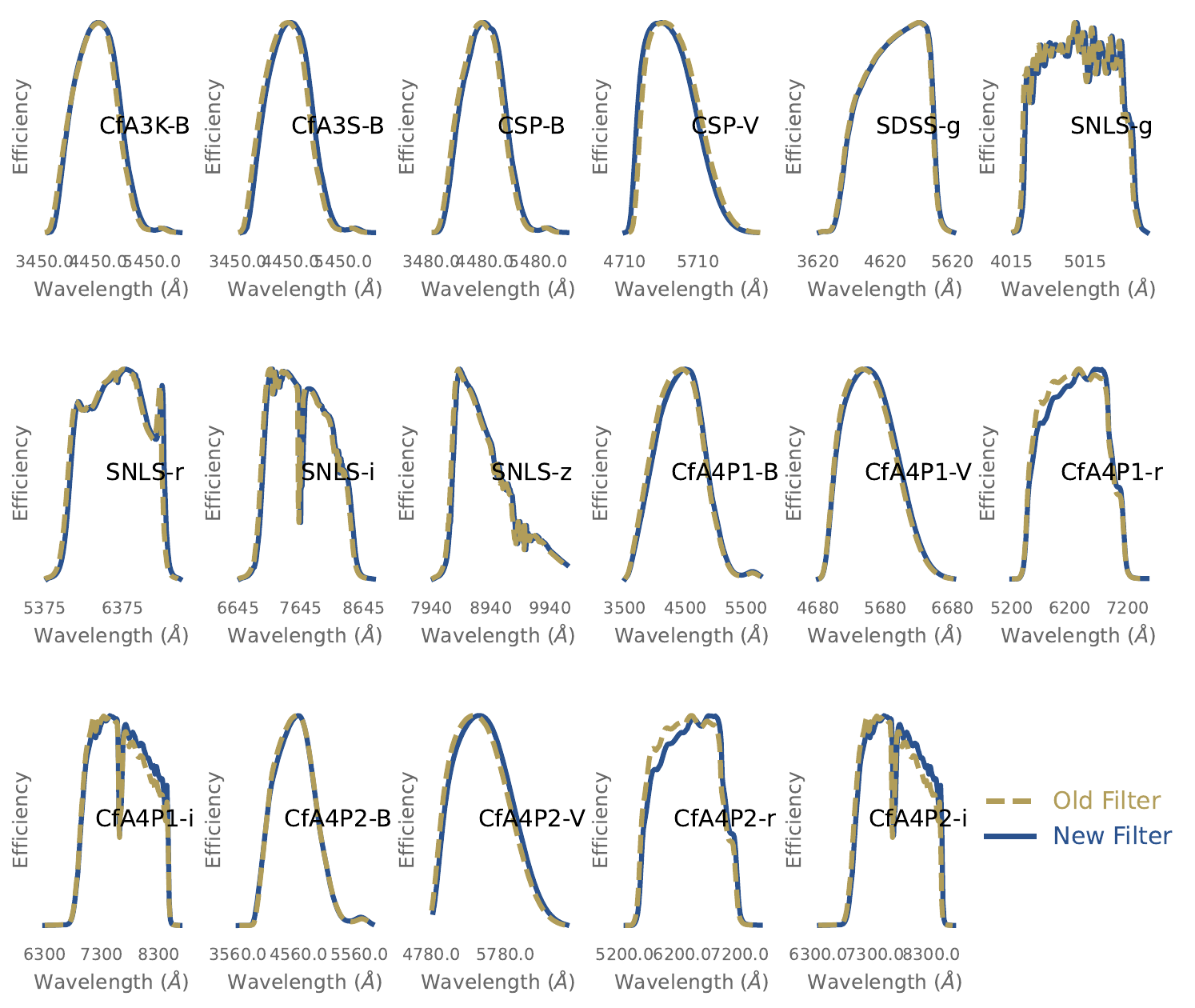}
    \caption{A visualisation of the filter changes performed in this analysis. The original published filter is presented in gold dashed line; the modified version used in this analysis is presented in blue line. A summary of the changes performed is given in Table \ref{tab:bandpasses}; both the original and modified bandpasses are available in the Dovekie github.}
    \label{fig:changedfilters}
\end{figure*}

In Figure \ref{fig:changedfilters}, we show changes to the filters that were done over the course of the analysis. These changes are listed in Table \ref{tab:bandpasses}, and updates to the associated uncertainties are given in Table \ref{tab:ShiftUncertainties}.

\begin{figure*}
    \centering
    \includegraphics[width=18cm]{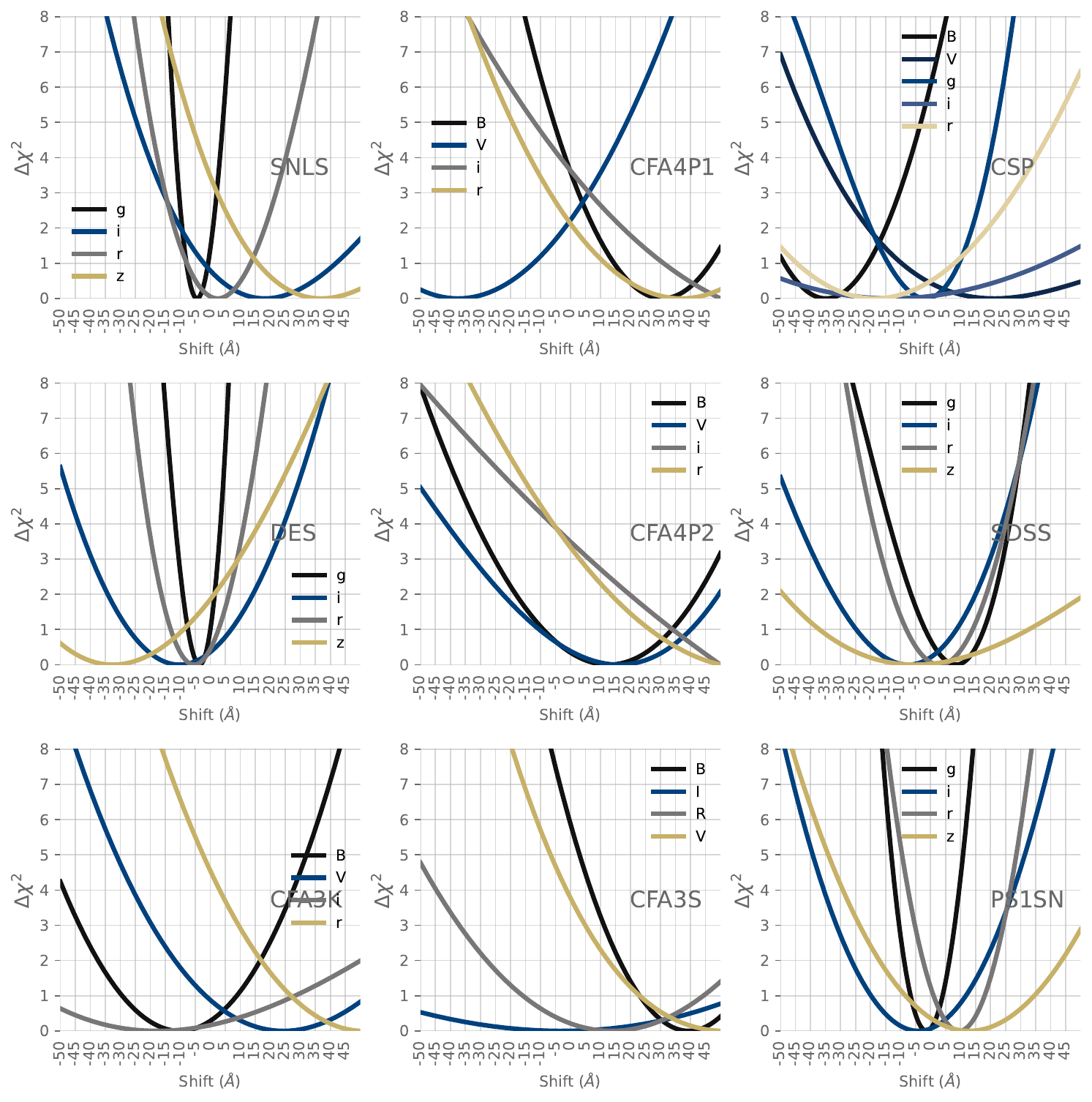}
    \caption{Posteriors for each filter in the surveys cross-calibrated in Dovekie, from PS1 and Gaia filter characterisation methods. For the g band, we only use the PS1 constraints. `PS1SN' represents Foundation as well.}
    \label{fig:Filters1sigma}
\end{figure*}

\begin{figure*}
    \centering
    \includegraphics[width=18cm]{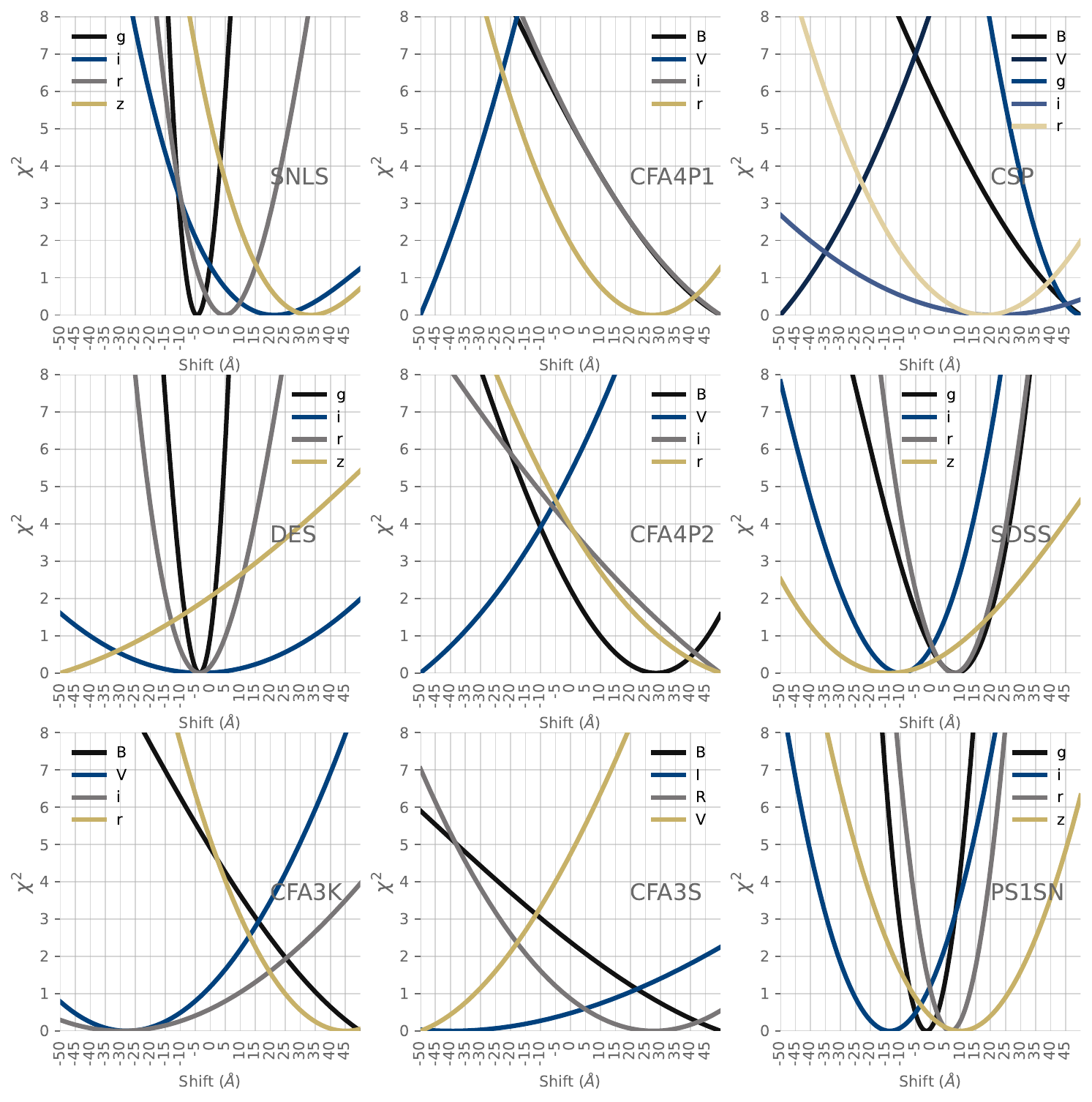}
    \caption{Posteriors for each filter in the surveys cross-calibrated in Dovekie, using PS1 as the reference survey.  }
    \label{fig:1sigmaPS1}
\end{figure*}

As mentioned in Section \ref{sec:Data:subsec:Real:subsubsec:CfA3}, the filters used in \citetalias{fragilistic} for the CfA3 surveys do not appear to match any published provenance. While these filters appear quite accurate, we nonetheless use the original published filters, with the above quoted MODTRAN modifications. Curiously, the \citetalias{fragilistic} filters include telluric absorption, which is not included in the MODTRAN function. Nonetheless, we include both sets of filters in the github for preservation. 

\subsection{Additional Low-$z$ Surveys}

Here we discuss additional low-redshift surveys that were included in \citetalias{fragilistic}, but not included in the nominal Dovekie analysis. These surveys contain SNe Ia hosted in nearby galaxies where distances have been measured with stellar indicators, hence their inclusion in the combined SH0ES and Pantheon+ analysis. We include the data in our data release with the option for users to cross-calibrate these surveys.

\subsubsection{SOUSA}

As of yet, there has not been a published data release of SOUSA stars or SNe, though they are available at \href{https://pbrown801.github.io/SOUSA/}{here}  and \href{https://github.com/pbrown801/SOUSA/tree/master/data}{here}. We include the calibrator stars from the SOUSA stellar photometry pipeline that were used in \citetalias{fragilistic}. We follow \citetalias{fragilistic} in using the most recent release of the CALSPEC VEGA measurement of Alpha Lyr stis010 for initial calibration.

\subsubsection{LOSS 1 \& 2}

We include two separate data releases from the LOSS survey: LOSS1 \citep{Ganeshalingam10} and LOSS2 \citep{Stahl19}. These data releases contain SNe Ia observed by the Nickel and KAIT telescopes, which pass through a variation of different sets of filters. LOSS1 uses the KAIT1-4 and Nickel 2 filter set, and LOSS2 uses the KAIT3-4 and Nickel 2 filter set. LOSS1 provides tertiary standards in the Landolt system and their transformations to the natural system; though the transformations for Nickel 1 and 2 is only valid within a smaller colour range of $0.65< g-i < 0.85$. LOSS2 uses public PS1 stars for their tertiary standards, calibrating the PS1 photometry to Landolt system following \cite{Tonry12}. However, these transformations were applied to generic $BVRI$ filters rather than the PS1 filters, so we follow \cite{Stahl19} to recreate their tertiary catalogue here. 

\begin{table}
    \centering
    \begin{tabular}{c|c}
        Object & Source \\
        \hline
        \hline
        \textcolor{white}{a} & \textbf{PS1-SN}\\ 
        \hline
        Filters & \cite{Tonry12} \href{https://vizier.cfa.harvard.edu/viz-bin/VizieR?-source=J/ApJ/750/99}{($griz$)}. \\
        Atmosphere & Yes \\
        Stars   & \cite{Scolnic18} (Included in github) \\
        System  & AB \\
        Calibration & Multiple Standards \\
        Number of Stars & 656 \\
        Transformations & None \\
        \textcolor{white}{a}  & \textcolor{white}{a} \\
        
        \textcolor{white}{a} & \textbf{PS1-Foundation}\\
        \hline
        Filters & \cite{Tonry12} (\href{https://vizier.cfa.harvard.edu/viz-bin/VizieR?-source=J/ApJ/750/99}{$griz$}) \\
        Atmosphere & Yes \\
        Stars   & \cite{Foley18} (Included in github) \\
        System  & AB \\
        Calibration & Multiple Standards \\
        Number of Stars & 656 \\ 
        Transformations & None \\
        \textcolor{white}{a}  & \textcolor{white}{a} \\

        \textcolor{white}{a} & \textbf{PS1-Public}\\  
        \hline
        Filters &  \cite{Tonry12} \href{https://vizier.cfa.harvard.edu/viz-bin/VizieR?-source=J/ApJ/750/99}{($griz$)}. \\
        Atmosphere & Yes \\
        Stars   & PS1 DR2 \\
        System  & AB \\
        Calibration & Multiple Standards \\
        Number of Stars & N/A \\ 
        Transformations & None \\
        \textcolor{white}{a}  & \textcolor{white}{a} \\
        \textcolor{white}{a} & \textbf{CfA3K}\\  
        \hline
        Filters & \cite{Hicken09b} \href{https://lweb.cfa.harvard.edu/supernova/CfA3/}{($BVri$)} \\
        Atmosphere & No \\
        Stars   & \cite{Hicken09b} \href{https://lweb.cfa.harvard.edu/supernova/CfA3/}{(Link)} \\
        System  & Standard, Natural \\
        Calibration & Landolt, BD17 \\
        Number of Stars & 592 \\
        Transformations & \\
        \textcolor{white}{A} & $(u - b)/(U - B) = 1.0279(69)$ \\
        \textcolor{white}{A} & $(b - v)/(B - V) = 0.9212(29)$ \\
        \textcolor{white}{A} & $(v - V )/(B - V) = 0.0185(23)$ \\
        \textcolor{white}{A} & $(v - r)/(v - r'') = 1.0508(29)$ \\
        \textcolor{white}{A} & $(v - i)/(V - i'') = 1.0185(29)$ \\
        \textcolor{white}{a}  & \textcolor{white}{a} \\

        \textcolor{white}{a} & \textbf{CfA4p1}\\  
        \hline
        Filters & \cite{Hicken12} \href{https://www.cfa.harvard.edu/supernova/CfA4/}{$BVri$} \\
        Atmosphere & Yes \\
        Stars   & \cite{Hicken12} \href{https://www.cfa.harvard.edu/supernova/CfA4/}{(Link)} \\
        System  & Standard, Natural \\
        Calibration & Landolt, BD17 \\
        Number of Stars & 1110 \\ 
        Transformations & \textcolor{white}{a} \\
        \textcolor{white}{a} & $(u - b)/(U - B) = 0.9981(09)$ \\
        \textcolor{white}{a} & $(u - b)/(u' - B) = 0.9089(57)$ \\
        \textcolor{white}{a} & $(b - v)/(B - V ) = 0.9294(26)$ \\
        \textcolor{white}{a} & $(v - V )/(B - V ) = 0.0233(18)$ \\
        \textcolor{white}{a} & $(v - r)/(V - r') = 1.0684(28)$ \\
        \textcolor{white}{a} & $(v - i)/(V - i') = 1.0239(16)$ \\
        \textcolor{white}{a} & \textcolor{white}{a} \\
    \end{tabular}
    \caption{Photometry Sources}
    \label{tab:Photometries}
\end{table}

\begin{table}
    \centering
    \begin{tabular}{c|c}
        Object & Source \\
        \hline
        \hline
        \textcolor{white}{a} & \textbf{SDSS}\\  
        \hline
        Filters & \cite{Doi10} \href{https://iopscience.iop.org/article/10.1088/0004-6256/139/4/1628/pdf}{$griz$} \\
        Atmosphere & Yes \\
        Stars   & \cite{Betoule13} \href{https://cdsarc.u-strasbg.fr/cgi-bin/qcat?J/A+A/552/A124}{(Link)} \\
        System  & AB \\
        Calibration & Multiple Standards \\
        Number of Stars & 953 \\
        Transformations & None \\
        \textcolor{white}{a}  & \textcolor{white}{a} \\

        \textcolor{white}{a} & \textbf{SNLS}\\  
        \hline
        Filters & \cite{Betoule13} \href{https://cdsarc.u-strasbg.fr/cgi-bin/qcat?J/A+A/552/A124}{$griz$} \\
        Atmosphere & Yes \\
        Stars   & \cite{Betoule13} \href{https://cdsarc.u-strasbg.fr/cgi-bin/qcat?J/A+A/552/A124}{(Link)} \\
        System  & AB \\
        Calibration & Multiple Standards \\
        Number of Stars & 819 \\
        Transformations & None\\
        \textcolor{white}{a}  & \textcolor{white}{a} \\

        \textcolor{white}{a} & \textbf{CfA4p2}\\  
        \hline
        Filters & Included in github \\
        Atmosphere & Yes \\
        Stars   & \cite{Hicken12} \href{https://www.cfa.harvard.edu/supernova/CfA4/}{(Link)} \\
        System  & Standard, Natural \\
        Calibration & Landolt, BD17 \\
        Number of Stars & 254 \\
        Transformations & \textcolor{white}{a} \\
        \textcolor{white}{a} & $(u - b)/(U - B) = 0.9981(09)$ \\
        \textcolor{white}{a} & $(u - b)/(u' - B) = 0.9089(57)$ \\
        \textcolor{white}{a} & $(b - v)/(B - V ) = 0.9294(26)$ \\
        \textcolor{white}{a} & $(v - V )/(B - V ) = 0.0233(18)$ \\
        \textcolor{white}{a} & $(v - r)/(V - r') = 1.0684(28)$ \\
        \textcolor{white}{a} & $(v - i)/(V - i') = 1.0239(16)$ \\
        \textcolor{white}{a} & \textcolor{white}{a} \\

    \end{tabular}
    \caption{Photometry Sources (Cont.)}
    \label{tab:Photometries2}
\end{table}

\begin{table}
    \centering
    \begin{tabular}{c|c}
        Object & Source \\
        \hline
        \hline
        \textcolor{white}{a} & \textbf{CFA3S}\\  
        Filters & \cite{Jha06} \href{ https://iopscience.iop.org/article/10.1086/497989/fulltext/204512.tables.html}{BVRI} \\
        Atmosphere & No \\
        Stars   & \cite{Hicken12}  \href{https://lweb.cfa.harvard.edu/supernova/CfA3/}{(Link)} \\
        System  & Standard, Natural  \\
        Calibration & Landolt,BD17  \\
        Number of Stars & 377 \\
        Transformations & \\
        \textcolor{white}{a} & $(u - b)/(U - B) = 0.9912(78)$ \\
        \textcolor{white}{a} & $(b - v)/(B - V) = 0.8928(19)$ \\
        \textcolor{white}{a} & $(v - V )/(B - V) = 0.0336(20)$ \\
        \textcolor{white}{a} & $(v - r)/(V - R) = 1.0855(58)$ \\
        \textcolor{white}{a} & $(v - i)/(V - I) = 1.0166(67)$ \\
        \textcolor{white}{a}  & \textcolor{white}{a} \\

        \hline
        \textcolor{white}{a} & \textbf{CSP}\\  
        Filters & \cite{Krisciunas20} \href{https://csp.obs.carnegiescience.edu/data/filters}{$ugriBV$} \\
        Atmosphere & Yes \\
        Stars   & \cite{Krisciunas17} \href{https://iopscience.iop.org/1538-3881/160/6/289/suppdata/ajabc431t5}{(link)} \\
        System  & Natural\\
        Calibration & Landolt,Smith,BD17  \\
        Number of Stars & 501 \\
        Transformations & \textcolor{white}{a} \\
        \textcolor{white}{a} & $u = u' -cu \times (u' - g') = 0.046(17)$ \\
        \textcolor{white}{a} & $g = g' -cg \times (g' - r') = -0.014(11)$ \\
        \textcolor{white}{a} & $r = r' -cr \times (r' - i') = -0.016(15)$ \\
        \textcolor{white}{a} & $i = i' -ci \times (r' - i') = -0.002(15)$ \\
        \textcolor{white}{a} & $B = B' -cB \times (B' - V') = 0.061(12)$ \\
        \textcolor{white}{a} & $V = V' -cV \times (V' - i') = -0.058(11)$ \\
        \textcolor{white}{a} & \textcolor{white}{a} \\
        
        \textcolor{white}{a} & \textbf{DES}\\  
        \hline
        Filters & \cite{Burke18} (included in github) \\
        Atmosphere & Yes \\
        Stars   & \cite{Rykoff23} \href{https://des.ncsa.illinois.edu/releases/other}{(Link)} \\
        System  & Natural, Landolt, AB \\
        Calibration & C26202 \\
        Number of Stars & 1259 \\
        Transformations & None\\
        \textcolor{white}{a}  & \textcolor{white}{a} \\

    \end{tabular}
    \caption{Photometry Sources (Cont.)}
    \label{tab:Photometries3}
\end{table}

\begin{table}
\centering
\begin{adjustbox}{width=8cm}
\begin{tabular}{c|c}
        Object & Source \\
        \hline
        \hline
        \textcolor{white}{a} & \textbf{LOSS2}\\  
        \hline
        Filters & \cite{Ganeshalingam10} ($bvri$) \\
        Atmosphere & Yes \\
        Stars & \href{https://cfn-live-content-bucket-iop-org.s3.amazonaws.com/journals/0067-0049/190/2/418/1/apjs341025t7_mrt.txt?AWSAccessKeyId=AKIAYDKQL6LTV7YY2HIK&Expires=1639449948&Signature=g\%2B\%2BlR\%2Brz3qogjr\%2FOiE5zsWjSkiU\%3D}{\cite{Ganeshalingam10}}  \\
        System & Landolt  \\
        Calibration & BD17 \\
        Transformations & \textcolor{white}{a} \\
        \textcolor{white}{a} & $b = B + C_B \times (B - V ) = -0.095$ \\
        \textcolor{white}{a} & $v = V + C_V \times (B - V ) = 0.027$ \\
        \textcolor{white}{a} & $r = R + C_R \times (V - R) = -0.181$ \\
        \textcolor{white}{a} & $i = I + C_I \times (V - I) = -0.071$ \\
        \textcolor{white}{a} & \textcolor{white}{a} \\

        \textcolor{white}{a} & \textbf{LOSS1}\\  
        \hline
        Filters & \cite{Stahl19} ($bvri$) \\
        Atmosphere & Yes \\
        Stars  & included in github \\
        System  & AB \\
        Calibration & PS1 \\
        Transformations & \textcolor{white}{a} \\
        \textcolor{white}{a} & $b = B + C_B \times (B - V ) = -0.095$ \\
        \textcolor{white}{a} & $v = V + C_V \times (B - V ) = 0.027$ \\
        \textcolor{white}{a} & $r = R + C_R \times (V - R) = -0.181$ \\
        \textcolor{white}{a} & $i = I + C_I \times (V - I) = -0.071$ \\
        \textcolor{white}{a} & \textcolor{white}{a} \\

        \textcolor{white}{a} & \textbf{SOUSA}\\  
        \hline
        Filters & \cite{Poole08} ($UBV$) \\
        Atmosphere & No \\
        Stars  & included in github, \href{https://github.com/pbrown801/SOUSA/tree/master}{here} \\
        System  & Natural, Vega \\
        Calibration & BD17 \\
        Transformations & \textcolor{white}{a} \\
        \textcolor{white}{a} & $U - V = 0.087 + $ \\
         & $0.8926\times(u - v) + 0.0274\times(u - v)^2$ \\
        \textcolor{white}{a} & $B - V = 0.0148 + 1.0184\times(b - v)$ \\
        \textcolor{white}{a} & $B = b + 0.0173 + $ \\
         & $ 0.0187\times(u - b) + 0.013\times(u - b)^2$ \\
         & $- 0.0108\times(u - b)^3 - $ \\
         & $0.0058\times(u - b)^4 + 0.0026(u - b)^5$ \\
        \textcolor{white}{a} & $V = v + 0.0006 - 0.0113\times(b - v) + $ \\
         & $0.0097\times(b - v)^2 - 0.0036\times(b - v)^3$ \\
        \textcolor{white}{a} & \textcolor{white}{a} \\
\end{tabular}
\end{adjustbox}
\caption{Additional Low-$z$ Photometric Systems }
\label{tab:PhotometriesAp}
\end{table}

\subsection{DA WD Residuals}
\label{sec:App:subsec:wdresids}

\begin{figure}
    \centering
    \includegraphics[width=8cm]{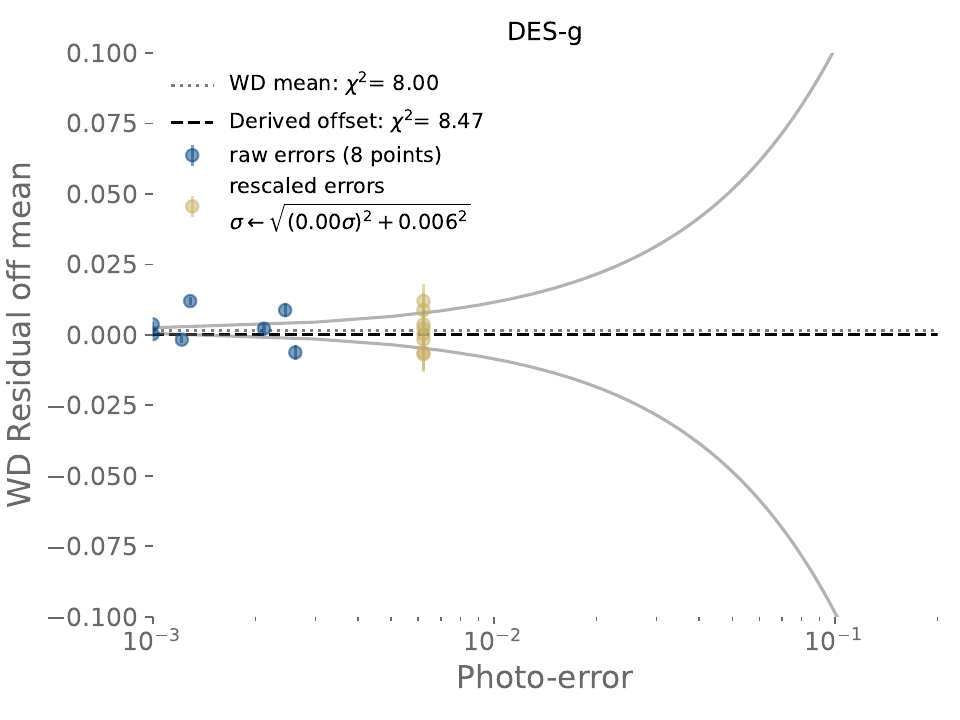}
    \caption{Comparison of residuals for eight white dwarfs between modelled and observed DES-$g$ photometry }
    \label{fig:DESDAg}
\end{figure}

\begin{figure}
    \centering
    \includegraphics[width=8cm]{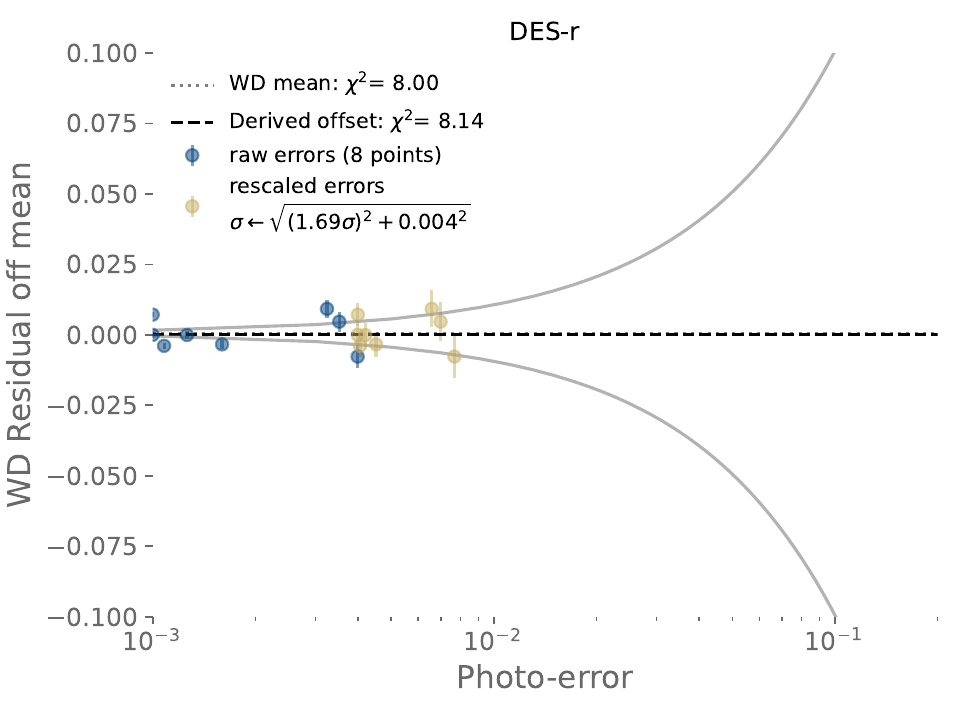}
    \caption{Comparison of residuals for eight white dwarfs between modelled and observed DES-$r$ photometry}
    \label{fig:DESDAr}
\end{figure}

\begin{figure}
    \centering
    \includegraphics[width=8cm]{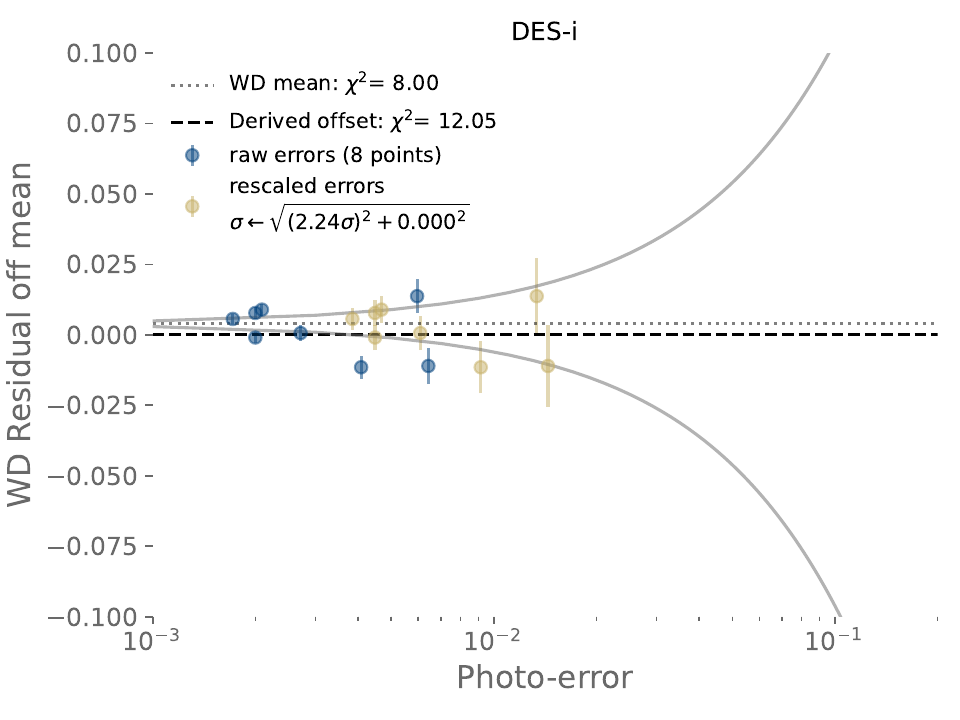}
    \caption{Comparison of residuals for eight white dwarfs between modelled and observed DES-$i$ photometry}
    \label{fig:DESDAi}
\end{figure}

\begin{figure}
    \centering
    \includegraphics[width=8cm]{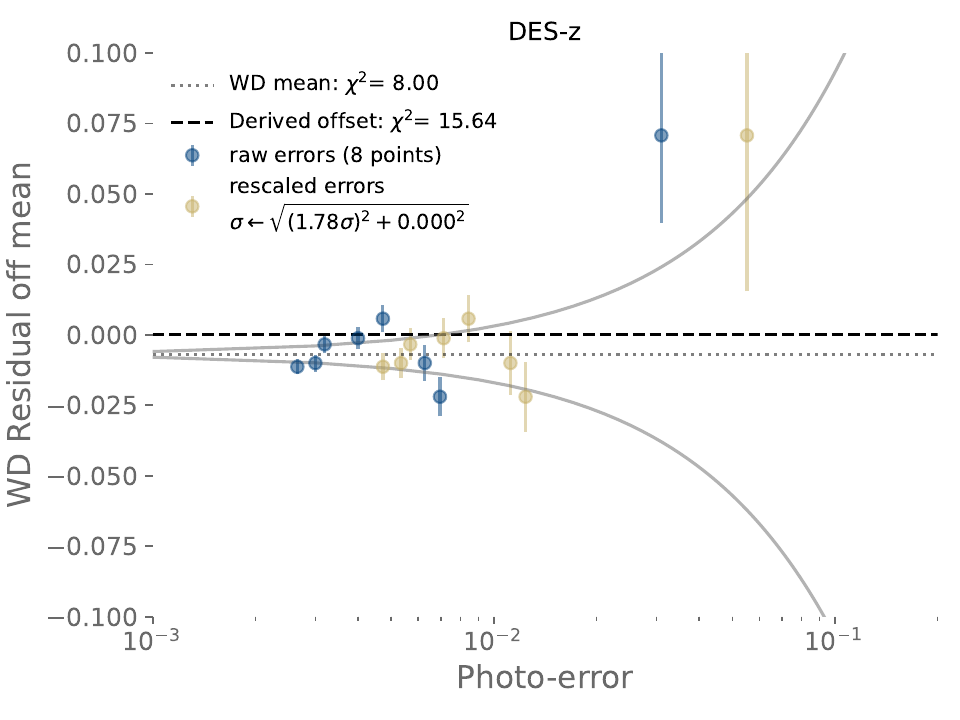}
    \caption{Comparison of residuals for eight white dwarfs between modelled and observed DES-$z$ photometry}
    \label{fig:DESDAz}
\end{figure}

As discussed in Section \ref{sec:Data:subsec:WDs}, we include white dwarf observations from \textit{HST} and associated modelling as constraints on our calibration. We show additional diagnostic plots in Figures \ref{fig:DESDAg}-\ref{fig:DESDAz}, which show the residuals between predicted and observed DES photometric measurements of these WD's, as a function of catalogue photometric uncertainties. We find that these uncertainties are typically understating the scatter of the points, and include empirical parameters to rescale and inflate these uncertainties to the amplitude of observed variation, which are listed in the legend of these figures. These figures also contrast the offset derived from the white dwarfs alone with the offset derived from the entire ensemble as a check of the internal consistency of the dataset. These values are  consistent within $2\sigma$.

\subsection{Recovery of Simulated Offsets}\label{sec:App:subsec:simbiases}

As a test of our pipeline, we simulate 100 samples using the methodology described in Section \ref{sec:Method:sims}. We take the offset values input to the simulation and check that they are recovered by the Dovekie code. We show in Table \ref{tab:simoffsetbias} the mean residual difference in input and recovered values, the RMS scatter of the residual difference, and the corresponding Z-score. Historic low-$z$ $B$ bands appear to be the most difficult to recover, with biases $\sim0.6\sigma$; most filters are recovered with biases lower than $0.1\sigma$. 

\begin{table}[htbp]
    \centering
    \caption{A summary of the biases in simulated offsets from 100 Dovekie simulations.}
    \label{tab:simoffsetbias}
        \begin{tabular}{llll}
                Filter & Bias & Scatter & Z-score \\
                \hline
                CFA3K-B & -0.0064 & 0.0215 & -0.5951 \\
                CFA3K-V & 0.0011 & 0.0116 & 0.1017 \\
                CFA3K-i & -0.0001 & 0.0039 & -0.0060 \\
                CFA3K-r & -0.0001 & 0.0039 & -0.0069 \\
                CFA3S-B & -0.0027 & 0.0171 & -0.2499 \\
                CFA3S-I & 0.0023 & 0.0047 & 0.2102 \\
                CFA3S-R & 0.0011 & 0.0045 & 0.0966 \\
                CFA3S-V & -0.0008 & 0.0114 & -0.0660 \\
                CFA4P1-B & -0.0074 & 0.0163 & -0.6983 \\
                CFA4P1-V & -0.0030 & 0.0108 & -0.2698 \\
                CFA4P1-i & 0.0023 & 0.0041 & 0.2159 \\
                CFA4P1-r & 0.0002 & 0.0039 & 0.0213 \\
                CFA4P2-B & -0.0037 & 0.0123 & -0.3398 \\
                CFA4P2-V & -0.0010 & 0.0106 & -0.0836 \\
                CFA4P2-i & 0.0005 & 0.0050 & 0.0478 \\
                CFA4P2-r & -0.0002 & 0.0053 & -0.0190 \\
                CSP-B & -0.0055 & 0.0188 & -0.5144 \\
                CSP-V & -0.0011 & 0.0123 & -0.0958 \\
                CSP-g & -0.0032 & 0.0059 & -0.2970 \\
                CSP-i & 0.0021 & 0.0045 & 0.1907 \\
                CSP-m & 0.0039 & 0.0123 & 0.3485 \\
                CSP-n & -0.0011 & 0.0123 & -0.0921 \\
                CSP-o & -0.0001 & 0.0123 & -0.0051 \\
                CSP-r & -0.0001 & 0.0044 & -0.0089 \\
                DES-g & -0.0005 & 0.0035 & -0.1784 \\
                DES-i & 0.0004 & 0.0033 & 0.1066 \\
                DES-r & 0.0000 & 0.0037 & 0.0300 \\
                DES-z & -0.0001 & 0.0039 & -0.0086 \\
                Foundation-g & 0.0051 & 0.0038 & 0.7631 \\
                Foundation-i & 0.0001 & 0.0035 & 0.0235 \\
                Foundation-r & -0.0001 & 0.0036 & -0.0134 \\
                Foundation-z & -0.0001 & 0.0046 & -0.0179 \\
                PS1-g & -0.0006 & 0.0038 & -0.1410 \\
                PS1-i & 0.0002 & 0.0033 & 0.0452 \\
                PS1-r & -0.0001 & 0.0032 & -0.0156 \\
                PS1-z & -0.0001 & 0.0043 & -0.0294 \\
                PS1SN-g & 0.0039 & 0.0046 & 0.6700 \\
                PS1SN-i & 0.0002 & 0.0041 & 0.0334 \\
                PS1SN-r & -0.0003 & 0.0040 & -0.0433 \\
                PS1SN-z & 0.0002 & 0.0048 & 0.0370 \\
                SDSS-g & -0.0002 & 0.0038 & -0.0718 \\
                SDSS-i & 0.0001 & 0.0040 & 0.0504 \\
                SDSS-r & -0.0002 & 0.0034 & -0.0420 \\
                SDSS-z & -0.0004 & 0.0050 & -0.0847 \\
                SNLS-g & -0.0005 & 0.0059 & -0.0797 \\
                SNLS-i & -0.0000 & 0.0035 & -0.0014 \\
                SNLS-r & -0.0003 & 0.0032 & -0.0410 \\
                SNLS-z & -0.0001 & 0.0046 & -0.0134 \\
        \hline
        \end{tabular}
\end{table}
\bibliography{research2.bib}

\begin{thebibliography}{80}
\expandafter\ifx\csname natexlab\endcsname\relax\def\natexlab#1{#1}\fi

\bibitem[{{Adame} {et~al.}(2025){Adame}, {Aguilar}, {Ahlen}, {Alam},
  {Alexander}, {Alvarez}, {Alves}, {Anand}, {Andrade}, {Armengaud}, {Avila},
  {Aviles}, {Awan}, {Bahr-Kalus}, {Bailey}, {Baltay}, {Bault}, {Behera},
  {BenZvi}, {Bera}, {Beutler}, {Bianchi}, {Blake}, {Blum}, {Brieden},
  {Brodzeller}, {Brooks}, {Buckley-Geer}, {Burtin}, {Calderon}, {Canning},
  {Carnero Rosell}, {Cereskaite}, {Cervantes-Cota}, {Chabanier}, {Chaussidon},
  {Chaves-Montero}, {Chen}, {Chen}, {Claybaugh}, {Cole}, {Cuceu}, {Davis},
  {Dawson}, {de la Macorra}, {de Mattia}, {Deiosso}, {Dey}, {Dey}, {Ding},
  {Doel}, {Edelstein}, {Eftekharzadeh}, {Eisenstein}, {Elliott}, {Fagrelius},
  {Fanning}, {Ferraro}, {Ereza}, {Findlay}, {Flaugher}, {Font-Ribera},
  {Forero-S{\'a}nchez}, {Forero-Romero}, {Frenk}, {Garcia-Quintero},
  {Gazta{\~n}aga}, {Gil-Mar{\'\i}n}, {Gontcho a Gontcho}, {Gonzalez-Morales},
  {Gonzalez-Perez}, {Gordon}, {Green}, {Gruen}, {Gsponer}, {Gutierrez}, {Guy},
  {Hadzhiyska}, {Hahn}, {Hanif}, {Herrera-Alcantar}, {Honscheid}, {Howlett},
  {Huterer}, {Ir{\v{s}}i{\v{c}}}, {Ishak}, {Juneau}, {Kara{\c{c}}ayl{\i}},
  {Kehoe}, {Kent}, {Kirkby}, {Kremin}, {Krolewski}, {Lai}, {Lan}, {Landriau},
  {Lang}, {Lasker}, {Le Goff}, {Le Guillou}, {Leauthaud}, {Levi}, {Li},
  {Linder}, {Lodha}, {Magneville}, {Manera}, {Margala}, {Martini}, {Maus},
  {McDonald}, {Medina-Varela}, {Meisner}, {Mena-Fern{\'a}ndez}, {Miquel},
  {Moon}, {Moore}, {Moustakas}, {Mueller}, {Mu{\~n}oz-Guti{\'e}rrez}, {Myers},
  {Nadathur}, {Napolitano}, {Neveux}, {Newman}, {Nguyen}, {Nie}, {Niz},
  {Noriega}, {Padmanabhan}, {Paillas}, {Palanque-Delabrouille}, {Pan},
  {Penmetsa}, {Percival}, {Pieri}, {Pinon}, {Poppett}, {Porredon}, {Prada},
  {P{\'e}rez-Fern{\'a}ndez}, {P{\'e}rez-R{\`a}fols}, {Rabinowitz}, {Raichoor},
  {Ram{\'\i}rez-P{\'e}rez}, {Ramirez-Solano}, {Rashkovetskyi}, {Ravoux},
  {Rezaie}, {Rich}, {Rocher}, {Rockosi}, {Roe}, {Rosado-Marin}, {Ross},
  {Rossi}, {Ruggeri}, {Ruhlmann-Kleider}, {Samushia}, {Sanchez}, {Saulder},
  {Schlafly}, {Schlegel}, {Schubnell}, {Seo}, {Shafieloo}, {Sharples},
  {Silber}, {Slosar}, {Smith}, {Sprayberry}, {Tan}, {Tarl{\'e}}, {Taylor},
  {Trusov}, {Ure{\~n}a-L{\'o}pez}, {Vaisakh}, {Valcin}, {Valdes},
  {Vargas-Maga{\~n}a}, {Verde}, {Walther}, {Wang}, {Wang}, {Weaver},
  {Weaverdyck}, {Wechsler}, {Weinberg}, {White}, {Yu}, {Yu}, {Yuan},
  {Y{\`e}che}, {Zaborowski}, {Zarrouk}, {Zhang}, {Zhao}, {Zhao}, {Zhou}, \&
  {Zhuang}}]{DESI_DR1}
{Adame}, A.~G., {Aguilar}, J., {Ahlen}, S., {et~al.} 2025, \jcap, 2025, 021

\bibitem[{{Axelrod} {et~al.}(2023){Axelrod}, {Saha}, {Matheson}, {Olszewski},
  {Bohlin}, {Calamida}, {Claver}, {Deustua}, {Holberg}, {Hubeny}, {Mackenty},
  {Malanchev}, {Narayan}, {Points}, {Rest}, {Sabbi}, \& {Stubbs}}]{axelrod2023}
{Axelrod}, T., {Saha}, A., {Matheson}, T., {et~al.} 2023, \apj, 951, 78

\bibitem[{{Babusiaux} {et~al.}(2023){Babusiaux}, {Fabricius}, {Khanna},
  {Muraveva}, {Reyl{\'e}}, {Spoto}, {Vallenari}, {Luri}, {Arenou},
  {{\'A}lvarez}, {Anders}, {Antoja}, {Balbinot}, {Barache}, {Bauchet},
  {Bossini}, {Busonero}, {Cantat-Gaudin}, {Carrasco}, {Dafonte}, {Diakit{\'e}},
  {Figueras}, {Garcia-Gutierrez}, {Garofalo}, {Helmi}, {Jim{\'e}nez-Arranz},
  {Jordi}, {Kervella}, {Kostrzewa-Rutkowska}, {Leclerc}, {Licata}, {Manteiga},
  {Masip}, {Mongui{\'o}}, {Ramos}, {Robichon}, {Robin}, {Romero-G{\'o}mez},
  {S{\'a}ez}, {Santove{\~n}a}, {Spina}, {Torralba Elipe}, \&
  {Weiler}}]{GAIATHREE}
{Babusiaux}, C., {Fabricius}, C., {Khanna}, S., {et~al.} 2023, \aap, 674, A32

\bibitem[{Barbary {et~al.}(2025)Barbary, Bailey, Barentsen, Barclay, Biswas,
  Boone, Craig, Feindt, Friesen, Goldstein, Jha, Jones, Mondon,
  Papadogiannakis, Perrefort, Pierel, Rodney, Rose, Saunders, Sipőcz,
  Sofiatti, Thomas, van Santen, Vincenzi, Wang, \& Wood-Vasey}]{SNCosmo}
Barbary, K., Bailey, S., Barentsen, G., {et~al.} 2025, SNCosmo

\bibitem[{{Betoule} {et~al.}(2014){Betoule}, {Kessler}, {Guy}, {Mosher},
  {Hardin}, {Biswas}, {Astier}, {El-Hage}, {Konig}, {Kuhlmann}, {Marriner},
  {Pain}, {Regnault}, {Balland}, {Bassett}, {Brown}, {Campbell}, {Carlberg},
  {Cellier-Holzem}, {Cinabro}, {Conley}, {D'Andrea}, {DePoy}, {Doi}, {Ellis},
  {Fabbro}, {Filippenko}, {Foley}, {Frieman}, {Fouchez}, {Galbany}, {Goobar},
  {Gupta}, {Hill}, {Hlozek}, {Hogan}, {Hook}, {Howell}, {Jha}, {Le Guillou},
  {Leloudas}, {Lidman}, {Marshall}, {M{\"o}ller}, {Mour{\~a}o}, {Neveu},
  {Nichol}, {Olmstead}, {Palanque-Delabrouille}, {Perlmutter}, {Prieto},
  {Pritchet}, {Richmond}, {Riess}, {Ruhlmann-Kleider}, {Sako}, {Schahmaneche},
  {Schneider}, {Smith}, {Sollerman}, {Sullivan}, {Walton}, \&
  {Wheeler}}]{Betoule14}
{Betoule}, M., {Kessler}, R., {Guy}, J., {et~al.} 2014, \aap, 568, A22

\bibitem[{{Betoule} {et~al.}(2013){Betoule}, {Marriner}, {Regnault},
  {Cuillandre}, {Astier}, {Guy}, {Balland}, {El Hage}, {Hardin}, {Kessler}, {Le
  Guillou}, {Mosher}, {Pain}, {Rocci}, {Sako}, \& {Schahmaneche}}]{Betoule13}
{Betoule}, M., {Marriner}, J., {Regnault}, N., {et~al.} 2013, \aap, 552, A124

\bibitem[{{Bohlin}(1996)}]{Bohlin96}
{Bohlin}, R.~C. 1996, \aj, 111, 1743

\bibitem[{{Bohlin}(2014)}]{bohlin2014b}
{Bohlin}, R.~C. 2014, \aj, 147, 127

\bibitem[{{Bohlin} {et~al.}(2025){Bohlin}, {Deustua}, {Narayan}, {Saha},
  {Calamida}, {Gordon}, {Holberg}, {Hubeny}, {Matheson}, \& {Rest}}]{Bohlin25}
{Bohlin}, R.~C., {Deustua}, S., {Narayan}, G., {et~al.} 2025, \aj, 169, 40

\bibitem[{{Bohlin} {et~al.}(2014{\natexlab{a}}){Bohlin}, {Gordon}, \&
  {Tremblay}}]{Bohlin14}
{Bohlin}, R.~C., {Gordon}, K.~D., \& {Tremblay}, P.-E. 2014{\natexlab{a}},
  Publications of the Astronomical Society of the Pacific, 126, 711

\bibitem[{{Bohlin} {et~al.}(2014{\natexlab{b}}){Bohlin}, {Gordon}, \&
  {Tremblay}}]{bohlin2014a}
{Bohlin}, R.~C., {Gordon}, K.~D., \& {Tremblay}, P.~E. 2014{\natexlab{b}},
  \pasp, 126, 711

\bibitem[{{Bohlin} {et~al.}(2020{\natexlab{a}}){Bohlin}, {Hubeny}, \&
  {Rauch}}]{Bohlin2020}
{Bohlin}, R.~C., {Hubeny}, I., \& {Rauch}, T. 2020{\natexlab{a}}, \aj, 160, 21

\bibitem[{{Bohlin} {et~al.}(2020{\natexlab{b}}){Bohlin}, {Hubeny}, \&
  {Rauch}}]{Bohlin21}
{Bohlin}, R.~C., {Hubeny}, I., \& {Rauch}, T. 2020{\natexlab{b}}, \aj, 160, 21

\bibitem[{{Bond} {et~al.}(1997){Bond}, {Efstathiou}, \& {Tegmark}}]{Bond1997}
{Bond}, J.~R., {Efstathiou}, G., \& {Tegmark}, M. 1997, \mnras, 291, L33

\bibitem[{Boyd {et~al.}(2025)Boyd, Narayan, Mandel, Grayling, Saha, Axelrod,
  Matheson, Olszewski, Calamida, Do, Bohlin, Holberg, Hubeny, Deustua, Rest,
  Stubbs, Berres, Li, Mackenty, \& Sabbi}]{Boyd25}
Boyd, B.~M., Narayan, G., Mandel, K.~S., {et~al.} 2025, \mnras, 540, 385

\bibitem[{{Brout} {et~al.}(2021){Brout}, {Hinton}, \& {Scolnic}}]{Binning}
{Brout}, D., {Hinton}, S.~R., \& {Scolnic}, D. 2021, \apjl, 912, L26

\bibitem[{{Brout} {et~al.}(2022){Brout}, {Scolnic}, {Popovic}, {Riess},
  {Zuntz}, {Kessler}, {Carr}, {Davis}, {Hinton}, {Jones}, {Kenworthy},
  {Peterson}, {Said}, {Taylor}, {Ali}, {Armstrong}, {Charvu}, {Dwomoh},
  {Palmese}, {Qu}, {Rose}, {Stubbs}, {Vincenzi}, {Wood}, {Brown}, {Chen},
  {Chambers}, {Coulter}, {Dai}, {Dimitriadis}, {Filippenko}, {Foley}, {Jha},
  {Kelsey}, {Kirshner}, {M{\"o}ller}, {Muir}, {Nadathur}, {Pan}, {Rest},
  {Rojas-Bravo}, {Sako}, {Siebert}, {Smith}, {Stahl}, \& {Wiseman}}]{Brout22}
{Brout}, D., {Scolnic}, D., {Popovic}, B., {et~al.} 2022, arXiv e-prints,
  arXiv:2202.04077

\bibitem[{Brout {et~al.}(2021)Brout, Taylor, Scolnic, Wood, Rose, Vincenzi,
  Dwomoh, Lidman, Riess, Ali, Qu, Dai, \& Stubbs}]{fragilistic}
Brout, D., Taylor, G., Scolnic, D., {et~al.} 2021, The Pantheon+ Analysis:
  SuperCal-Fragilistic Cross Calibration, Retrained SALT2 Light Curve Model,
  and Calibration Systematic Uncertainty

\bibitem[{{Brownsberger} {et~al.}(2023){Brownsberger}, {Brout}, {Scolnic},
  {Stubbs}, \& {Riess}}]{Brownsberger23Uncertainties}
{Brownsberger}, S.~R., {Brout}, D., {Scolnic}, D., {Stubbs}, C.~W., \& {Riess},
  A.~G. 2023, \apj, 944, 188

\bibitem[{{Burke} {et~al.}(2018){Burke}, {Rykoff}, {Allam}, {Annis}, {Bechtol},
  {Bernstein}, {Drlica-Wagner}, {Finley}, {Gruendl}, {James}, {Kent},
  {Kessler}, {Kuhlmann}, {Lasker}, {Li}, {Scolnic}, {Smith}, {Tucker},
  {Wester}, {Yanny}, {Abbott}, {Abdalla}, {Benoit-L{\'e}vy}, {Bertin}, {Carnero
  Rosell}, {Carrasco Kind}, {Carretero}, {Cunha}, {D'Andrea}, {da Costa},
  {Desai}, {Diehl}, {Doel}, {Estrada}, {Garc{\'\i}a-Bellido}, {Gruen},
  {Gutierrez}, {Honscheid}, {Kuehn}, {Kuropatkin}, {Maia}, {March}, {Marshall},
  {Melchior}, {Menanteau}, {Miquel}, {Plazas}, {Sako}, {Sanchez}, {Scarpine},
  {Schindler}, {Sevilla-Noarbe}, {Smith}, {Smith}, {Soares-Santos}, {Sobreira},
  {Suchyta}, {Tarle}, {Walker}, \& {DES Collaboration}}]{Burke18}
{Burke}, D.~L., {Rykoff}, E.~S., {Allam}, S., {et~al.} 2018, \aj, 155, 41

\bibitem[{{Calamida} {et~al.}(2019){Calamida}, {Matheson}, {Saha}, {Olszewski},
  {Narayan}, {Claver}, {Shanahan}, {Holberg}, {Axelrod}, {Bohlin}, {Stubbs},
  {Deustua}, {Hubeny}, {Mackenty}, {Points}, {Rest}, \& {Sabbi}}]{calamida2019}
{Calamida}, A., {Matheson}, T., {Saha}, A., {et~al.} 2019, \apj, 872, 199

\bibitem[{{Camilleri} {et~al.}(2024){Camilleri}, {Davis}, {Vincenzi}, {Shah},
  {Frieman}, {Kessler}, {Armstrong}, {Brout}, {Carr}, {Chen}, {Galbany},
  {Glazebrook}, {Hinton}, {Lee}, {Lidman}, {M{\"o}ller}, {Popovic}, {Qu},
  {Sako}, {Scolnic}, {Smith}, {Sullivan}, {S{\'a}nchez}, {Taylor}, {Toy},
  {Wiseman}, {Abbott}, {Aguena}, {Allam}, {Alves}, {Annis}, {Avila}, {Bacon},
  {Bertin}, {Bocquet}, {Brooks}, {Burke}, {Carnero Rosell}, {Carretero},
  {Castander}, {da Costa}, {Pereira}, {Desai}, {Diehl}, {Doel}, {Doux},
  {Everett}, {Ferrero}, {Flaugher}, {Fosalba}, {Garc{\'\i}a-Bellido}, {Gatti},
  {Gaztanaga}, {Giannini}, {Gruen}, {Hollowood}, {Honscheid}, {James}, {Kuehn},
  {Lahav}, {Lee}, {Lewis}, {Marshall}, {Mena-Fern{\'a}ndez}, {Miquel}, {Muir},
  {Myles}, {Ogando}, {Pieres}, {Malag{\'o}n}, {Porredon}, {Rodriguez-Monroy},
  {Sanchez}, {Sanchez Cid}, {Schubnell}, {Sevilla-Noarbe}, {Suchyta},
  {Swanson}, {Tarle}, {Walker}, {Weaverdyck}, \& {DES
  Collaboration}}]{Camilleri24}
{Camilleri}, R., {Davis}, T.~M., {Vincenzi}, M., {et~al.} 2024, \mnras, 533,
  2615

\bibitem[{{Chambers} {et~al.}(2016){Chambers}, {Magnier}, {Metcalfe},
  {Flewelling}, {Huber}, {Waters}, {Denneau}, {Draper}, {Farrow}, {Finkbeiner},
  {Holmberg}, {Koppenhoefer}, {Price}, {Saglia}, {Schlafly}, {Smartt},
  {Sweeney}, {Wainscoat}, {Burgett}, {Grav}, {Heasley}, {Hodapp}, {Jedicke},
  {Kaiser}, {Kudritzki}, {Luppino}, {Lupton}, {Monet}, {Morgan}, {Onaka},
  {Stubbs}, {Tonry}, {Banados}, {Bell}, {Bender}, {Bernard}, {Botticella},
  {Casertano}, {Chastel}, {Chen}, {Chen}, {Cole}, {Deacon}, {Frenk},
  {Fitzsimmons}, {Gezari}, {Goessl}, {Goggia}, {Goldman}, {Grebel}, {Hambly},
  {Hasinger}, {Heavens}, {Heckman}, {Henderson}, {Henning}, {Holman}, {Hopp},
  {Ip}, {Isani}, {Keyes}, {Koekemoer}, {Kotak}, {Long}, {Lucey}, {Liu},
  {Martin}, {McLean}, {Morganson}, {Murphy}, {Nieto-Santisteban}, {Norberg},
  {Peacock}, {Pier}, {Postman}, {Primak}, {Rae}, {Rest}, {Riess}, {Riffeser},
  {Rix}, {Roser}, {Schilbach}, {Schultz}, {Scolnic}, {Szalay}, {Seitz},
  {Shiao}, {Small}, {Smith}, {Soderblom}, {Taylor}, {Thakar}, {Thiel},
  {Thilker}, {Urata}, {Valenti}, {Walter}, {Watters}, {Werner}, {White},
  {Wood-Vasey}, \& {Wyse}}]{Chambers16}
{Chambers}, K.~C., {Magnier}, E.~A., {Metcalfe}, N., {et~al.} 2016, ArXiv
  e-prints [\eprint[arXiv]{1612.05560}]

\bibitem[{{Chevallier} \& {Polarski}(2001)}]{Chevallier2001}
{Chevallier}, M. \& {Polarski}, D. 2001, International Journal of Modern
  Physics D, 10, 213

\bibitem[{{Colina} \& {Bohlin}(1994)}]{Colina94}
{Colina}, L. \& {Bohlin}, R.~C. 1994, \aj, 108, 1931

\bibitem[{{Currie} {et~al.}(2020){Currie}, {Rubin}, {Aldering}, {Deustua},
  {Fruchter}, \& {Perlmutter}}]{Currie20}
{Currie}, M., {Rubin}, D., {Aldering}, G., {et~al.} 2020, arXiv e-prints,
  arXiv:2007.02458

\bibitem[{{Dhawan} {et~al.}(2020){Dhawan}, {Brout}, {Scolnic}, {Goobar},
  {Riess}, \& {Miranda}}]{Dhawan2020}
{Dhawan}, S., {Brout}, D., {Scolnic}, D., {et~al.} 2020, \apj, 894, 54

\bibitem[{{Doi} {et~al.}(2010){Doi}, {Tanaka}, {Fukugita}, {Gunn}, {Yasuda},
  {Ivezi{\'c}}, {Brinkmann}, {de Haars}, {Kleinman}, {Krzesinski}, \& {French
  Leger}}]{Doi10}
{Doi}, M., {Tanaka}, M., {Fukugita}, M., {et~al.} 2010, \aj, 139, 1628

\bibitem[{{Foley} {et~al.}(2018){Foley}, {Scolnic}, {Rest}, {Jha}, {Pan},
  {Riess}, {Challis}, {Chambers}, {Coulter}, {Dettman}, {Foley}, {Fox},
  {Huber}, {Jones}, {Kilpatrick}, {Kirshner}, {Schultz}, {Siebert},
  {Flewelling}, {Gibson}, {Magnier}, {Miller}, {Primak}, {Smartt}, {Smith},
  {Wainscoat}, {Waters}, \& {Willman}}]{Foley18}
{Foley}, R.~J., {Scolnic}, D., {Rest}, A., {et~al.} 2018, \mnras, 475, 193

\bibitem[{{Fukugita} {et~al.}(1996){Fukugita}, {Ichikawa}, {Gunn}, {Doi},
  {Shimasaku}, \& {Schneider}}]{Fukugita96}
{Fukugita}, M., {Ichikawa}, T., {Gunn}, J.~E., {et~al.} 1996, \aj, 111, 1748

\bibitem[{{Gaia Collaboration} {et~al.}(2016){Gaia Collaboration}, {Prusti},
  {de Bruijne}, {Brown}, {Vallenari}, {Babusiaux}, {Bailer-Jones}, {Bastian},
  {Biermann}, {Evans}, {Eyer}, {Jansen}, {Jordi}, {Klioner}, {Lammers},
  {Lindegren}, {Luri}, {Mignard}, {Milligan}, {Panem}, {Poinsignon},
  {Pourbaix}, {Randich}, {Sarri}, {Sartoretti}, {Siddiqui}, {Soubiran},
  {Valette}, {van Leeuwen}, {Walton}, {Aerts}, {Arenou}, {Cropper}, {Drimmel},
  {H{\o}g}, {Katz}, {Lattanzi}, {O'Mullane}, {Grebel}, {Holland}, {Huc},
  {Passot}, {Bramante}, {Cacciari}, {Casta{\~n}eda}, {Chaoul}, {Cheek}, {De
  Angeli}, {Fabricius}, {Guerra}, {Hern{\'a}ndez}, {Jean-Antoine-Piccolo},
  {Masana}, {Messineo}, {Mowlavi}, {Nienartowicz}, {Ord{\'o}{\~n}ez-Blanco},
  {Panuzzo}, {Portell}, {Richards}, {Riello}, {Seabroke}, {Tanga},
  {Th{\'e}venin}, {Torra}, {Els}, {Gracia-Abril}, {Comoretto},
  {Garcia-Reinaldos}, {Lock}, {Mercier}, {Altmann}, {Andrae}, {Astraatmadja},
  {Bellas-Velidis}, {Benson}, {Berthier}, {Blomme}, {Busso}, {Carry},
  {Cellino}, {Clementini}, {Cowell}, {Creevey}, {Cuypers}, {Davidson}, {De
  Ridder}, {de Torres}, {Delchambre}, {Dell'Oro}, {Ducourant}, {Fr{\'e}mat},
  {Garc{\'\i}a-Torres}, {Gosset}, {Halbwachs}, {Hambly}, {Harrison}, {Hauser},
  {Hestroffer}, {Hodgkin}, {Huckle}, {Hutton}, {Jasniewicz}, {Jordan},
  {Kontizas}, {Korn}, {Lanzafame}, {Manteiga}, {Moitinho}, {Muinonen},
  {Osinde}, {Pancino}, {Pauwels}, {Petit}, {Recio-Blanco}, {Robin}, {Sarro},
  {Siopis}, {Smith}, {Smith}, {Sozzetti}, {Thuillot}, {van Reeven}, {Viala},
  {Abbas}, {Abreu Aramburu}, {Accart}, {Aguado}, {Allan}, {Allasia},
  {Altavilla}, {{\'A}lvarez}, {Alves}, {Anderson}, {Andrei}, {Anglada Varela},
  {Antiche}, {Antoja}, {Ant{\'o}n}, {Arcay}, {Atzei}, {Ayache}, {Bach},
  {Baker}, {Balaguer-N{\'u}{\~n}ez}, {Barache}, {Barata}, {Barbier}, {Barblan},
  {Baroni}, {Barrado y Navascu{\'e}s}, {Barros}, {Barstow}, {Becciani},
  {Bellazzini}, {Bellei}, {Bello Garc{\'\i}a}, {Belokurov}, {Bendjoya},
  {Berihuete}, {Bianchi}, {Bienaym{\'e}}, {Billebaud}, {Blagorodnova},
  {Blanco-Cuaresma}, {Boch}, {Bombrun}, {Borrachero}, {Bouquillon}, {Bourda},
  {Bouy}, {Bragaglia}, {Breddels}, {Brouillet}, {Br{\"u}semeister},
  {Bucciarelli}, {Budnik}, {Burgess}, {Burgon}, {Burlacu}, {Busonero}, {Buzzi},
  {Caffau}, {Cambras}, {Campbell}, {Cancelliere}, {Cantat-Gaudin}, {Carlucci},
  {Carrasco}, {Castellani}, {Charlot}, {Charnas}, {Charvet}, {Chassat},
  {Chiavassa}, {Clotet}, {Cocozza}, {Collins}, {Collins}, {Costigan}, {Crifo},
  {Cross}, {Crosta}, {Crowley}, {Dafonte}, {Damerdji}, {Dapergolas}, {David},
  {David}, {De Cat}, {de Felice}, {de Laverny}, {De Luise}, {De March}, {de
  Martino}, {de Souza}, {Debosscher}, {del Pozo}, {Delbo}, {Delgado},
  {Delgado}, {di Marco}, {Di Matteo}, {Diakite}, {Distefano}, {Dolding}, {Dos
  Anjos}, {Drazinos}, {Dur{\'a}n}, {Dzigan}, {Ecale}, {Edvardsson}, {Enke},
  {Erdmann}, {Escolar}, {Espina}, {Evans}, {Eynard Bontemps}, {Fabre},
  {Fabrizio}, {Faigler}, {Falc{\~a}o}, {Farr{\`a}s Casas}, {Faye}, {Federici},
  {Fedorets}, {Fern{\'a}ndez-Hern{\'a}ndez}, {Fernique}, {Fienga}, {Figueras},
  {Filippi}, {Findeisen}, {Fonti}, {Fouesneau}, {Fraile}, {Fraser}, {Fuchs},
  {Furnell}, {Gai}, {Galleti}, {Galluccio}, {Garabato}, {Garc{\'\i}a-Sedano},
  {Gar{\'e}}, {Garofalo}, {Garralda}, {Gavras}, {Gerssen}, {Geyer}, {Gilmore},
  {Girona}, {Giuffrida}, {Gomes}, {Gonz{\'a}lez-Marcos},
  {Gonz{\'a}lez-N{\'u}{\~n}ez}, {Gonz{\'a}lez-Vidal}, {Granvik}, {Guerrier},
  {Guillout}, {Guiraud}, {G{\'u}rpide}, {Guti{\'e}rrez-S{\'a}nchez}, {Guy},
  {Haigron}, {Hatzidimitriou}, {Haywood}, {Heiter}, {Helmi}, {Hobbs},
  {Hofmann}, {Holl}, {Holland}, {Hunt}, {Hypki}, {Icardi}, {Irwin}, {Jevardat
  de Fombelle}, {Jofr{\'e}}, {Jonker}, {Jorissen}, {Julbe}, {Karampelas},
  {Kochoska}, {Kohley}, {Kolenberg}, {Kontizas}, {Koposov}, {Kordopatis},
  {Koubsky}, {Kowalczyk}, {Krone-Martins}, {Kudryashova}, {Kull}, {Bachchan},
  {Lacoste-Seris}, {Lanza}, {Lavigne}, {Le Poncin-Lafitte}, {Lebreton},
  {Lebzelter}, {Leccia}, {Leclerc}, {Lecoeur-Taibi}, {Lemaitre}, {Lenhardt},
  {Leroux}, {Liao}, {Licata}, {Lindstr{\o}m}, {Lister}, {Livanou}, {Lobel},
  {L{\"o}ffler}, {L{\'o}pez}, {Lopez-Lozano}, {Lorenz}, {Loureiro},
  {MacDonald}, {Magalh{\~a}es Fernandes}, {Managau}, {Mann}, {Mantelet},
  {Marchal}, {Marchant}, {Marconi}, {Marie}, {Marinoni}, {Marrese},
  {Marschalk{\'o}}, {Marshall}, {Mart{\'\i}n-Fleitas}, {Martino}, {Mary},
  {Matijevi{\v{c}}}, {Mazeh}, {McMillan}, {Messina}, {Mestre}, {Michalik},
  {Millar}, {Miranda}, {Molina}, {Molinaro}, {Molinaro}, {Moln{\'a}r},
  {Moniez}, {Montegriffo}, {Monteiro}, {Mor}, {Mora}, {Morbidelli}, {Morel},
  {Morgenthaler}, {Morley}, {Morris}, {Mulone}, {Muraveva}, {Musella},
  {Narbonne}, {Nelemans}, {Nicastro}, {Noval}, {Ord{\'e}novic},
  {Ordieres-Mer{\'e}}, {Osborne}, {Pagani}, {Pagano}, {Pailler}, {Palacin},
  {Palaversa}, {Parsons}, {Paulsen}, {Pecoraro}, {Pedrosa}, {Pentik{\"a}inen},
  {Pereira}, {Pichon}, {Piersimoni}, {Pineau}, {Plachy}, {Plum}, {Poujoulet},
  {Pr{\v{s}}a}, {Pulone}, {Ragaini}, {Rago}, {Rambaux}, {Ramos-Lerate},
  {Ranalli}, {Rauw}, {Read}, {Regibo}, {Renk}, {Reyl{\'e}}, {Ribeiro},
  {Rimoldini}, {Ripepi}, {Riva}, {Rixon}, {Roelens}, {Romero-G{\'o}mez},
  {Rowell}, {Royer}, {Rudolph}, {Ruiz-Dern}, {Sadowski}, {Sagrist{\`a}
  Sell{\'e}s}, {Sahlmann}, {Salgado}, {Salguero}, {Sarasso}, {Savietto},
  {Schnorhk}, {Schultheis}, {Sciacca}, {Segol}, {Segovia}, {Segransan},
  {Serpell}, {Shih}, {Smareglia}, {Smart}, {Smith}, {Solano}, {Solitro},
  {Sordo}, {Soria Nieto}, {Souchay}, {Spagna}, {Spoto}, {Stampa}, {Steele},
  {Steidelm{\"u}ller}, {Stephenson}, {Stoev}, {Suess}, {S{\"u}veges}, {Surdej},
  {Szabados}, {Szegedi-Elek}, {Tapiador}, {Taris}, {Tauran}, {Taylor},
  {Teixeira}, {Terrett}, {Tingley}, {Trager}, {Turon}, {Ulla}, {Utrilla},
  {Valentini}, {van Elteren}, {Van Hemelryck}, {van Leeuwen}, {Varadi},
  {Vecchiato}, {Veljanoski}, {Via}, {Vicente}, {Vogt}, {Voss}, {Votruba},
  {Voutsinas}, {Walmsley}, {Weiler}, {Weingrill}, {Werner}, {Wevers},
  {Whitehead}, {Wyrzykowski}, {Yoldas}, {{\v{Z}}erjal}, {Zucker}, {Zurbach},
  {Zwitter}, {Alecu}, {Allen}, {Allende Prieto}, {Amorim},
  {Anglada-Escud{\'e}}, {Arsenijevic}, {Azaz}, {Balm}, {Beck}, {Bernstein},
  {Bigot}, {Bijaoui}, {Blasco}, {Bonfigli}, {Bono}, {Boudreault}, {Bressan},
  {Brown}, {Brunet}, {Bunclark}, {Buonanno}, {Butkevich}, {Carret}, {Carrion},
  {Chemin}, {Ch{\'e}reau}, {Corcione}, {Darmigny}, {de Boer}, {de Teodoro}, {de
  Zeeuw}, {Delle Luche}, {Domingues}, {Dubath}, {Fodor}, {Fr{\'e}zouls},
  {Fries}, {Fustes}, {Fyfe}, {Gallardo}, {Gallegos}, {Gardiol}, {Gebran},
  {Gomboc}, {G{\'o}mez}, {Grux}, {Gueguen}, {Heyrovsky}, {Hoar}, {Iannicola},
  {Isasi Parache}, {Janotto}, {Joliet}, {Jonckheere}, {Keil}, {Kim},
  {Klagyivik}, {Klar}, {Knude}, {Kochukhov}, {Kolka}, {Kos}, {Kutka}, {Lainey},
  {LeBouquin}, {Liu}, {Loreggia}, {Makarov}, {Marseille}, {Martayan},
  {Martinez-Rubi}, {Massart}, {Meynadier}, {Mignot}, {Munari}, {Nguyen},
  {Nordlander}, {Ocvirk}, {O'Flaherty}, {Olias Sanz}, {Ortiz}, {Osorio},
  {Oszkiewicz}, {Ouzounis}, {Palmer}, {Park}, {Pasquato}, {Peltzer}, {Peralta},
  {P{\'e}turaud}, {Pieniluoma}, {Pigozzi}, {Poels}, {Prat}, {Prod'homme},
  {Raison}, {Rebordao}, {Risquez}, {Rocca-Volmerange}, {Rosen}, {Ruiz-Fuertes},
  {Russo}, {Sembay}, {Serraller Vizcaino}, {Short}, {Siebert}, {Silva},
  {Sinachopoulos}, {Slezak}, {Soffel}, {Sosnowska}, {Strai{\v{z}}ys}, {ter
  Linden}, {Terrell}, {Theil}, {Tiede}, {Troisi}, {Tsalmantza}, {Tur},
  {Vaccari}, {Vachier}, {Valles}, {Van Hamme}, {Veltz}, {Virtanen}, {Wallut},
  {Wichmann}, {Wilkinson}, {Ziaeepour}, \& {Zschocke}}]{GAIAONE}
{Gaia Collaboration}, {Prusti}, T., {de Bruijne}, J.~H.~J., {et~al.} 2016,
  \aap, 595, A1

\bibitem[{{Gaia Collaboration} {et~al.}(2023){Gaia Collaboration}, {Vallenari},
  {Brown}, {Prusti}, {de Bruijne}, {Arenou}, {Babusiaux}, {Biermann},
  {Creevey}, {Ducourant}, {Evans}, {Eyer}, {Guerra}, {Hutton}, {Jordi},
  {Klioner}, {Lammers}, {Lindegren}, {Luri}, {Mignard}, {Panem}, {Pourbaix},
  {Randich}, {Sartoretti}, {Soubiran}, {Tanga}, {Walton}, {Bailer-Jones},
  {Bastian}, {Drimmel}, {Jansen}, {Katz}, {Lattanzi}, {van Leeuwen}, {Bakker},
  {Cacciari}, {Casta{\~n}eda}, {De Angeli}, {Fabricius}, {Fouesneau},
  {Fr{\'e}mat}, {Galluccio}, {Guerrier}, {Heiter}, {Masana}, {Messineo},
  {Mowlavi}, {Nicolas}, {Nienartowicz}, {Pailler}, {Panuzzo}, {Riclet}, {Roux},
  {Seabroke}, {Sordo}, {Th{\'e}venin}, {Gracia-Abril}, {Portell}, {Teyssier},
  {Altmann}, {Andrae}, {Audard}, {Bellas-Velidis}, {Benson}, {Berthier},
  {Blomme}, {Burgess}, {Busonero}, {Busso}, {C{\'a}novas}, {Carry}, {Cellino},
  {Cheek}, {Clementini}, {Damerdji}, {Davidson}, {de Teodoro}, {Nu{\~n}ez
  Campos}, {Delchambre}, {Dell'Oro}, {Esquej}, {Fern{\'a}ndez-Hern{\'a}ndez},
  {Fraile}, {Garabato}, {Garc{\'\i}a-Lario}, {Gosset}, {Haigron}, {Halbwachs},
  {Hambly}, {Harrison}, {Hern{\'a}ndez}, {Hestroffer}, {Hodgkin}, {Holl},
  {Jan{\ss}en}, {Jevardat de Fombelle}, {Jordan}, {Krone-Martins}, {Lanzafame},
  {L{\"o}ffler}, {Marchal}, {Marrese}, {Moitinho}, {Muinonen}, {Osborne},
  {Pancino}, {Pauwels}, {Recio-Blanco}, {Reyl{\'e}}, {Riello}, {Rimoldini},
  {Roegiers}, {Rybizki}, {Sarro}, {Siopis}, {Smith}, {Sozzetti}, {Utrilla},
  {van Leeuwen}, {Abbas}, {{\'A}brah{\'a}m}, {Abreu Aramburu}, {Aerts},
  {Aguado}, {Ajaj}, {Aldea-Montero}, {Altavilla}, {{\'A}lvarez}, {Alves},
  {Anders}, {Anderson}, {Anglada Varela}, {Antoja}, {Baines}, {Baker},
  {Balaguer-N{\'u}{\~n}ez}, {Balbinot}, {Balog}, {Barache}, {Barbato},
  {Barros}, {Barstow}, {Bartolom{\'e}}, {Bassilana}, {Bauchet}, {Becciani},
  {Bellazzini}, {Berihuete}, {Bernet}, {Bertone}, {Bianchi}, {Binnenfeld},
  {Blanco-Cuaresma}, {Blazere}, {Boch}, {Bombrun}, {Bossini}, {Bouquillon},
  {Bragaglia}, {Bramante}, {Breedt}, {Bressan}, {Brouillet}, {Brugaletta},
  {Bucciarelli}, {Burlacu}, {Butkevich}, {Buzzi}, {Caffau}, {Cancelliere},
  {Cantat-Gaudin}, {Carballo}, {Carlucci}, {Carnerero}, {Carrasco},
  {Casamiquela}, {Castellani}, {Castro-Ginard}, {Chaoul}, {Charlot}, {Chemin},
  {Chiaramida}, {Chiavassa}, {Chornay}, {Comoretto}, {Contursi}, {Cooper},
  {Cornez}, {Cowell}, {Crifo}, {Cropper}, {Crosta}, {Crowley}, {Dafonte},
  {Dapergolas}, {David}, {David}, {de Laverny}, {De Luise}, {De March}, {De
  Ridder}, {de Souza}, {de Torres}, {del Peloso}, {del Pozo}, {Delbo},
  {Delgado}, {Delisle}, {Demouchy}, {Dharmawardena}, {Di Matteo}, {Diakite},
  {Diener}, {Distefano}, {Dolding}, {Edvardsson}, {Enke}, {Fabre}, {Fabrizio},
  {Faigler}, {Fedorets}, {Fernique}, {Fienga}, {Figueras}, {Fournier},
  {Fouron}, {Fragkoudi}, {Gai}, {Garcia-Gutierrez}, {Garcia-Reinaldos},
  {Garc{\'\i}a-Torres}, {Garofalo}, {Gavel}, {Gavras}, {Gerlach}, {Geyer},
  {Giacobbe}, {Gilmore}, {Girona}, {Giuffrida}, {Gomel}, {Gomez},
  {Gonz{\'a}lez-N{\'u}{\~n}ez}, {Gonz{\'a}lez-Santamar{\'\i}a},
  {Gonz{\'a}lez-Vidal}, {Granvik}, {Guillout}, {Guiraud},
  {Guti{\'e}rrez-S{\'a}nchez}, {Guy}, {Hatzidimitriou}, {Hauser}, {Haywood},
  {Helmer}, {Helmi}, {Sarmiento}, {Hidalgo}, {Hilger}, {H{\l}adczuk}, {Hobbs},
  {Holland}, {Huckle}, {Jardine}, {Jasniewicz}, {Jean-Antoine Piccolo},
  {Jim{\'e}nez-Arranz}, {Jorissen}, {Juaristi Campillo}, {Julbe}, {Karbevska},
  {Kervella}, {Khanna}, {Kontizas}, {Kordopatis}, {Korn}, {K{\'o}sp{\'a}l},
  {Kostrzewa-Rutkowska}, {Kruszy{\'n}ska}, {Kun}, {Laizeau}, {Lambert},
  {Lanza}, {Lasne}, {Le Campion}, {Lebreton}, {Lebzelter}, {Leccia}, {Leclerc},
  {Lecoeur-Taibi}, {Liao}, {Licata}, {Lindstr{\o}m}, {Lister}, {Livanou},
  {Lobel}, {Lorca}, {Loup}, {Madrero Pardo}, {Magdaleno Romeo}, {Managau},
  {Mann}, {Manteiga}, {Marchant}, {Marconi}, {Marcos}, {Marcos Santos},
  {Mar{\'\i}n Pina}, {Marinoni}, {Marocco}, {Marshall}, {Martin Polo},
  {Mart{\'\i}n-Fleitas}, {Marton}, {Mary}, {Masip}, {Massari},
  {Mastrobuono-Battisti}, {Mazeh}, {McMillan}, {Messina}, {Michalik}, {Millar},
  {Mints}, {Molina}, {Molinaro}, {Moln{\'a}r}, {Monari}, {Mongui{\'o}},
  {Montegriffo}, {Montero}, {Mor}, {Mora}, {Morbidelli}, {Morel}, {Morris},
  {Muraveva}, {Murphy}, {Musella}, {Nagy}, {Noval}, {Oca{\~n}a}, {Ogden},
  {Ordenovic}, {Osinde}, {Pagani}, {Pagano}, {Palaversa}, {Palicio},
  {Pallas-Quintela}, {Panahi}, {Payne-Wardenaar}, {Pe{\~n}alosa Esteller},
  {Penttil{\"a}}, {Pichon}, {Piersimoni}, {Pineau}, {Plachy}, {Plum}, {Poggio},
  {Pr{\v{s}}a}, {Pulone}, {Racero}, {Ragaini}, {Rainer}, {Raiteri}, {Rambaux},
  {Ramos}, {Ramos-Lerate}, {Re Fiorentin}, {Regibo}, {Richards}, {Rios Diaz},
  {Ripepi}, {Riva}, {Rix}, {Rixon}, {Robichon}, {Robin}, {Robin}, {Roelens},
  {Rogues}, {Rohrbasser}, {Romero-G{\'o}mez}, {Rowell}, {Royer}, {Ruz Mieres},
  {Rybicki}, {Sadowski}, {S{\'a}ez N{\'u}{\~n}ez}, {Sagrist{\`a} Sell{\'e}s},
  {Sahlmann}, {Salguero}, {Samaras}, {Sanchez Gimenez}, {Sanna},
  {Santove{\~n}a}, {Sarasso}, {Schultheis}, {Sciacca}, {Segol}, {Segovia},
  {S{\'e}gransan}, {Semeux}, {Shahaf}, {Siddiqui}, {Siebert}, {Siltala},
  {Silvelo}, {Slezak}, {Slezak}, {Smart}, {Snaith}, {Solano}, {Solitro},
  {Souami}, {Souchay}, {Spagna}, {Spina}, {Spoto}, {Steele},
  {Steidelm{\"u}ller}, {Stephenson}, {S{\"u}veges}, {Surdej}, {Szabados},
  {Szegedi-Elek}, {Taris}, {Taylor}, {Teixeira}, {Tolomei}, {Tonello}, {Torra},
  {Torra}, {Torralba Elipe}, {Trabucchi}, {Tsounis}, {Turon}, {Ulla}, {Unger},
  {Vaillant}, {van Dillen}, {van Reeven}, {Vanel}, {Vecchiato}, {Viala},
  {Vicente}, {Voutsinas}, {Weiler}, {Wevers}, {Wyrzykowski}, {Yoldas}, {Yvard},
  {Zhao}, {Zorec}, {Zucker}, \& {Zwitter}}]{GAIATWO}
{Gaia Collaboration}, {Vallenari}, A., {Brown}, A.~G.~A., {et~al.} 2023, \aap,
  674, A1

\bibitem[{{Ganeshalingam} {et~al.}(2010){Ganeshalingam}, {Li}, {Filippenko},
  {Anderson}, {Foster}, {Gates}, {Griffith}, {Grigsby}, {Joubert}, {Leja},
  {Lowe}, {Macomber}, {Pritchard}, {Thrasher}, \& {Winslow}}]{Ganeshalingam10}
{Ganeshalingam}, M., {Li}, W., {Filippenko}, A.~V., {et~al.} 2010, \apjs, 190,
  418

\bibitem[{Handley {et~al.}(2015{\natexlab{a}})Handley, Hobson, \&
  Lasenby}]{Handley15b}
Handley, W.~J., Hobson, M.~P., \& Lasenby, A.~N. 2015{\natexlab{a}}, Monthly
  Notices of the Royal Astronomical Society: Letters, 450, L61–L65

\bibitem[{Handley {et~al.}(2015{\natexlab{b}})Handley, Hobson, \&
  Lasenby}]{Handley15a}
Handley, W.~J., Hobson, M.~P., \& Lasenby, A.~N. 2015{\natexlab{b}}, Monthly
  Notices of the Royal Astronomical Society, 453, 4385–4399

\bibitem[{{Hicken} {et~al.}(2009{\natexlab{a}}){Hicken}, {Challis}, {Jha},
  {Kirshner}, {Matheson}, {Modjaz}, {Rest}, {Wood-Vasey}, {Bakos}, {Barton},
  {Berlind}, {Bragg}, {Brice{\~n}o}, {Brown}, {Caldwell}, {Calkins}, {Cho},
  {Ciupik}, {Contreras}, {Dendy}, {Dosaj}, {Durham}, {Eriksen}, {Esquerdo},
  {Everett}, {Falco}, {Fernandez}, {Gaba}, {Garnavich}, {Graves}, {Green},
  {Groner}, {Hergenrother}, {Holman}, {Hradecky}, {Huchra}, {Hutchison},
  {Jerius}, {Jordan}, {Kilgard}, {Krauss}, {Luhman}, {Macri}, {Marrone},
  {McDowell}, {McIntosh}, {McNamara}, {Megeath}, {Mochejska}, {Munoz},
  {Muzerolle}, {Naranjo}, {Narayan}, {Pahre}, {Peters}, {Peterson}, {Rines},
  {Ripman}, {Roussanova}, {Schild}, {Sicilia-Aguilar}, {Sokoloski}, {Smalley},
  {Smith}, {Spahr}, {Stanek}, {Barmby}, {Blondin}, {Stubbs}, {Szentgyorgyi},
  {Torres}, {Vaz}, {Vikhlinin}, {Wang}, {Westover}, {Woods}, \&
  {Zhao}}]{Hicken09a}
{Hicken}, M., {Challis}, P., {Jha}, S., {et~al.} 2009{\natexlab{a}}, \apj, 700,
  331

\bibitem[{{Hicken} {et~al.}(2012){Hicken}, {Challis}, {Kirshner}, {Rest},
  {Cramer}, {Wood-Vasey}, {Bakos}, {Berlind}, {Brown}, {Caldwell}, {Calkins},
  {Currie}, {de Kleer}, {Esquerdo}, {Everett}, {Falco}, {Fernandez},
  {Friedman}, {Groner}, {Hartman}, {Holman}, {Hutchins}, {Keys}, {Kipping},
  {Latham}, {Marion}, {Narayan}, {Pahre}, {Pal}, {Peters}, {Perumpilly},
  {Ripman}, {Sipocz}, {Szentgyorgyi}, {Tang}, {Torres}, {Vaz}, {Wolk}, \&
  {Zezas}}]{Hicken12}
{Hicken}, M., {Challis}, P., {Kirshner}, R.~P., {et~al.} 2012, \apjs, 200, 12

\bibitem[{{Hicken} {et~al.}(2009{\natexlab{b}}){Hicken}, {Wood-Vasey},
  {Blondin}, {Challis}, {Jha}, {Kelly}, {Rest}, \& {Kirshner}}]{Hicken09b}
{Hicken}, M., {Wood-Vasey}, W.~M., {Blondin}, S., {et~al.} 2009{\natexlab{b}},
  \apj, 700, 1097

\bibitem[{Hinton \& Brout(2020)}]{PIPPIN}
Hinton, S. \& Brout, D. 2020, Journal of Open Source Software, 5, 2122

\bibitem[{{Hlozek} {et~al.}(2012){Hlozek}, {Kunz}, {Bassett}, {Smith},
  {Newling}, {Varughese}, {Kessler}, {Bernstein}, {Campbell}, {Dilday},
  {Falck}, {Frieman}, {Kuhlmann}, {Lampeitl}, {Marriner}, {Nichol}, {Riess},
  {Sako}, \& {Schneider}}]{Hlozek12}
{Hlozek}, R., {Kunz}, M., {Bassett}, B., {et~al.} 2012, \apj, 752, 79

\bibitem[{{Hoffman} \& {Gelman}(2011)}]{NUTS}
{Hoffman}, M.~D. \& {Gelman}, A. 2011, arXiv e-prints, arXiv:1111.4246

\bibitem[{{Hounsell} {et~al.}(2018){Hounsell}, {Scolnic}, {Foley}, {Kessler},
  {Miranda}, {Avelino}, {Bohlin}, {Filippenko}, {Frieman}, {Jha}, {Kelly},
  {Kirshner}, {Mandel}, {Rest}, {Riess}, {Rodney}, \&
  {Strolger}}]{Hounsell2018}
{Hounsell}, R., {Scolnic}, D., {Foley}, R.~J., {et~al.} 2018, \apj, 867, 23

\bibitem[{{Jha} {et~al.}(2006){Jha}, {Kirshner}, {Challis}, {Garnavich},
  {Matheson}, {Soderberg}, {Graves}, {Hicken}, {Alves}, {Arce}, {Balog},
  {Barmby}, {Barton}, {Berlind}, {Bragg}, {Brice{\~n}o}, {Brown}, {Buckley},
  {Caldwell}, {Calkins}, {Carter}, {Concannon}, {Donnelly}, {Eriksen},
  {Fabricant}, {Falco}, {Fiore}, {Garcia}, {G{\'o}mez}, {Grogin}, {Groner},
  {Groot}, {Haisch}, {Hartmann}, {Hergenrother}, {Holman}, {Huchra},
  {Jayawardhana}, {Jerius}, {Kannappan}, {Kim}, {Kleyna}, {Kochanek},
  {Koranyi}, {Krockenberger}, {Lada}, {Luhman}, {Luu}, {Macri}, {Mader},
  {Mahdavi}, {Marengo}, {Marsden}, {McLeod}, {McNamara}, {Megeath}, {Moraru},
  {Mossman}, {Muench}, {Mu{\~n}oz}, {Muzerolle}, {Naranjo}, {Nelson-Patel},
  {Pahre}, {Patten}, {Peters}, {Peters}, {Raymond}, {Rines}, {Schild},
  {Sobczak}, {Spahr}, {Stauffer}, {Stefanik}, {Szentgyorgyi}, {Tollestrup},
  {V{\"a}is{\"a}nen}, {Vikhlinin}, {Wang}, {Willner}, {Wolk}, {Zajac}, {Zhao},
  \& {Stanek}}]{Jha06}
{Jha}, S., {Kirshner}, R.~P., {Challis}, P., {et~al.} 2006, \aj, 131, 527

\bibitem[{{Kenworthy} {et~al.}(2025){Kenworthy}, {Goobar}, {Jones},
  {Johansson}, {Thorp}, {Kessler}, {Burgaz}, {Dhawan}, {Dimitriadis},
  {Galbany}, {Ginolin}, {Kim}, {Maguire}, {M{\"u}ller-Bravo}, {Nugent},
  {Nordin}, {Popovic}, {Pessi}, {Rigault}, {Rosnet}, {Sollerman}, {Terwel},
  {Townsend}, {Laher}, {Purdum}, {Rosselli}, \& {Rusholme}}]{Kenworthy2025}
{Kenworthy}, W.~D., {Goobar}, A., {Jones}, D.~O., {et~al.} 2025, arXiv
  e-prints, arXiv:2502.09713

\bibitem[{{Kenworthy} {et~al.}(2021){Kenworthy}, {Jones}, {Dai}, {Kessler},
  {Scolnic}, {Brout}, {Siebert}, {Pierel}, {Dettman}, {Dimitriadis}, {Foley},
  {Jha}, {Pan}, {Riess}, {Rodney}, \& {Rojas-Bravo}}]{Kenworthy21}
{Kenworthy}, W.~D., {Jones}, D.~O., {Dai}, M., {et~al.} 2021, \apj, 923, 265

\bibitem[{Kessler {et~al.}(2024)Kessler, Brout, \& Jones}]{SNDATA_ROOT}
Kessler, R., Brout, D., \& Jones, D. 2024, SNDATA\_ROOT for SNANA software

\bibitem[{{Kessler} \& {Scolnic}(2017)}]{Kessler16}
{Kessler}, R. \& {Scolnic}, D. 2017, \apj, 836, 56

\bibitem[{{Koleva} \& {Vazdekis}(2012)}]{Koleva12}
{Koleva}, M. \& {Vazdekis}, A. 2012, \aap, 538, A143

\bibitem[{Krisciunas {et~al.}(2020)Krisciunas, Contreras, Burns, Phillips,
  Hamuy, Stritzinger, Anais, Boldt, Busta, Campillay, Castellón, Folatelli,
  Freedman, González, Hsiao, Krzeminski, Morrell, Persson, Roth, Salgado,
  Serón, Suntzeff, Torres, Filippenko, Li, Madore, Depoy, Marshall, Rheault,
  \& Villanueva}]{Krisciunas20}
Krisciunas, K., Contreras, C., Burns, C.~R., {et~al.} 2020, The Astronomical
  Journal, 160, 289

\bibitem[{{Krisciunas} {et~al.}(2017){Krisciunas}, {Contreras}, {Burns},
  {Phillips}, {Stritzinger}, {Morrell}, {Hamuy}, {Anais}, {Boldt}, {Busta},
  {Campillay}, {Castell{\'o}n}, {Folatelli}, {Freedman}, {Gonz{\'a}lez},
  {Hsiao}, {Krzeminski}, {Persson}, {Roth}, {Salgado}, {Ser{\'o}n}, {Suntzeff},
  {Torres}, {Filippenko}, {Li}, {Madore}, {DePoy}, {Marshall}, {Rheault}, \&
  {Villanueva}}]{Krisciunas17}
{Krisciunas}, K., {Contreras}, C., {Burns}, C.~R., {et~al.} 2017, \aj, 154, 211

\bibitem[{{Landolt}(1992)}]{Landolt92}
{Landolt}, A.~U. 1992, \aj, 104, 340

\bibitem[{{Marriner} {et~al.}(2011){Marriner}, {Bernstein}, {Kessler},
  {Lampeitl}, {Miquel}, {Mosher}, {Nichol}, {Sako}, {Schneider}, \&
  {Smith}}]{Marriner11}
{Marriner}, J., {Bernstein}, J.~P., {Kessler}, R., {et~al.} 2011, \apj, 740, 72

\bibitem[{{M{\"o}ller} \& {de Boissi{\`e}re}(2019)}]{Moller19}
{M{\"o}ller}, A. \& {de Boissi{\`e}re}, T. 2019, arXiv e-prints,
  arXiv:1901.06384

\bibitem[{{Narayan} {et~al.}(2019){Narayan}, {Matheson}, {Saha}, {Axelrod},
  {Calamida}, {Olszewski}, {Claver}, {Mandel}, {Bohlin}, {Holberg}, {Deustua},
  {Rest}, {Stubbs}, {Shanahan}, {Vaz}, {Zenteno}, {Strampelli}, {Hubeny},
  {Points}, {Sabbi}, \& {Mackenty}}]{narayan2019}
{Narayan}, G., {Matheson}, T., {Saha}, A., {et~al.} 2019, \apjs, 241, 20

\bibitem[{{Oke} \& {Gunn}(1983)}]{Oke83}
{Oke}, J.~B. \& {Gunn}, J.~E. 1983, \apj, 266, 713

\bibitem[{{Perlmutter} {et~al.}(1999){Perlmutter}, {Aldering}, {Goldhaber},
  {Knop}, {Nugent}, {Castro}, {Deustua}, {Fabbro}, {Goobar}, {Groom}, {Hook},
  {Kim}, {Kim}, {Lee}, {Nunes}, {Pain}, {Pennypacker}, {Quimby}, {Lidman},
  {Ellis}, {Irwin}, {McMahon}, {Ruiz-Lapuente}, {Walton}, {Schaefer}, {Boyle},
  {Filippenko}, {Matheson}, {Fruchter}, {Panagia}, {Newberg}, {Couch}, \&
  {Project}}]{Perlmutter99}
{Perlmutter}, S., {Aldering}, G., {Goldhaber}, G., {et~al.} 1999, \apj, 517,
  565

\bibitem[{{Poole} {et~al.}(2008){Poole}, {Breeveld}, {Page}, {Landsman},
  {Holland}, {Roming}, {Kuin}, {Brown}, {Gronwall}, {Hunsberger}, {Koch},
  {Mason}, {Schady}, {vanden Berk}, {Blustin}, {Boyd}, {Broos}, {Carter},
  {Chester}, {Cucchiara}, {Hancock}, {Huckle}, {Immler}, {Ivanushkina},
  {Kennedy}, {Marshall}, {Morgan}, {Pandey}, {de Pasquale}, {Smith}, \&
  {Still}}]{Poole08}
{Poole}, T.~S., {Breeveld}, A.~A., {Page}, M.~J., {et~al.} 2008, \mnras, 383,
  627

\bibitem[{{Popovic} {et~al.}(2023){Popovic}, {Scolnic}, {Vincenzi}, {Sullivan},
  {Brout}, {Sanchez}, {Chen}, {Patel}, {Peterson}, {Kessler}, {Kelsey},
  {Bailey}, {Wiseman}, \& {Toy}}]{Amalgame}
{Popovic}, B., {Scolnic}, D., {Vincenzi}, M., {et~al.} 2023, arXiv e-prints,
  arXiv:2309.05654

\bibitem[{{Rest} {et~al.}(2014){Rest}, {Scolnic}, {Foley}, {Huber}, {Chornock},
  {Narayan}, {Tonry}, {Berger}, {Soderberg}, {Stubbs}, {Riess}, {Kirshner},
  {Smartt}, {Schlafly}, {Rodney}, {Botticella}, {Brout}, {Challis}, {Czekala},
  {Drout}, {Hudson}, {Kotak}, {Leibler}, {Lunnan}, {Marion}, {McCrum},
  {Milisavljevic}, {Pastorello}, {Sanders}, {Smith}, {Stafford}, {Thilker},
  {Valenti}, {Wood-Vasey}, {Zheng}, {Burgett}, {Chambers}, {Denneau}, {Draper},
  {Flewelling}, {Hodapp}, {Kaiser}, {Kudritzki}, {Magnier}, {Metcalfe},
  {Price}, {Sweeney}, {Wainscoat}, \& {Waters}}]{Rest14}
{Rest}, A., {Scolnic}, D., {Foley}, R.~J., {et~al.} 2014, \apj, 795, 44

\bibitem[{{Riess} {et~al.}(1998){Riess}, {Filippenko}, {Challis},
  {Clocchiatti}, {Diercks}, {Garnavich}, {Gilliland}, {Hogan}, {Jha},
  {Kirshner}, {Leibundgut}, {Phillips}, {Reiss}, {Schmidt}, {Schommer},
  {Smith}, {Spyromilio}, {Stubbs}, {Suntzeff}, \& {Tonry}}]{Riess98}
{Riess}, A.~G., {Filippenko}, A.~V., {Challis}, P., {et~al.} 1998, \aj, 116,
  1009

\bibitem[{{Riess} {et~al.}(1999){Riess}, {Kirshner}, {Schmidt}, {Jha},
  {Challis}, {Garnavich}, {Esin}, {Carpenter}, {Grashius}, {Schild}, {Berlind},
  {Huchra}, {Prosser}, {Falco}, {Benson}, {Brice{\~n}o}, {Brown}, {Caldwell},
  {dell'Antonio}, {Filippenko}, {Goodman}, {Grogin}, {Groner}, {Hughes},
  {Green}, {Jansen}, {Kleyna}, {Luu}, {Macri}, {McLeod}, {McLeod}, {McNamara},
  {McLean}, {Milone}, {Mohr}, {Moraru}, {Peng}, {Peters}, {Prestwich},
  {Stanek}, {Szentgyorgyi}, \& {Zhao}}]{Riess99}
{Riess}, A.~G., {Kirshner}, R.~P., {Schmidt}, B.~P., {et~al.} 1999, \aj, 117,
  707

\bibitem[{{Rigault} {et~al.}(2025){Rigault}, {Smith}, {Goobar}, {Maguire},
  {Dimitriadis}, {Johansson}, {Nordin}, {Burgaz}, {Dhawan}, {Sollerman},
  {Regnault}, {Kowalski}, {Nugent}, {Andreoni}, {Amenouche}, {Aubert},
  {Barjou-Delayre}, {Bautista}, {Bellm}, {Betoule}, {Bloom}, {Carreres},
  {Chen}, {Copin}, {Deckers}, {de Jaeger}, {Feinstein}, {Fouchez}, {Fremling},
  {Galbany}, {Ginolin}, {Graham}, {Groom}, {Harvey}, {Kasliwal}, {Kenworthy},
  {Kim}, {Kuhn}, {Kulkarni}, {Lacroix}, {Laher}, {Masci}, {M{\"u}ller-Bravo},
  {Miller}, {Osman}, {Perley}, {Popovic}, {Purdum}, {Qin}, {Racine}, {Reusch},
  {Riddle}, {Rosnet}, {Rosselli}, {Ruppin}, {Senzel}, {Rusholme}, {Schweyer},
  {Terwel}, {Townsend}, {Tzanidakis}, {Wold}, \& {Yan}}]{Rigault24}
{Rigault}, M., {Smith}, M., {Goobar}, A., {et~al.} 2025, \aap, 694, A1

\bibitem[{{Rubin} {et~al.}(2023){Rubin}, {Aldering}, {Betoule}, {Fruchter},
  {Huang}, {Kim}, {Lidman}, {Linder}, {Perlmutter}, {Ruiz-Lapuente}, \&
  {Suzuki}}]{Rubin23}
{Rubin}, D., {Aldering}, G., {Betoule}, M., {et~al.} 2023, arXiv e-prints,
  arXiv:2311.12098

\bibitem[{{Rykoff} {et~al.}(2023){Rykoff}, {Tucker}, {Burke}, {Allam},
  {Bechtol}, {Bernstein}, {Brout}, {Gruendl}, {Lasker}, {Smith}, {Wester},
  {Yanny}, {Abbott}, {Aguena}, {Alves}, {Andrade-Oliveira}, {Annis}, {Bacon},
  {Bertin}, {Brooks}, {Carnero Rosell}, {Carretero}, {Castander}, {Choi}, {da
  Costa}, {Pereira}, {Davis}, {De Vicente}, {Diehl}, {Doel}, {Drlica-Wagner},
  {Everett}, {Ferrero}, {Frieman}, {Garc{\'\i}a-Bellido}, {Giannini}, {Gruen},
  {Gutierrez}, {Hinton}, {Hollowood}, {James}, {Kuehn}, {Lahav}, {Marshall},
  {Mena-Fern{\'a}ndez}, {Menanteau}, {Myles}, {Nord}, {Ogando}, {Palmese},
  {Pieres}, {Plazas Malag{\'o}n}, {Raveri}, {Rodgr{\'\i}guez-Monroy},
  {Sanchez}, {Santiago}, {Schubnell}, {Sevilla-Noarbe}, {Smith},
  {Soares-Santos}, {Suchyta}, {Swanson}, {Varga}, {Vincenzi}, {Walker},
  {Weaverdyck}, \& {Wiseman}}]{Rykoff23}
{Rykoff}, E.~S., {Tucker}, D.~L., {Burke}, D.~L., {et~al.} 2023, arXiv
  e-prints, arXiv:2305.01695

\bibitem[{{S{\'a}nchez} {et~al.}(2021){S{\'a}nchez}, {Kessler}, {Scolnic},
  {Armstrong}, {Biswas}, {Bogart}, {Chiang}, {Cohen-Tanugi}, {Fouchez}, {Gris},
  {Heitmann}, {Hlo{\v{z}}ek}, {Jha}, {Kelly}, {Liu}, {Narayan}, {Racine},
  {Rykoff}, {Sullivan}, {Walter}, {Wood-Vasey}, \& {The LSST Dark Energy
  Science Collaboration}}]{Sanchez21}
{S{\'a}nchez}, B., {Kessler}, R., {Scolnic}, D., {et~al.} 2021, arXiv e-prints,
  arXiv:2111.06858

\bibitem[{{Schlafly} \& {Finkbeiner}(2011)}]{Schlafly11}
{Schlafly}, E.~F. \& {Finkbeiner}, D.~P. 2011, \apj, 737, 103

\bibitem[{{Schlafly} {et~al.}(2016){Schlafly}, {Meisner}, {Stutz},
  {Kainulainen}, {Peek}, {Tchernyshyov}, {Rix}, {Finkbeiner}, {Covey}, {Green},
  {Bell}, {Burgett}, {Chambers}, {Draper}, {Flewelling}, {Hodapp}, {Kaiser},
  {Magnier}, {Martin}, {Metcalfe}, {Wainscoat}, \& {Waters}}]{Schlafly16}
{Schlafly}, E.~F., {Meisner}, A.~M., {Stutz}, A.~M., {et~al.} 2016, \apj, 821,
  78

\bibitem[{{Scolnic} {et~al.}(2015){Scolnic}, {Casertano}, {Riess}, {Rest},
  {Schlafly}, {Foley}, {Finkbeiner}, {Tang}, {Burgett}, {Chambers}, {Draper},
  {Flewelling}, {Hodapp}, {Huber}, {Kaiser}, {Kudritzki}, {Magnier},
  {Metcalfe}, \& {Stubbs}}]{Scolnic15}
{Scolnic}, D., {Casertano}, S., {Riess}, A., {et~al.} 2015, \apj, 815, 117

\bibitem[{{Scolnic} {et~al.}(2018){Scolnic}, {Kessler}, {Brout},
  {Cowperthwaite}, {Soares-Santos}, {Annis}, {Herner}, {Chen}, {Sako},
  {Doctor}, {Butler}, {Palmese}, {Diehl}, {Frieman}, {Holz}, {Berger},
  {Chornock}, {Villar}, {Nicholl}, {Biswas}, {Hounsell}, {Foley}, {Metzger},
  {Rest}, {Garc{\'{\i}}a-Bellido}, {M{\"o}ller}, {Nugent}, {Abbott}, {Abdalla},
  {Allam}, {Bechtol}, {Benoit-L{\'e}vy}, {Bertin}, {Brooks}, {Buckley-Geer},
  {Carnero Rosell}, {Carrasco Kind}, {Carretero}, {Castander}, {Cunha},
  {D{'}Andrea}, {da Costa}, {Davis}, {Doel}, {Drlica-Wagner}, {Eifler},
  {Flaugher}, {Fosalba}, {Gaztanaga}, {Gerdes}, {Gruen}, {Gruendl}, {Gschwend},
  {Gutierrez}, {Hartley}, {Honscheid}, {James}, {Johnson}, {Johnson}, {Krause},
  {Kuehn}, {Kuhlmann}, {Lahav}, {Li}, {Lima}, {Maia}, {March}, {Marshall},
  {Menanteau}, {Miquel}, {Neilsen}, {Plazas}, {Sanchez}, {Scarpine},
  {Schubnell}, {Sevilla-Noarbe}, {Smith}, {Smith}, {Sobreira}, {Suchyta},
  {Swanson}, {Tarle}, {Thomas}, {Tucker}, {Walker}, \& {DES
  Collaboration}}]{Scolnic18}
{Scolnic}, D., {Kessler}, R., {Brout}, D., {et~al.} 2018, \apjl, 852, L3

\bibitem[{{Scolnic} {et~al.}(2014){Scolnic}, {Riess}, {Foley}, {Rest},
  {Rodney}, {Brout}, \& {Jones}}]{Scolnic14a}
{Scolnic}, D.~M., {Riess}, A.~G., {Foley}, R.~J., {et~al.} 2014, \apj, 780, 37

\bibitem[{{Smith} {et~al.}(2002){Smith}, {Tucker}, {Kent}, {Richmond},
  {Fukugita}, {Ichikawa}, {Ichikawa}, {Jorgensen}, {Uomoto}, {Gunn}, {Hamabe},
  {Watanabe}, {Tolea}, {Henden}, {Annis}, {Pier}, {McKay}, {Brinkmann}, {Chen},
  {Holtzman}, {Shimasaku}, \& {York}}]{Smith02}
{Smith}, J.~A., {Tucker}, D.~L., {Kent}, S., {et~al.} 2002, \aj, 123, 2121

\bibitem[{Stahl {et~al.}(2019)Stahl, Zheng, de~Jaeger, Filippenko, Bigley,
  Blanchard, Blanchar\~d, Brink, Cargill, Casper, Channa, Choi, Choksi, Chu,
  Cohen, Ellison, Falcon, Fazeli, Fuller, Ganeshalingam, Gates, Gould, Halevi,
  Hayakawa, Hestenes, Jeffers, Joubert, Kandrashoff, Kim, Kim, Kislak, Kleiser,
  Kong, de~Kouchkovsky, Krishnan, Leja, Leonard, Li, Li, Lu, Mason, Molloy,
  Pina, Rex, Ross, Stegman, Tang, Thrasher, Wang, Wilkins, Yuk, Yunus, \&
  Zhang}]{Stahl19}
Stahl, B.~E., Zheng, W., de~Jaeger, T., {et~al.} 2019, Monthly Notices of the
  Royal Astronomical Society, 490, 3882

\bibitem[{{Stubbs} \& {Tonry}(2012)}]{StubbsTonry12}
{Stubbs}, C.~W. \& {Tonry}, J.~L. 2012, arXiv e-prints, arXiv:1206.6695

\bibitem[{{Tang} {et~al.}(2025){Tang}, {Brout}, {Karwal}, {Chang}, {Miranda},
  \& {Vincenzi}}]{Tang25}
{Tang}, X.~T., {Brout}, D., {Karwal}, T., {et~al.} 2025, \apjl, 983, L27

\bibitem[{{Taylor} {et~al.}(2021){Taylor}, {Lidman}, {Tucker}, {Brout},
  {Hinton}, \& {Kessler}}]{Taylor21}
{Taylor}, G., {Lidman}, C., {Tucker}, B.~E., {et~al.} 2021, \mnras, 504, 4111

\bibitem[{{The LSST Dark Energy Science Collaboration} {et~al.}(2018){The LSST
  Dark Energy Science Collaboration}, {Mandelbaum}, {Eifler}, {Hlo{\v{z}}ek},
  {Collett}, {Gawiser}, {Scolnic}, {Alonso}, {Awan}, {Biswas}, {Blazek},
  {Burchat}, {Chisari}, {Dell'Antonio}, {Digel}, {Frieman}, {Goldstein},
  {Hook}, {Ivezi{\'c}}, {Kahn}, {Kamath}, {Kirkby}, {Kitching}, {Krause},
  {Leget}, {Marshall}, {Meyers}, {Miyatake}, {Newman}, {Nichol}, {Rykoff},
  {Sanchez}, {Slosar}, {Sullivan}, \& {Troxel}}]{LSSTSRD}
{The LSST Dark Energy Science Collaboration}, {Mandelbaum}, R., {Eifler}, T.,
  {et~al.} 2018, arXiv e-prints, arXiv:1809.01669

\bibitem[{{Tonry} {et~al.}(2012){Tonry}, {Stubbs}, {Lykke}, {Doherty},
  {Shivvers}, {Burgett}, {Chambers}, {Hodapp}, {Kaiser}, {Kudritzki},
  {Magnier}, {Morgan}, {Price}, \& {Wainscoat}}]{Tonry12}
{Tonry}, J.~L., {Stubbs}, C.~W., {Lykke}, K.~R., {et~al.} 2012, \apj, 750, 99

\bibitem[{{Tripp}(1998)}]{Tripp98}
{Tripp}, R. 1998, \aap, 331, 815

\bibitem[{{Vincenzi} {et~al.}(2024){Vincenzi}, {Brout}, {Armstrong}, {Popovic},
  {Taylor}, {Acevedo}, {Camilleri}, {Chen}, {Davis}, {Lee}, {Lidman}, {Hinton},
  {Kelsey}, {Kessler}, {M{\"o}ller}, {Qu}, {Sako}, {Sanchez}, {Scolnic},
  {Smith}, {Sullivan}, {Wiseman}, {Asorey}, {Bassett}, {Carollo}, {Carr},
  {Foley}, {Frohmaier}, {Galbany}, {Glazebrook}, {Graur}, {Kovacs}, {Kuehn},
  {Malik}, {Nichol}, {Rose}, {Tucker}, {Toy}, {Tucker}, {Yuan}, {Abbott},
  {Aguena}, {Alves}, {Allam}, {Andrade-Oliveira}, {Annis}, {Bacon}, {Bechtol},
  {Bernstein}, {Brooks}, {Burke}, {Carnero Rosell}, {Carretero}, {Castander},
  {Conselice}, {da Costa}, {Pereira}, {Desai}, {Diehl}, {Doel}, {Ferrero},
  {Flaugher}, {Friedel}, {Frieman}, {Garc{\'\i}a-Bellido}, {Gatti}, {Giannini},
  {Gruen}, {Gruendl}, {Hollowood}, {Honscheid}, {Huterer}, {James},
  {Kuropatkin}, {Lahav}, {Lee}, {Lin}, {Marshall}, {Mena-Fern{\'a}ndez},
  {Menanteau}, {Miquel}, {Palmese}, {Pieres}, {Plazas Malag{\'o}n}, {Porredon},
  {Romer}, {Roodman}, {Sanchez}, {Sanchez Cid}, {Schubnell}, {Sevilla-Noarbe},
  {Suchyta}, {Swanson}, {Tarle}, {To}, {Walker}, {Weaverdyck}, \&
  {Yamamoto}}]{DES5YR}
{Vincenzi}, M., {Brout}, D., {Armstrong}, P., {et~al.} 2024, \apj, 975, 86

\bibitem[{{Xiao} {et~al.}(2023){Xiao}, {Yuan}, {Huang}, {Zhang}, {Yang}, \&
  {Xu}}]{Xiao23}
{Xiao}, K., {Yuan}, H., {Huang}, B., {et~al.} 2023, \apjs, 268, 53

\end{thebibliography}

\bibliographystyle{aa.bst}

\end{document}